# Systematic comparison of ISOLDE-SC yields with calculated in-target production rates

**S. Lukić,[1] F. Gevaert, A. Kelić, M. V. Ricciardi, K.-H. Schmidt, O. Yordanov**

*GSI, Planckstr. 1, 64291 Darmstadt, Germany*

***Abstract***

Recently, a series of dedicated inverse-kinematics experiments performed at GSI, Darmstadt, has brought an important progress in our understanding of proton and heavy-ion induced reactions at relativistic energies. The nuclear reaction code ABRABLA that has been developed and benchmarked against the results of these experiments has been used to calculate nuclide production cross sections at different energies and with different targets and beams. These calculations are used to estimate nuclide production rates by protons in thick targets, taking into account the energy loss and the attenuation of the proton beam in the target, as well as the low-energy fission induced by the secondary neutrons. The results are compared to the yields of isotopes of various elements obtained from different targets at CERN-ISOLDE with 600 MeV protons, and the overall extraction efficiencies are deduced. The dependence of these extraction efficiencies on the nuclide half-life is found to follow a simple pattern in many different cases. A simple function is proposed to parameterize this behavior in a way that quantifies the essential properties of the extraction efficiency for the element and the target – ion-source system in question.

---

[1] Corresponding author.
Strahinja Lukić
GSI
Planckstr. 1
64291 Darmstadt
GERMANY
Phone: +49 (0)6159 71 24 30
Fax: +49 (0)6159 71 29 02
E-mail: s.lukic@gsi.de

## 1. Introduction

The ISOL (Isotope Separation On-Line) method is widely used for producing nuclear beams of radioactive isotopes. In this method, radioactive nuclides are produced by nuclear reactions in thick targets and released from these by thermal diffusion. The research with ISOL beams has a long tradition at CERN ISOLDE [1]. Within the scope of the EURISOL project [2], the design of a large-scale European secondary-beam facility based on the ISOL method is actually being worked out. One of the tasks defined in the project is devoted to the predictions of the nuclide yields of the future facility, based on measured reaction data and theoretical reaction and transport calculations.

There are several different processes that determine the final yields of an ISOL facility:

- Nuclide production in nuclear reactions in the target
- Diffusion of the reaction products to the surface of the target
- Desorption from the surface
- Effusion through the transfer line to the ion source
- Ionization
- Transport of the ion to the experimental site

The produced residues diffuse from the target and are ionized in order to separate them using the standard mass-spectrometry techniques. Part of the produced nuclides is lost during diffusion, effusion, ionization and transport of the reaction products, e.g. due to chemical reactions, sticking to the walls, condensation, leaks in the assembly or because atoms escape without being ionized. Some of these mechanisms are chemically selective and can be used to suppress impurities [3].

During the time required for the reaction products to reach the experimental setup, beta decay leads to additional losses of short-lived isotopes and to additional production of the corresponding isobars. All these processes disturb a direct relation between the in-target production rates and the final yields. It is of interest to establish information on systematic tendencies in their overall effect, particularly in function of the isotope half-life. The magnitude of the overall losses of nuclides in processes like chemical reactions, sticking to the walls or condensation is difficult to measure or to estimate independently while, on the other hand, the most important information for practical applications is their overall effect. Besides, the information on the behavior of the overall efficiencies with short half-lives can help identifying the issues that need most attention in the process of target and ion-source development. This information is of great value for the design of future secondary-beam facilities, based on the ISOL method, like the EURISOL [2] or the RIA [4,5] facility.

The overall extraction efficiency for individual nuclides has already been directly obtained by measuring the extracted yields of stable [6] or radioactive [7] tracers implanted into a target. However, this method requires a dedicated experimental effort for every nuclide and every target and ion-source system. On the other hand, a comparison of systematically measured ISOL yields, such as those that are documented in the ISOLDE database, with reliable estimates of the in-target production rates can provide an exhaustive overview of the efficiencies of, in principle, any available ISOL beam. In this way, it is possible to study the dependence of the extraction efficiency on the isotopic half-life over a large number of cases and to establish systematic tendencies. Such a study can help providing quantitative estimates on the achievements obtained in the long-term operation of the existing ISOL facilities and identifying the needs for further development of extraction methods for certain elements.

Before moving to its present setup with the Proton Synchrotron Booster (PSB) accelerator, ISOLDE has been successfully operated for almost 30 years with the 600 MeV continuous proton beam from the Synchrocyclotron (SC) accelerator in CERN impinging on a set of different production targets. During this time, an extended data base of nuclide yields has been established [1]. Recently, a series of dedicated inverse-kinematics experiments [8,9,10,11,12,13,14,15,16,17,18,19,20,21,22,23,24,25,26,27], performed at GSI, Darmstadt, have brought an important progress in our understanding of the nuclide production in reactions induced by protons and heavy ions at relativistic energies. Profiting from these data, improved nuclear reaction codes have been developed that allow performing reliable calculations on the nuclide production rates. In the present work, we combine all this information in order to investigate the relation between the in-target production and the yields recorded at the ISOLDE facility. The in-target production calculations rely mostly on the nuclide production cross sections induced by the primary protons, taking into account the energy loss and attenuation of the beam along the target. Additional nuclide production induced by secondary neutrons in fissile targets is also taken into account.

The data considered in this work refer to the nuclide production by the 600 MeV SC proton beam in CERN. The more recent ISOLDE yields obtained since 1992 with the PSB proton beams in the energy range of 1 to 1.4 GeV are generally higher than those obtained with the SC beam. That is partly due to the higher in-target production rates, but also to some more elaborate extraction techniques. These techniques are constantly being improved, and the work on the new yield database is underway. An analysis similar to this one could be done with the new yield data in the future.

## 2. Nuclide production and extraction mechanisms in ISOL targets

The primary reactions occurring in ISOL targets can be subdivided in two groups:

- Proton-induced fragmentation reactions
- Proton-induced fission reactions

In a proton-induced fragmentation reaction a few nucleons from the target nucleus are removed by the high-energy proton during the intra-nuclear-cascade stage. Since the produced pre-fragment is highly excited, this stage is followed by the evaporation of protons, neutrons, clusters like alpha particles or by the emission of gamma rays. In the proton-induced fragmentation of non-fissile stable nuclei a large variety of mainly neutron-deficient radioactive isotopes of different masses are produced.

In the proton-induced fission, two fragments are produced from the excited pre-fragment. In the case of high-energy fission, the symmetric split of the pre-fragment is the most probable. However, the distribution of the products covers a large part of the nuclide chart and is extended down to very light elements as Na and Ne [19]. In general, fission of heavy fissile nuclides as $^{238}$U, $^{232}$Th is a powerful tool that gives access to very neutron-rich isotopes [28].

In addition, reactions induced by secondary protons and neutrons may also contribute to the nuclide production in the target. Since the average energies of the secondary particles are significantly lower than those of the primary projectiles, the distribution of secondary reaction products on the chart of the nuclides differs considerably from the primary production.

Obviously, the in-target nuclide production can be optimized by choosing a target material and the incident particle type and energy so that the production cross sections for the desired nuclide are high. Besides, the type of the target, i.e. its chemistry, composition and density, also affects the extraction. That is why different kinds of targets have been elaborated: carbides, metal powder, thin foils, liquid metal targets etc. Low-density and porous targets provide higher yields for the short-lived isotopes, because the diffusion process is much faster. A disadvantage of low-density targets is the lower primary production rate per target volume. On the other hand, dense targets are better suited for the production of long-lived nuclides, since the extracted amount is less affected by the slower diffusion process. Furthermore, to provide a faster diffusion of the residues, the temperature of the target must be high. Assuring stable physical properties of the target at high temperatures represents a special technical challenge. Often composite targets like carbides are used to provide the thermal stability up to high temperatures. Therefore, extraction optimization is another important consideration in the choice of the target material [29,30,31,32,33,34,35].

Depending on the element which has to be extracted, different ion sources are used [3,35,36]:

- Surface ion sources for elements with low ionization potential (positive) or for elements with high electron affinity (negative)
- Plasma ion sources for the elements that cannot be surface ionized
- Laser ion sources for the elements that require chemical selectivity at the ionization step because of the presence of a strong isobaric contamination.

After the ionization, the particles are accelerated in an electrostatic field and are separated using a mass separator. Using the selectivity of the ion source and the mass separator, isotopes of a certain element are extracted and can be used for producing the exotic beam.

During the extraction process, a part of the short-lived reaction products is lost due to radioactive decays. Fast extraction is, therefore, crucial for the work with very short-lived nuclides. On the other hand, the final yield of a nucleus can also be enhanced by the decay of corresponding precursors. This effect, commonly known as the side feeding, depends on both the mother and daughter half-lives, but also on the ratio of the respective production rates. When the production reactions are induced by high-energy protons, the influence of the side feeding from β decay is usually not large because the precursors are nuclei that are situated further away from the stability line and produced with smaller cross sections.[2] On the other hand, the side feeding from α decay can have very large influence when the nuclide in question is produced by spallation-evaporation of uranium or thorium. Many of the different α emitters that are produced in these reactions have production cross sections that are comparable or much higher than those of their respective α-decay daughters. Some examples from this region will be treated in this work.

These facts help us understand the difference between the useful yields at the experimental area and the in-target production rates.

## 3. Model calculations

Experimental nuclide production cross-section data have been measured at GSI in inverse kinematics for several different projectile and target types, mostly at the energy of 1 *A* GeV [8-27]. Since experimental cross-section data do not exist for all the ISOLDE target materials studied in this work, and the energy of the SC proton beam was 600 MeV, we have to rely on model calculations to account for the variation of the prefragment properties with decreasing beam energy and different types of target.

The evaluation of the in-target production rates requires employing computational programs capable of describing proton-nucleus collisions with high predictive power. Among all possible models, intra-nuclear-cascade (INC) models, followed by statistical de-excitation models, can rather well describe nucleon-nucleus collisions in the energy range we are interested in.

---

[2] In low-energy neutron-induced reactions, the peak of the production by fission is shifted to the neutron-rich side, and the side-feeding of more stable nuclei by β decay can be significant.

In this work, we used the ABRABLA [37, 38, 39, 40] nuclear reaction code to evaluate the in-target production rates. ABRABLA has continuously been developed at GSI during the last 10 years. It is a Monte-Carlo code that simulates both the nucleus-nucleus and the nucleon-nucleus collisions at relativistic energies assuming that the reaction can be divided in two stages. The first stage is an interaction stage, where the target nucleus loses part of its nucleons and is left in an excited state. This stage is followed by a deexcitation cascade where evaporation and fission are in competition. In the nucleus-nucleus collisions the interaction stage is described in terms of a geometrical "abrasion" picture [37]. The interaction stage in nucleon-nucleus collisions is treated by an analytical parameterization of the results of an intranuclear-cascade (INC) model. In the following sections we will shortly describe the physics contained in ABRABLA. Since in this work we are interested in proton-nucleus collisions only, we will not describe the abrasion model.

### *3.1 The nucleon-nucleus interaction stage in ABRABLA*

The description of the nucleon-nucleus interaction stage in ABRABLA is based on the INCL3 Monte Carlo model developed at Liège, Belgium [41]. Hereafter, we shortly recall the principal characteristics of INCL3.

In this model, the intranuclear-cascade stage is described as a sequence of nucleon-nucleon collisions and decays well separated in space and time. The impinging particle hits the target nucleus at a certain point at the nucleus surface, which is randomly sampled from the distribution of all possible values of the impact parameter. Coulomb deflection is taken into account. The initial positions of the target nucleons are set randomly in the spherical nuclear target volume with a sharp surface. The nucleus is considered as a degenerated Fermi gas. The nucleons travel along straight lines until two of them reach their minimum distance of approach, in which case they are scattered, or until they hit the border of the potential well, described by the nuclear mean field. Relativistic kinematics is used to determine the trajectories. The Pauli principle is considered, due to which the occupation of a fully filled state is forbidden for the nucleons emerging from a collision ("blocking effect"). When a nucleon collides with another nucleon, the gender of the latter is sampled according to the $N/Z$-ratio of the target nucleus and the momentum according to the corresponding Fermi distribution. If the two nucleons scatter, the conservation laws must be respected for energy, momenta and angular momenta. If a nucleon is kicked out of the nucleus, the mass and the charge also change. The diffusion angles are chosen using the experimental free-nucleon differential cross sections ($\mathrm{d}\sigma/\mathrm{d}\Omega$) corrected for in-medium effects. Inelastic collisions, leading to the formation of pions or other hadrons are also considered, and the following decay treated. The cascade stops after a certain time after which a thermally equilibrated system is formed.

Instead of simulating the INC stage, the state of the nucleus at the end of the INC stage can be sampled from a parameterized distribution in excitation energy, angular momentum, mass number and atomic number depending on the proton energy and the mass of the target nucleus. This approach makes the code much faster, the INC being its most time-consuming part. The speed of the calculation is a crucial issue for our systematic investigation. For this reason, the analytical code BURST was developed based on the parameterization of the output results of the intranuclear-cascade stage predicted by INCL3 [41].

### *3.2 The de-excitation stage in ABRABLA*

The deexcitation part of ABRABLA, named ABLA, is a dynamical code that describes the de-excitation of the compound nucleus through the evaporation of light particles and fission. The particle evaporation is considered in the framework of the statistical model, where the probability for the emission of a certain particle is essentially given by the ratio of the available phase space in the daughter and in the mother nucleus. In the description of fission, the ABLA code explicitly treats the relaxation process in deformation space. The resulting time-dependent fission width is calculated using an analytical approximation [42] to the solution of the Fokker-Planck equation. The calculation of the fission yields is done on the basis of a semi-empirical model as described in ref. [38].

### 3.3 The evaporation model

If the acquired excitation energy of the compound nucleus is higher than the separation energy of a given particle, there is a certain probability that this particle will escape the nucleus. Neutrons and the light charged particles ($Z<3$) constitute the most abundant part of the evaporated particles. The evaporation of light nuclei (the so called "intermediate-mass-fragments" (IMF)) is also considered in ABLA. In both cases, the treatment of evaporation is based on the Weisskopf theory [43].

The probability for a given initial nucleus with excitation energy $E_i$ to decay into a final nucleus with $E_f$ by emission of a particle of type ν with mass $m_\nu$, spin $s_\nu \cdot \hbar$, and kinetic energy $\varepsilon_\nu$, depends on the decay width $\Gamma_\nu$ that can be expressed as [43]:

$$\Gamma_\nu(E_i) = \frac{2 \cdot s_\nu + 1}{2 \cdot \pi \cdot \rho_m(E_i)} \cdot \frac{2 \cdot m_\nu}{\pi \cdot \hbar^2} \cdot \int_0^{E_i - S_\nu} \sigma_{inv}(\varepsilon_\nu) \cdot \rho_d(E_f) \cdot (\varepsilon_\nu - B_\nu) dE_f \qquad 1$$

with $\varepsilon_\nu = E_i - S_\nu - E_f$, where $S_\nu$ is the binding energy of the evaporated particle, $\rho_m(E_i)$ and $\rho_d(E_f)$ are the level density of the mother and daughter nuclei, respectively, and $\sigma_{inv}$ is the inverse cross section – i.e. the cross section for the capture of the particle by the daughter nucleus.

In ABLA, the density of excited states is calculated with the Fermi-gas formula [44]:

$$\rho(E) = \frac{\sqrt{\pi}}{12} \frac{\exp(S)}{\tilde{a}^{1/4} E^{5/4}} \qquad 2$$

where $S$ is the entropy calculated as described in reference [37], $\tilde{a}$ is the asymptotic level-density parameter, as given in ref. [45], that takes into account deviations from spherical nuclei.

The decay width for the evaporation of IMF differs from that for the emission of light particles (eq. 1) basically because the available excited states of the IMF enter into the balance of the available phase space, and, consequently have to be considered in calculating the decay width:

$$\Gamma = \frac{m_{imf}}{\pi^2 \hbar^2} \int_0^{E_{imf}^{\max}} \int_0^{E_d^{\max}} \sigma_{inv} \frac{\rho_{imf}(E_{imf}) \cdot \rho_d(E_d)}{\rho_m(E)} (\varepsilon_{imf} - B_{imf}) dE_d dE_{imf} \qquad 3$$

with the following relation that guarantees the energy conservation:

$$E = E_{imf} + E_d + Q + \varepsilon - B \qquad 4$$

Here $E$, $E_{imf}$ and $E_d$ represent the initial excitation energy of the mother nucleus, and the excitation energies of the IMF and of the daughter nucleus, respectively. $Q$ is the Q-value, $\varepsilon$ is the total kinetic energy in the centre of mass of the system, and $B$ is the barrier for the IMF emission.

Finally, also $\gamma$ radiation is included as a possible channel. As the statistical emission of $\gamma$ rays occurs predominantly via the giant dipole resonance, the $\gamma$ radiation rate was calculated according to ref. [46]. For high excitation energy the $\gamma$ emission is negligible compared to the particle emission and it becomes important only at the energies around and below the particle separation energies.

In summary, the emission probability of the particle of type $\nu$ from the fragment with neutron number $N$, atomic number $Z$, and excitation energy $E$ is given by:

$$W_\nu = \frac{\Gamma_\nu(N,Z,E)}{\sum_k \Gamma_k(N,Z,E)} \qquad 5$$

with $k$ denoting all possible decay channels. The decay channels are: light-particle evaporation (n, p, d, t, $^3$He and $\alpha$), IMF emission, gamma emission and fission.

### *3.4 The fission model*

The evaporation process competes with another process: fission. Modelling of the fission decay width at high excitation energies requires the treatment of the evolution of the fission degree of freedom as a diffusion process, determined by the interaction of the fission collective degree of freedom with the heat bath formed by the individual nucleons [47,48]. Such process can be described by the Fokker-Planck equation (FPE) [49], where the variable is the time-dependent probability distribution as a function of the deformation in fission direction and its canonically conjugate momentum. A dominant parameter is the dissipation coefficient $\beta$ that rules the influence of the nuclear viscosity on the time needed for the deformation. The solution of the FPE leads to a time-dependent fission width $\Gamma_f(t)$. However, these numerical calculations are too time-consuming to be used in nuclear-reaction codes. Recently, a new description of the fission width based on the analytical solution of the FPE has been developed. Details and description of this treatment can be found in ref. [42]. The angular-momentum dependent fission barriers are taken from the finite-range liquid-drop model predictions of Sierk [50].

After the system passes a certain point (the "scission point") in the deformation space, the two fission fragments separate from each other[3]. The characteristics of the fission fragments are described within the fission code PROFI [38], used as a subroutine of the ABLA code. PROFI is a semi-empirical Monte-Carlo code developed to calculate the nuclide distributions of fission fragments. In the model, for a given excitation energy, $E^*$, the yield of the fission fragments with the mass number $A$, $Y(E^*, A)$, is determined by the number of available transition states above the potential energy at the outer fission saddle point. It is assumed that the mass-asymmetric degree of freedom at the fission barrier is on average uniquely related to the neutron number $N$ of the fission fragments. The mean value and the fluctuations of the $N/Z$ degree of freedom are calculated at the scission point. The barrier as function of the mass asymmetry is defined by three components (or “channels”): The symmetric component, defined by the liquid-drop potential, is described by a parabola. Two other components, the asymmetric ones, describe the neutron shells at $N$=82 and around $N\approx 90$ (introduced by Brosa et al. [51] as “standard I” and “standard II”). They modulate the parabolic potential with four negative Gaussian functions. At high excitation energies, the effects of the shells diminish and, finally, only the parabolic potential is reflected in the mass distribution of the fragments. The excitation energies of the fission fragments are calculated from the excitation and deformation energy of the fissioning system at the scission point. After scission, the two excited fission fragments will undergo an evaporation/γ-radiation cascade until they are completely cooled down. A detailed description of the model is given in [38].

[3] Ternary fission is not considered in ABLA, however it is an extremely improbable process [19].

### *3.5 Validation of the code with experimental results from GSI*

The ABRABLA code is being continuously benchmarked with the systematic high-quality experimental data measured in inverse kinematics at GSI. In these experiments, the formation cross sections of the produced nuclides, in the almost entire production range, were measured down to values of 10 μb with a typical accuracy of 15%. The tabulated experimental data can be found in reference [27]; details of the experimental technique and of the physical meaning of the results in references therein.

In the figure 1, the results of the code for the reaction of $^{238}$U with protons at 1 *A* GeV are compared with the experimental data for the reaction of the same system [19-26,27]. In the upper part, the comparison is made on the chart of nuclides, with individual nuclide production cross sections represented by colors. In the lower part, the same comparison is made in a compact form of diagrams of the mean *N*/*Z* ratio and of the width of the isotopic distributions versus the element number. In this way, the essential characteristics of the isotopic distributions, obtained using the code, are compared to the experimental values. The agreement is very good over the whole range of the produced elements.

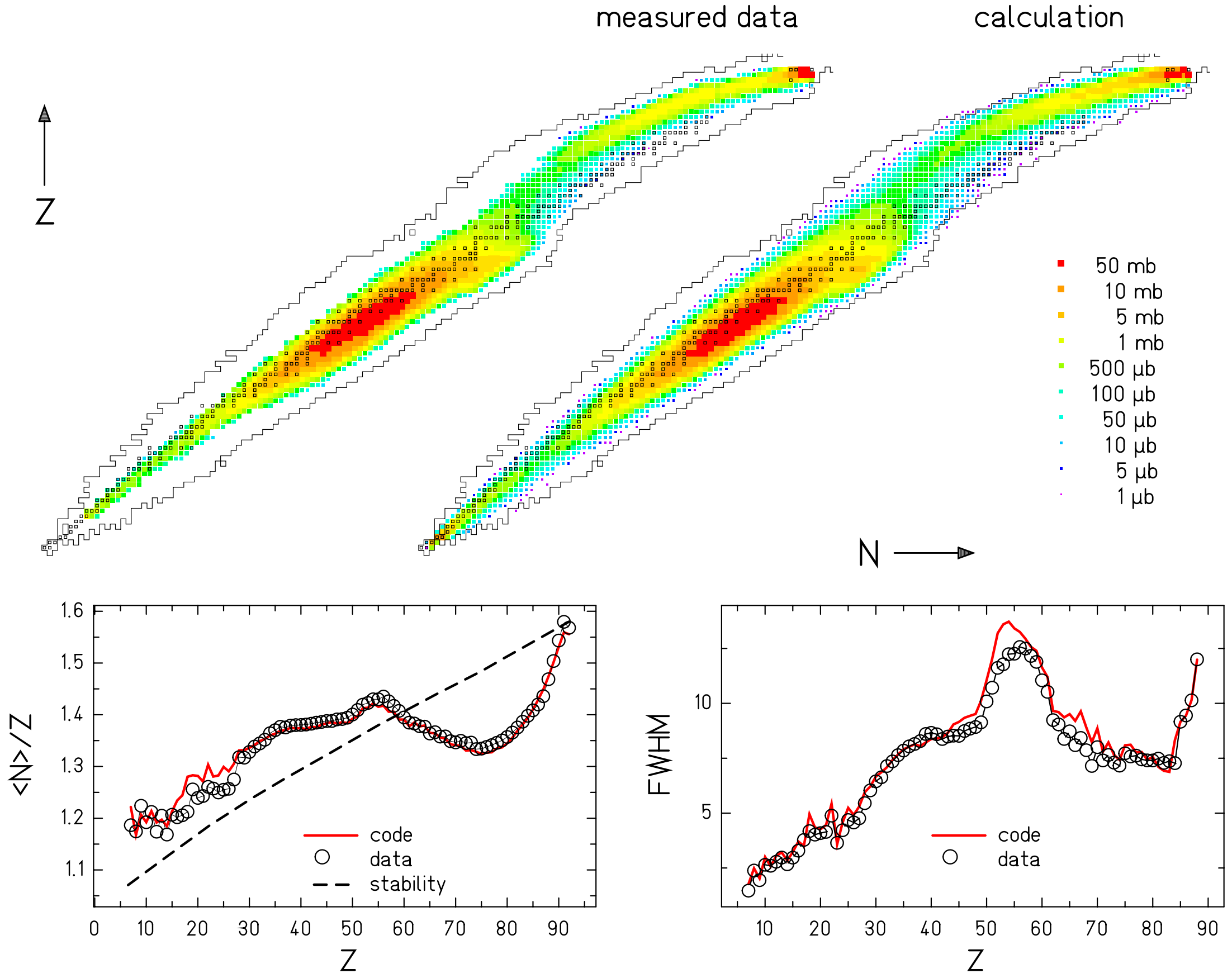

**Figure 1: Comparison of the experimental data with the ABRABLA code calculations for the nuclide production cross sections in the reaction $^{238}$U (1 *A* GeV) + p. Above: cross sections plotted as colors for each nuclide on the chart of nuclides. Note that the calculation reaches down to lower cross sections than the experimental data. Below: mean *N*/*Z* ratio and width of the isotopic distributions versus the element number. The dots represent the experimental data and the solid lines the predictions of ABRABLA.**

The ABRABLA code has also been benchmarked against the measured nuclide production cross sections in the following reactions: $^{208}$Pb + $^{1}$H at 500 *A* MeV and 1 *A* GeV, $^{179}$Au + $^{1}$H at 800 *A* MeV, $^{136}$Xe + $^{1}$H at 1 *A* GeV and $^{56}$Fe + $^{1}$H between 300 *A* MeV and 1.5 *A* GeV [8-17,27]

## 4. The thick target

When one wants to calculate production rates in a thick target, one must take into account the attenuation and the energy loss of the primary beam in the target, as well as the contribution of the reactions induced by the secondary particles.

The attenuation of the beam occurs via the nuclear reactions. Considering every particle that has undergone any nuclear reaction as removed from the beam, the intensity of the beam at a certain point $x$ along the target was calculated according to the following equation:

$$I(x) = I_0 \exp\left( -\frac{\rho_{tg} x \sigma_{tot} N_a}{A_{tg}} \right) \quad 6$$

$I_0$ : Intensity of the beam before the target
$\rho_{tg}$ : Target density in g/cm$^3$
$\sigma_{tot}$ : The total interaction cross section, calculated using a Glauber calculation [52,53]
$A_{tg}$ : Atomic mass of the target material
$N_a$ : Avogadro Number

The primary production rate for a nuclide with a production cross section $\sigma_i$ is given by the integral:

$$N_i^{primary} = \int_0^X \frac{\rho_{tg} \sigma_i N_a I_0}{A_{tg}} \cdot \exp\left( -\frac{\rho_{tg} x \sigma_{tot} N_a}{A_{tg}} \right) dx \quad 7$$

As the beam loses its energy via electromagnetic interactions, predominantly with the electrons, in the target material, the total reaction cross section and the nuclide production cross sections are not constant along the target. As for the total reaction cross section, its variation is usually negligible. For example, in the case of a 170 g/cm$^2$ lead target, which is one of the thickest targets used at ISOLDE, its overall variation is about 4%. On the other hand, the individual nuclide production cross sections can change by up to a factor of two and, in some cases, even more. Assuming that its variation is linear along the target, we solve the integral in equation 7 using a constant value of $\sigma_i$ obtained as the arithmetic average between its value at the initial proton energy and the energy of the protons at the end of the target. We assume that the error we introduce in this way is of acceptable size. However, in the case of some spallation-evaporation products that are close to the target, e.g. mercury produced in a 170 g/cm$^2$ molten lead target, the shape of the nuclide distribution shifts significantly to the neutron-rich side as the beam energy drops, and individual nuclide production cross sections on the neutron-deficient side can drop by up to one order of magnitude. In cases where individual nuclide production cross sections change by more than a factor of two along the target, we solve the integral by dividing the target in a sufficient number of slices. Thus we obtain:

$$N_i^{primary} = I_0 \frac{\overline{\sigma}_i}{\sigma_{tot}} \left( 1 - \exp\left( -\frac{\rho_{tg} x \sigma_{tot} N_a}{A_{tg}} \right) \right) \quad 8$$

Here $\overline{\sigma}_i$ stands for the effective nuclide production cross section obtained by averaging its value or, if necessary, by slicing the target. $d_{tg}$ is the target density in g/cm$^2$.

The energy loss of protons was calculated with the aid of the AMADEUS program as documented in [54].

### *4.1 Secondary reactions*

In the reaction of high-energy protons with a thick target, neutrons are produced with cross sections that are typically one order of magnitude higher than those for the production of the secondary charged particles [55,56]. The cross sections for the reactions induced by the secondary neutrons are also generally higher than those of the secondary charged particles. Moreover, due to electronic stopping, the range of the charged particles in the target is much shorter than that of the neutrons. Therefore, we assume that we can approximate the secondary nuclide production by that induced by neutrons only. Comparing GSI and ISOLDE data for the reaction of protons with uranium, U. Köster et al. have found that the difference in the shape of the produced isotopic distributions is mostly due to the low-energy fission in the thick target [57]. This is in agreement with the measured energy spectrum of secondary neutrons created by a 600 MeV proton beam in uranium [56]. The mean energy of the neutrons produced in this way is around 2 MeV.

Individual nuclide production rates in secondary reactions might be determined using dedicated transport codes. We have chosen an alternative approach based on ISOLDE measured data. We assume that the secondary reaction rate can be approximated by the neutron capture rate $N^{capture}$. The probability $P_{A,Z}^{fission}$ of producing the nucleus *A,Z* in a neutron-capture reaction can be estimated using the ABLA code to simulate the deexcitation stage of the compound nucleus formed in the neutron capture with fixed excitation energy equal to the sum of the binding energy of the neutron in the compound nucleus and the average secondary neutron energy. By multiplying this distribution by $N_{capture}$, one gets the individual nuclide production rates in low-energy fission:

$$N_{A,Z}^{fission} = N^{capture} P_{A,Z}^{fission} \tag{9}$$

Thus, the total nuclide production rate can be expressed as:

$$N_{A,Z} = N_{A,Z}^{primary} + N^{capture} P_{A,Z}^{fission} \tag{10}$$

The neutron capture rate depends on the target type and geometry. It can be obtained from equation 10, by comparing the calculated primary in-target production rates (eq. 8) to known total in-target production rates for an isotopic chain in the specified target. ISOLDE in-target production rates are available for the case of krypton in uranium-carbide target of the same geometry as in the SC yield database, but with higher thickness and for the primary beam energy of 1.4 GeV [58]. Krypton is close to the light peak of the low-energy fission-fragment distribution, which makes it a suitable choice for the estimation of the neutron capture rate.

We first compare our estimation of the primary production rates for krypton at 1.4 GeV to the ISOLDE total in-target production estimates from *A*=87 to *A*=99 in the figure 2.

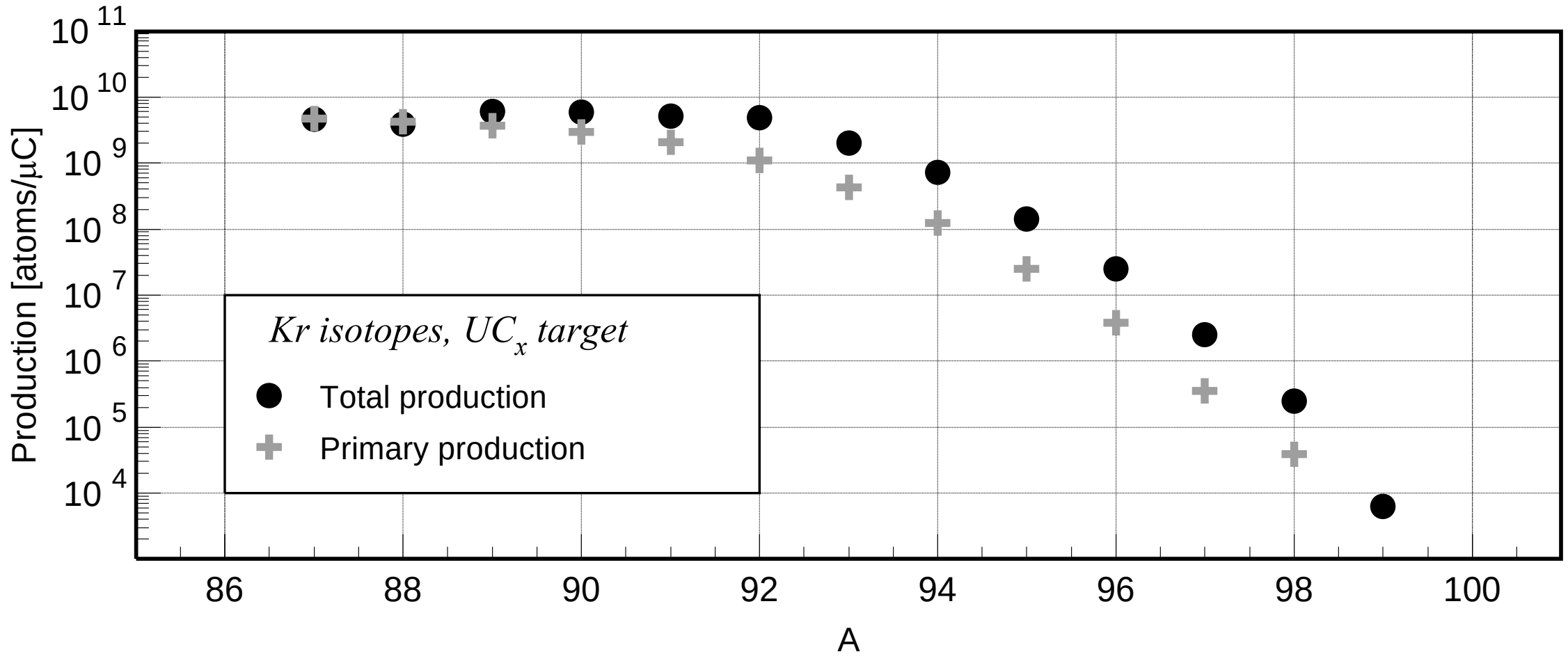

**Figure 2: Comparison of our calculation of primary production rates to ISOLDE total in-target production estimates for Kr isotopes in a $UC_x$ target bombarded by 1.4 GeV protons.**

From $A$=89 on, the total in-target production rate is significantly enhanced by the secondary reactions. The position of the enhancement in the very neutron-rich region confirms the hypothesis that it is mainly due to the low-energy fission. The ratio of the secondary to the primary production rate can be calculated for each isotope of the chain as:

$$\eta_A^{ISOLDE} = \frac{N_A^{ISOLDE} - N_A^{primary}}{N_A^{primary}} \qquad 11$$

$N_A^{ISOLDE}$ – Total in-target production rates obtained from krypton isotopic yields and release efficiencies [58].

In a similar way, we can denote by $\eta_A^{ABRABLA}$ the ratio of the secondary (low-energy fission) to the primary production rate obtained using ABRABLA for the isotope with mass $A$:[4]

$$\eta_A^{ABRABLA} = \frac{N^{capture} P_A^{fission}}{N_A^{primary}} \qquad 12$$

Under the simple condition that the average values of $\eta_A^{ISOLDE}$ and $\eta_A^{ABRABLA}$ in the mass range $A$=87 to $A$=99 should be equal, we obtain the expression for the neutron capture rate:

$$N^{capture} = \frac{\sum_A \eta_A^{ISOLDE}}{\sum_A \frac{P_A^{fission}}{N_A^{primary}}} \qquad 13$$

The obtained neutron capture rate per μC of incident protons is $2.6 \times 10^{11}$ $\mu C^{-1}$, which means 0.042 neutron captures per incident proton. One should note that in the reaction of 1.4 GeV protons with a $UC_x$ target of 46 $g/cm^2$, the probability for a proton to undergo a nuclear reaction is about 40%. Therefore, the ratio of the neutron capture rate to the primary reaction rate is about 0.1. However, one should recall that the distribution of the nuclides produced by the low-energy fission is concentrated in a relatively narrow area of the chart of the nuclides compared to the distribution of the primary-reaction residues. Thus, the contribution of the low-energy fission to the production rates for individual nuclides can be up to one order of magnitude larger than the contribution from the primary reactions (see figures 2,3).

[4] Here $P_A^{fission}$ and $N_A^{primary}$ are the same quantities as $P_{A,Z}^{fission}$ and $N_{A,Z}^{primary}$ in equations 9 and 10. We have deliberately omitted $Z$ in the notation because here we are discussing a specific isotopic chain.

Figure 3 represents a comparison of $\eta_A^{ISOLDE}$ and $\eta_A^{ABRABLA}$ from $A$=87 to $A$=99. One can notice that the overall distribution of $\eta_A^{ISOLDE}$ is slightly broader than that of $\eta_A^{ABRABLA}$, has a lower maximum and extends more toward the less neutron-rich isotopes. This is probably due to the reactions induced by the high-energy tail of the secondary-neutron energy distribution, as well as by the high-energy secondary protons. This effect introduces errors of less than 50% in the in-target production rates and will be neglected.

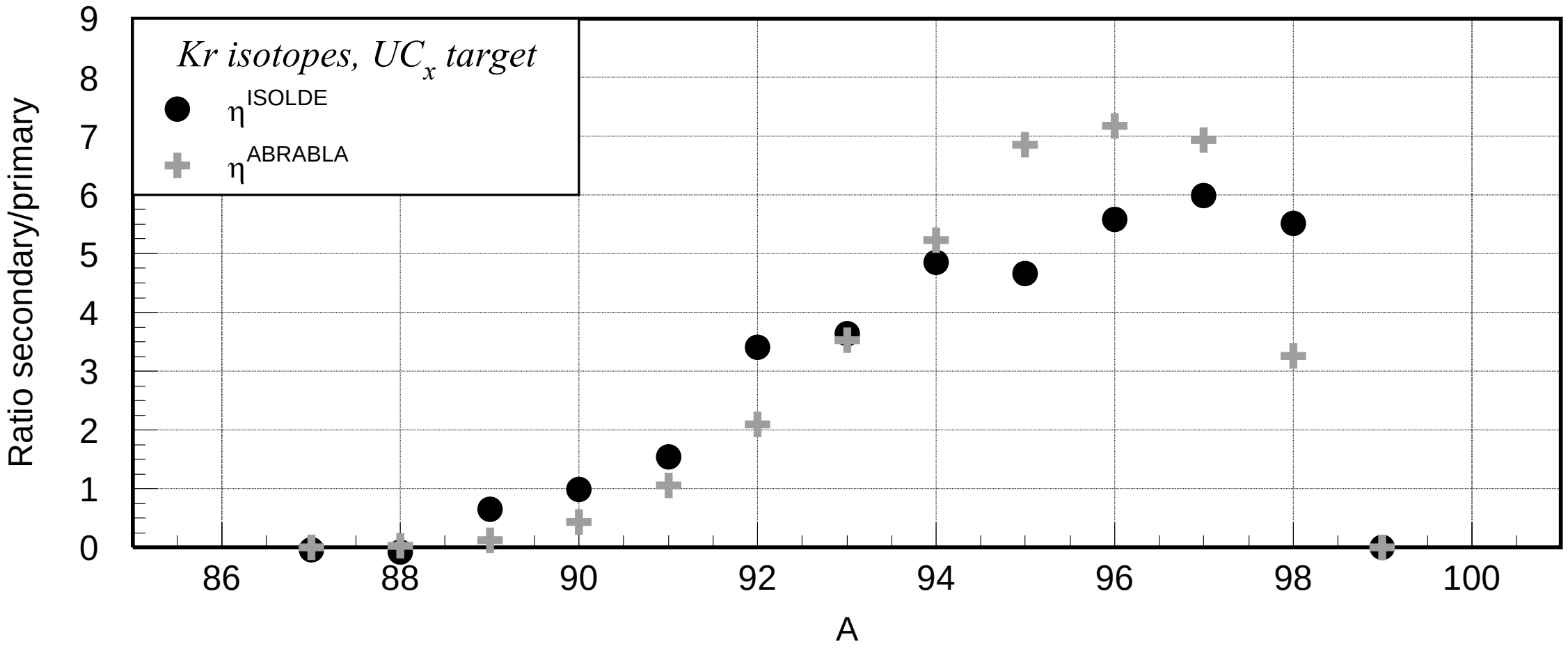

**Figure 3: Comparison of the secondary to the primary production ratios for Kr isotopes obtained from ISOLDE data and from ABRABLA calculations.**

In order to obtain the neutron-capture rate for the beam energy and target geometry from the SC yield database, we apply the following two considerations:

- According to ABRABLA calculations, at 1.4 GeV proton energy the secondary neutron production is higher by a factor of ~1.4 than at 600 MeV, but the neutron energy spectrum does not significantly change. To account for the difference in the proton beam energy, we divide the neutron capture rate by 1.4.
- The difference in geometry amounts to the difference in the target density, while the shape and the dimensions of the target remain the same [58,32]. The thickness of the $UC_x$ target in the EURISOL report [58] is 46 g/cm$^2$, while the thicknesses of the $UC_x$ targets in the ISOLDE yield database [1] range from 10 to 65 g/cm$^2$. In order to account for the difference in target densities, a scaling proportional to the target thickness was necessary.

The value for the neutron capture rate obtained in this way was used in this work to estimate the secondary reaction contribution to the production of any nuclide in uranium-carbide targets in the ISOLDE-SC database.

## 5. Dependence of the ratio of the yield to the in-target production rate on the isotope half-life

The ratio of the ISOLDE yield of a nuclide to its in-target production rate gives directly the overall extraction efficiency for that nuclide. The efficiency for an isotope of a specific element, extracted from a specific target-ion source system, is expected to depend, in the first order, only on the half-life of the isotope and not on its mass. It is our aim to study this dependence in cases of different elements produced and extracted from different targets and to establish and quantify any existing general tendencies. The way we proceed with this task will be illustrated on the example of francium produced in a uranium-carbide target. The francium isotopic chain is particularly suitable because the dependence of the half-lives on the mass number of its isotopes is rather non-uniform (see figure 5), which eliminates the possibility of confusing eventual mass-dependent systematic errors in the calculation of the in-target production rates with the real dependence of the extraction efficiencies on the isotope half-lives. Figure 4 represents the comparison of the francium isotopic production rates in a uranium-carbide target, calculated using the ABRABLA code, with the ISOLDE yields of these nuclides extracted from the same target. Both these quantities have been normalized to a proton beam intensity of 1 μC. The target contained 13 g of uranium per $cm^2$, and a tungsten-surface ion source was used for the extraction.

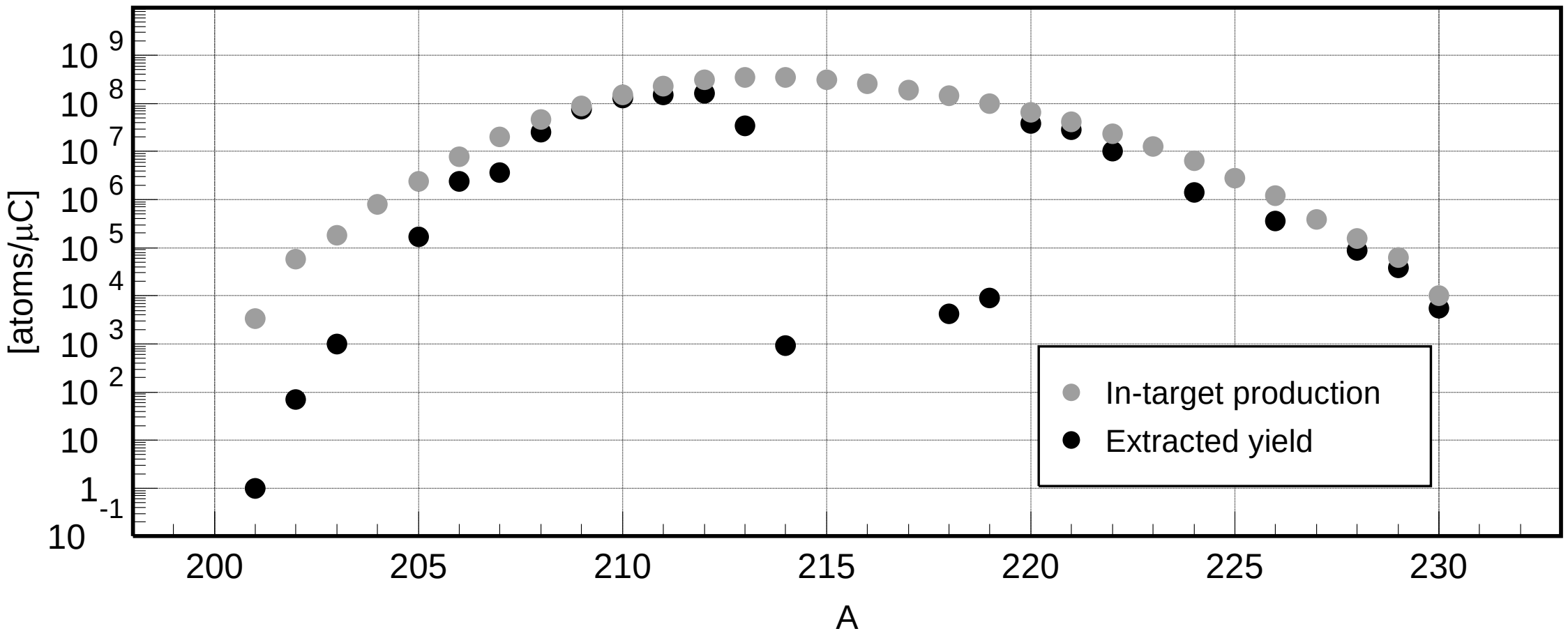

**Figure 4: Comparison of the Fr isotopic production rates in a $UC_x$ target, calculated using ABRABLA, with the ISOLDE yields of these nuclides extracted from the same target using a W-surface ion source. Both these quantities have been normalized to the proton beam intensity of 1 μC. The extraction efficiency of the isotopes is strongly affected by their half-lives (see figure 5).**

The ratio of the yields to the in-target production rate is compared to the respective isotopic half-lives in figure 5. The extraction efficiency is lower for the shorter-lived isotopes. For the half-lives longer than, roughly, 10 s the extraction efficiency reaches saturation. Note that for the extremely short-lived isotopes ($t_{½} < 1$ ms) there is no yield information, because the extraction of such isotopes in measurable quantities would be extremely difficult.

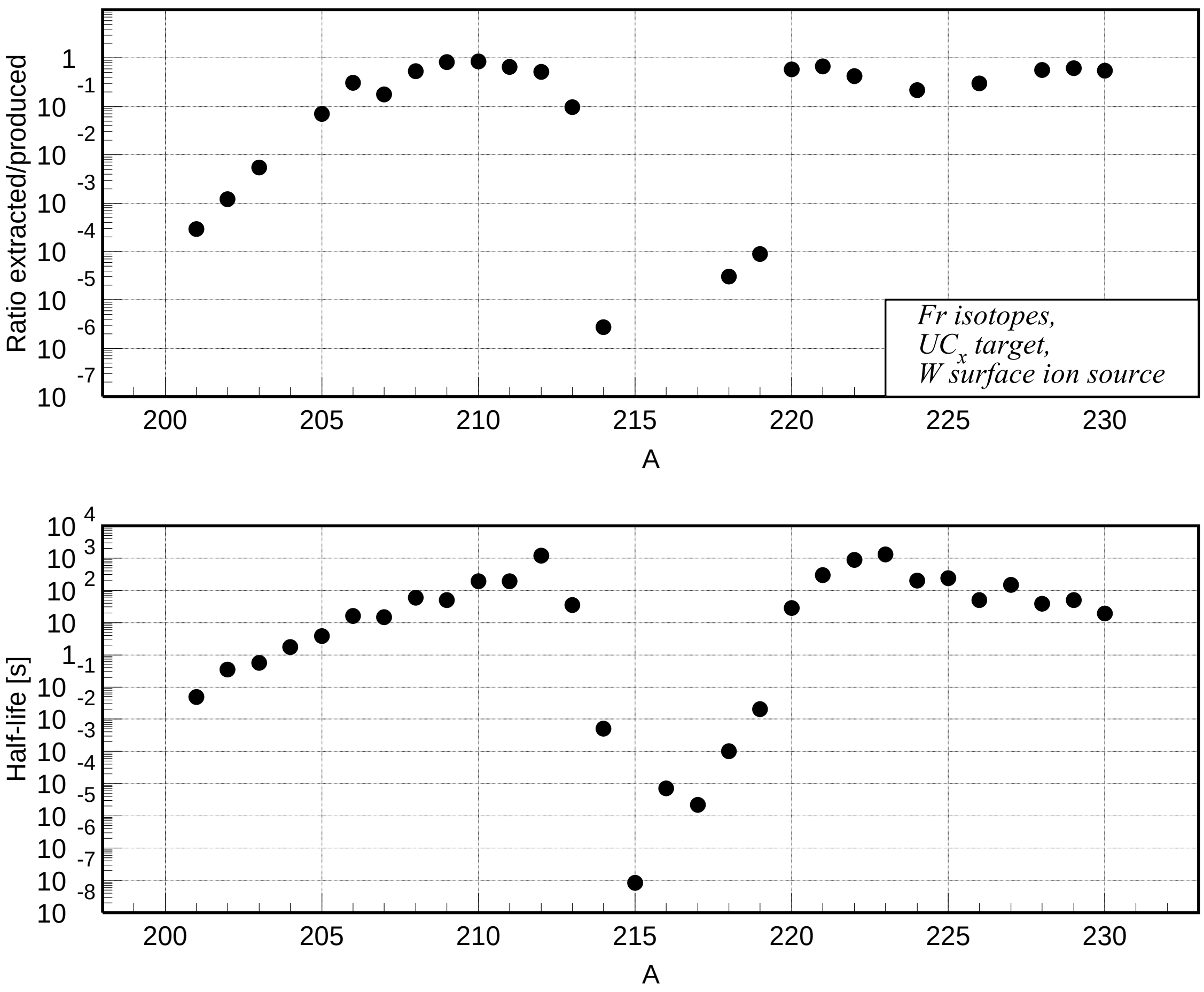

**Figure 5: Comparison of the overall extraction efficiency of Fr isotopes produced in a UC$_x$ target (up) to the half-lives of the respective isotopes (down). The extraction efficiency is lower for the shorter-lived isotopes. For the half-lives longer than ~10 s, the extraction efficiency reaches saturation.**

The extraction efficiencies are plotted in function of the francium isotope half-life in figure 6. For the shortest-lived isotopes, the extraction efficiency scales with a power function of the isotope half-life. For longer half-lives, the efficiency reaches a saturation value. This behavior is present in many other cases of various elements obtained from different target – ion source systems.

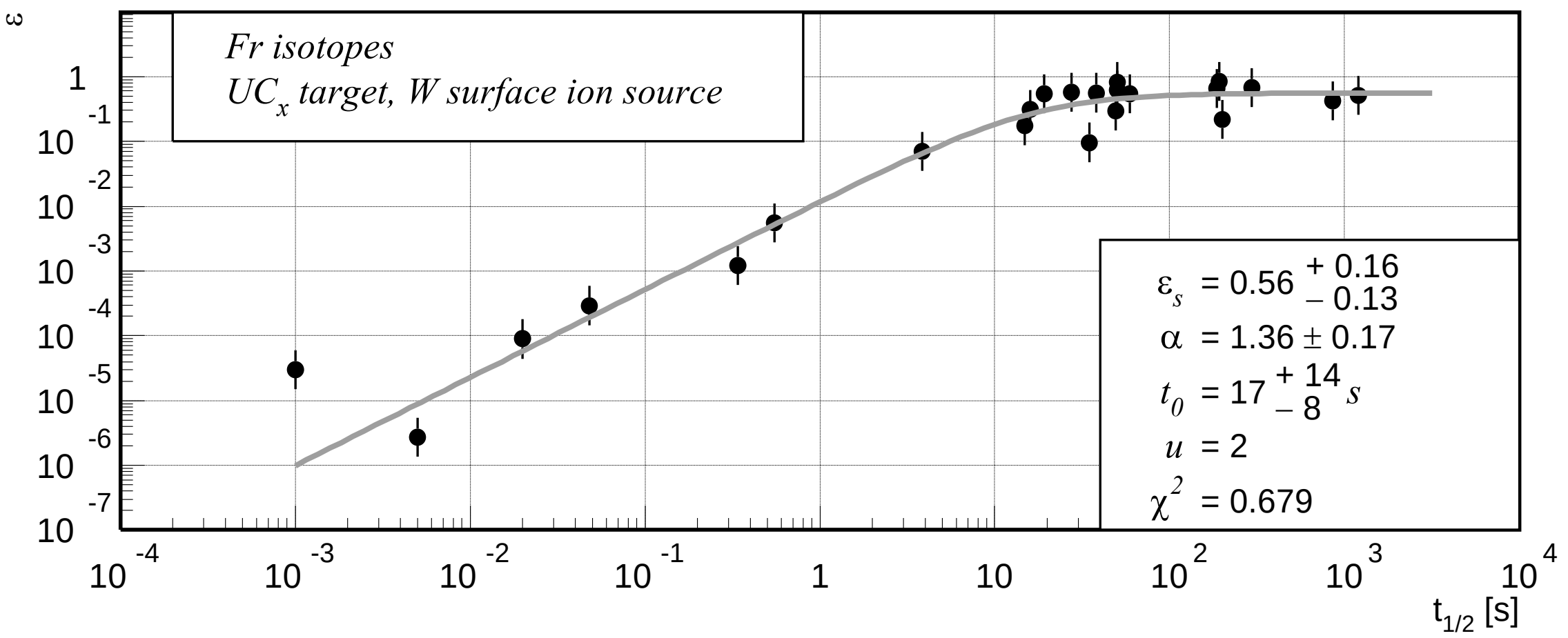

**Figure 6: Overall release efficiency in function of the half-life of Fr isotopes from a UC$_x$ target with a W-surface ion source. The values were obtained by comparing ISOLDE SC yields for this system with the in-target yields calculated using ABRABLA. The function described by equation 14 was fitted to the data.**

It would be of interest to find a curve that parameterizes this behavior with at least three parameters that summarize the essential properties of the system in question:

1. *The efficiency value $\varepsilon_s$ for long half-lives.* For isotopes with sufficiently long half lives the decay losses become negligible, and the extraction efficiency approaches the value valid for the stable isotopes. This parameter can take a value between zero and unity.
2. *The exponent α of the power-function behavior for the short half-lives.* For the short half lives, the extraction efficiency scales with the power function of the isotope half-life $\varepsilon \propto t_{1/2}^{\alpha}$. According to theoretical formalisms of the diffusion through solid matrices and of the effusion, for elements that are volatile at temperatures of the target and the transfer line, the value of the exponent $\alpha$ is expected to be 3/2 [59]. A higher value indicates higher decay losses with short half-lives and vice versa. In any case, it can not be negative.
3. The value $t_0$ of the half-life around which the transition from the power-function to the constant behavior occurs.

The following simple function meets these requirements:

$$\varepsilon\left(t_{1/2}\right) = \frac{\varepsilon_s}{1+\left(\frac{t_{1/2}}{t_0}\right)^{-\alpha}} \qquad 14$$

This function was fitted to the data shown in the figure 6.

In the ISOLDE database, yields are quoted without uncertainties, except for a general statement that they are subject to an uncertainty of typically a factor of two to three for the longest-lived isotopes, and up to an order of magnitude for the most unstable ones [1]. In our understanding, these uncertainties originate mainly in isobaric and molecular contaminations of ISOLDE beams, and possibly in unknown fraction of nuclides of a certain type that are produced in isomeric states. When calculating overall efficiencies, two additional sources of uncertainties come into play:

- difficulties to precisely calculate cross sections at the steep outer slopes of isotopic distributions far off stability
- unknown contributions from the side feeding

We have attributed to every efficiency point an uncertainty range from $1/u$ to $u$ of its value, where $u$ is the uncertainty factor that is fixed for the whole isotopic chain and that can, a priori, take a value between 2 and 3, depending on the element and the target-ion-source system. In this way, the error bars appear symmetric in the logarithmic presentation. The quantitative criterion for the selection of the uncertainty factor was the calculated chi-square parameter of the fit. It was required that it should be close to unity, with the restriction that the uncertainty factor can not be smaller than two. We are aware that this is a rather rough procedure that can severely underestimate the uncertainties of specific data points. Nevertheless, this assumption allows deducing crude estimates of the uncertainties of the parameters of the function 14. We used the same criterion for other elements and targets we studied (see sect. 6 - Results).

The point at $t_{½}$ = 1 ms refers to $^{218}$Fr. It lays more than one order of magnitude above the trend followed by the other short-lived isotopes. The possible physical reasons for this will be discussed in the section dedicated to the results, and arguments will be given for excluding $^{218}$Fr from the fitting procedure as it was done here.

From the fitted parameters we learn that, for short half-lives, the francium release efficiency from the uranium-carbide target scales with the power function of the isotopic half-life, with an exponent that is within the error bars from the theoretical value for solid matrices. The transition from the power-function to the constant behavior occurs for the half-lives of the order of 20 seconds, and the saturation efficiency is around 56%.

## 6. Results

This section is devoted to fits of the function 14 to the data for elements from several chemical groups that were extracted from different target-ion source systems.

### *6.1 Alkali elements*

Alkalis can be very efficiently ionized by surface ionization due to their low ionization potential. The ionization step separates them chemically from other elements. Their high vapour pressure allows for an extraction with high efficiencies [33,32].

#### *6.1.1 Sodium*

The data shown here refer to the titanium and uranium-carbide targets as found in the ISOLDE database. The thickness of the titanium target was 40 g/cm$^2$ and that of the uranium-carbide target was 13 g/cm$^2$.[5] The energy of the proton beam at the end of the titanium target is 526 MeV. According to ABRABLA calculations, the production cross sections for the relevant sodium isotopes at this energy drop to between 60 and 75% of their values at 600 MeV. Therefore, it is safe to assume that their variation is linear along the target. At the end of the uranium-carbide target, the proton beam energy is around 575 MeV. The change in the production cross sections for the sodium isotopes at this energy is below the level of the uncertainties of the experimental nuclide production cross-section data for the same reaction.

---

[5] The exact ratio of carbon to uranium atoms is not documented in the database. However, the influence of the exact ratio on the in-target production rate is rather small. At 13 g/cm$^2$, a change in the composition of 2 carbon atoms per one uranium atom changes the calculated in-target production rates by only about 1%. We assumed a composition with 6 carbon atoms per one uranium atom.

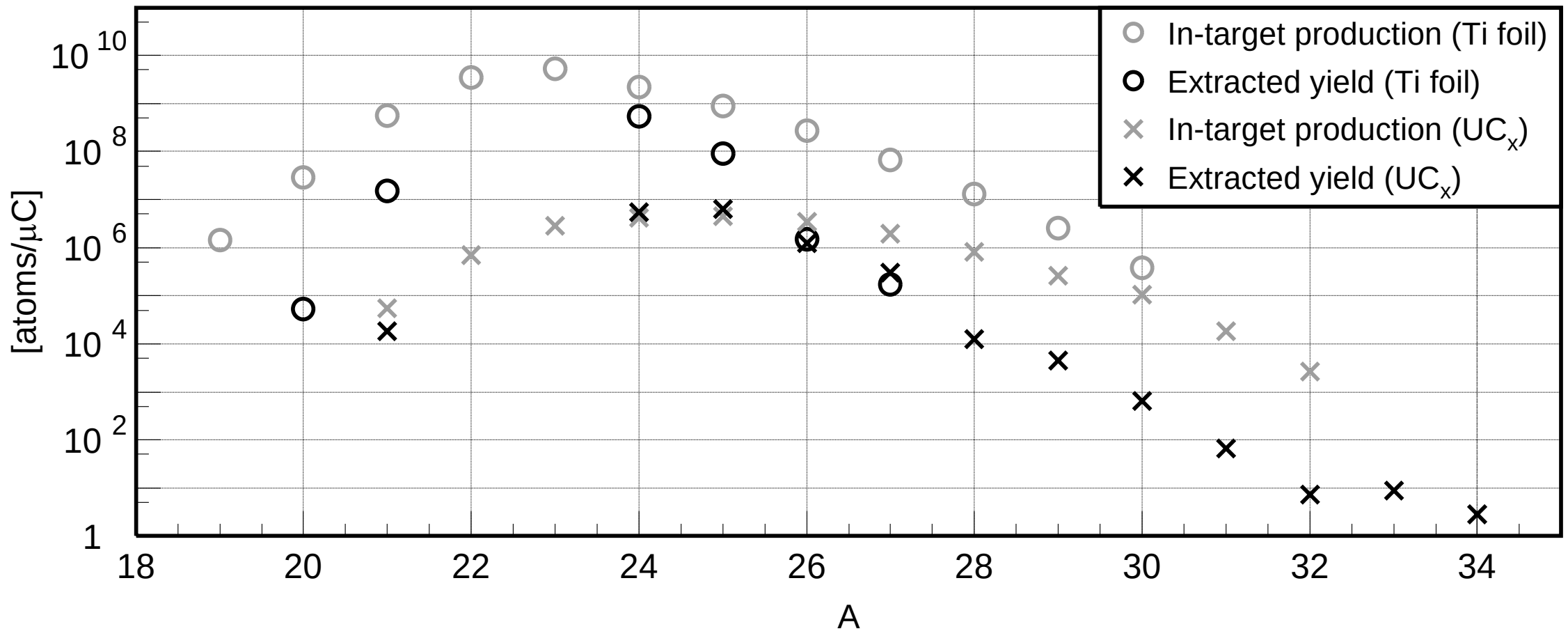

**Figure 7: Comparison of the Na isotopic production rates in Ti and $UC_x$ targets calculated using ABRABLA with the ISOLDE yields of these nuclides extracted from the same targets using a W-surface ion source.**

The comparison of sodium isotopic production rates in titanium and uranium-carbide targets calculated using ABRABLA with the ISOLDE yields of these nuclides extracted from the same targets using a W-surface ion source is shown in the figure 7. One can notice that the production rates are higher in the titanium target up to $A$=30. However, the distribution of the production rates in uranium carbide drops less steeply towards the neutron-rich side because sodium is produced in uranium by very asymmetric fission. Uranium carbide seems to provide faster extraction so that its yield distribution is broader in $A$. The yields from uranium carbide are of the same size as those from titanium already at $A$=26, and extend to much more neutron rich and shorter-lived isotopes. The yields of $^{33,34}$Na have been measured by the detection of the β-delayed neutrons, assigning all of the detected neutrons to Na. However, this procedure is vulnerable to contaminations from $^{33,34}$Al that also emit β-delayed neutrons. Therefore, the yields for these two isotopes are most probably overestimated [60].

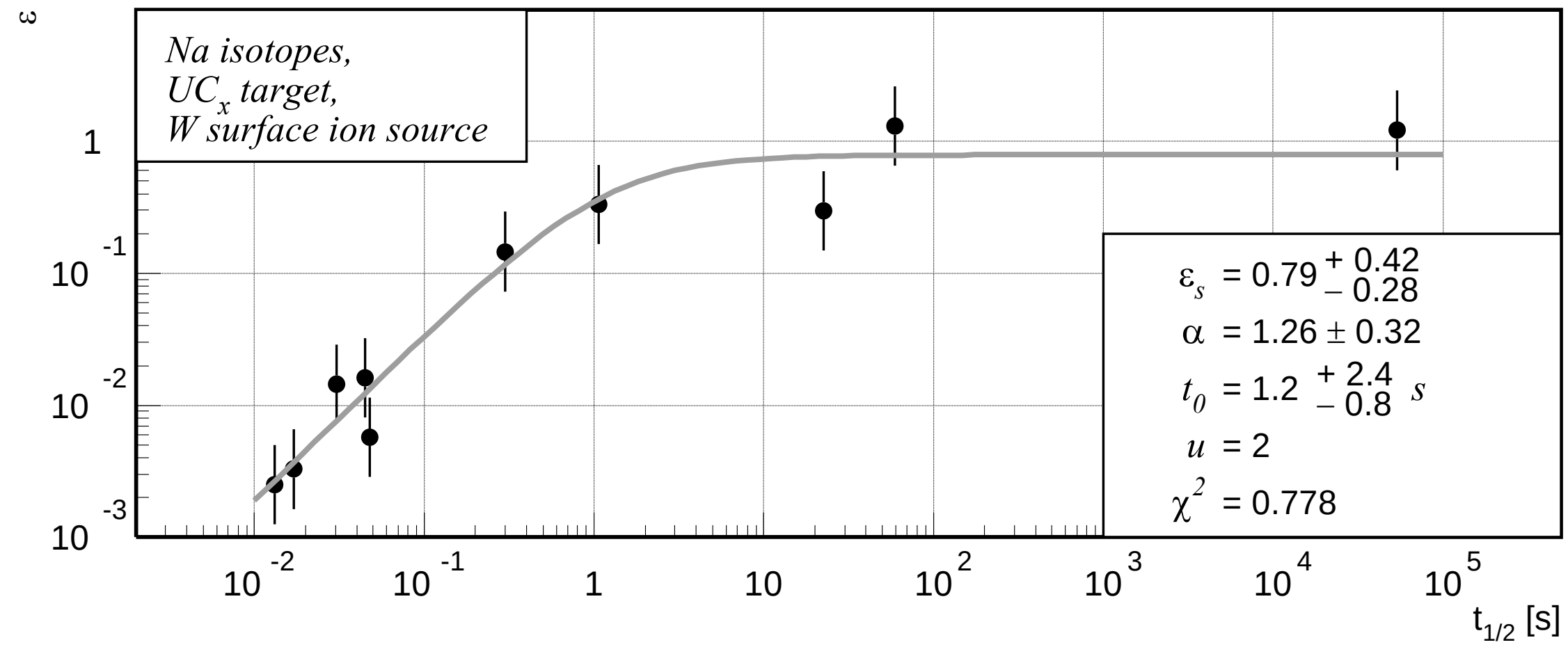

**Figure 8: The fit to the data for the Na isotopes produced by the SC beam in a $UC_x$ target and extracted using a W surface ion source.**

Figure 8 represents the fit to the data for the Na isotopes produced by the SC beam in a 13 g/cm$^2$ uranium-carbide target and extracted using a W surface ion source. The extraction efficiency in the long half-life limit is roughly 80%. The exponent of the power-function behavior for the short half-lives is within its uncertainty from the theoretical value, and the transition between the two regions occurs at half-lives of the order of one second, which indicates fast extraction.

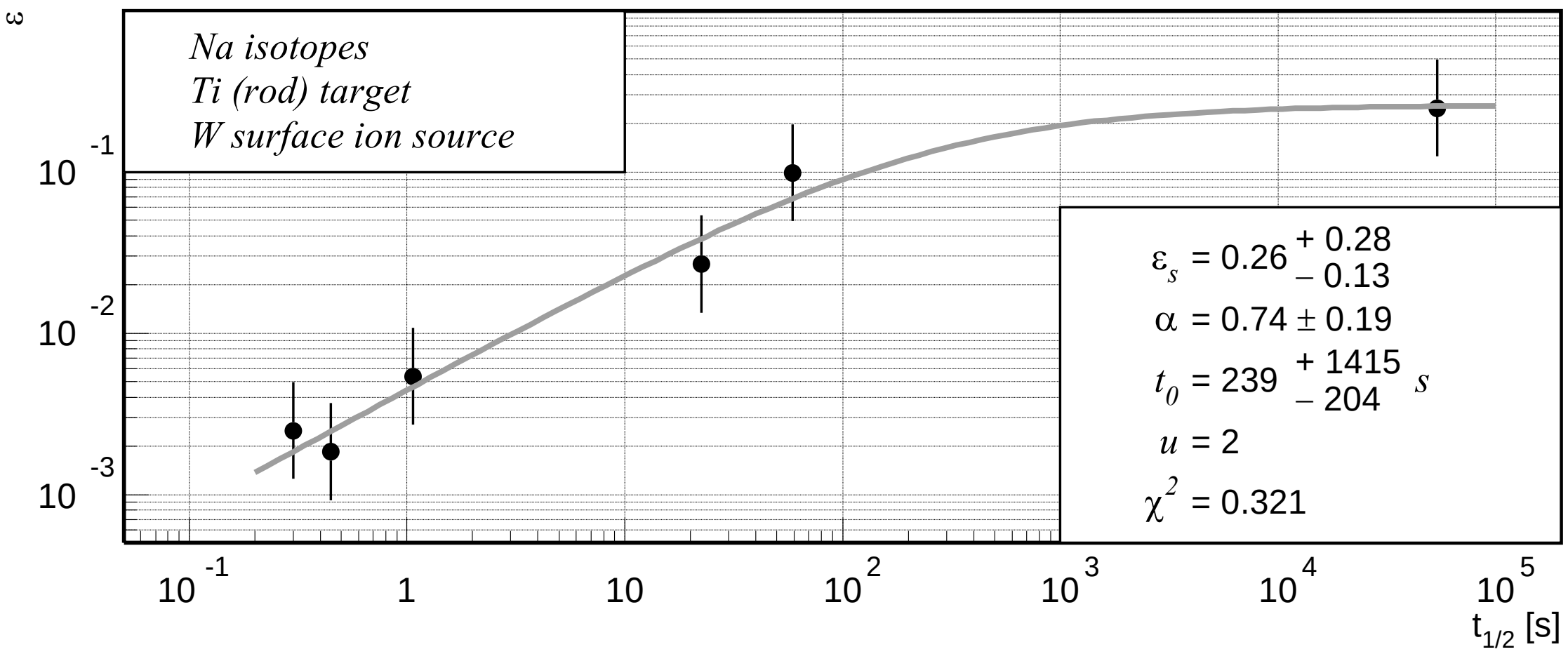

**Figure 9: The fit to the data for the Na isotopes produced by the SC beam in a Ti foil target and extracted using a W surface ion source.**

Figure 9 represents the fit to the data for the Na isotopes produced by the SC beam in a titanium rod target and extracted using a tungsten surface ion source. One notices that the transition from the power-function behavior to constant efficiencies happens at relatively long half-lives. At the same time, the efficiency drop in the power-function region appears to be much less steep than expected – the value of the exponent $\alpha$ is 0.74, which is significantly lower than 3/2. However, this curve has been fitted on rather few points, and both these parameters are very sensitive to the uncertainties of the data points in the transition region.

### *6.1.2 Potassium*

The data for potassium refer to the same two targets as the data for sodium. The energy loss of the beam is also the same. The variation of potassium isotope production cross sections with the beam energy loss along both targets is below the level of typical uncertainties of the experimental data used to benchmark the ABRABLA code.

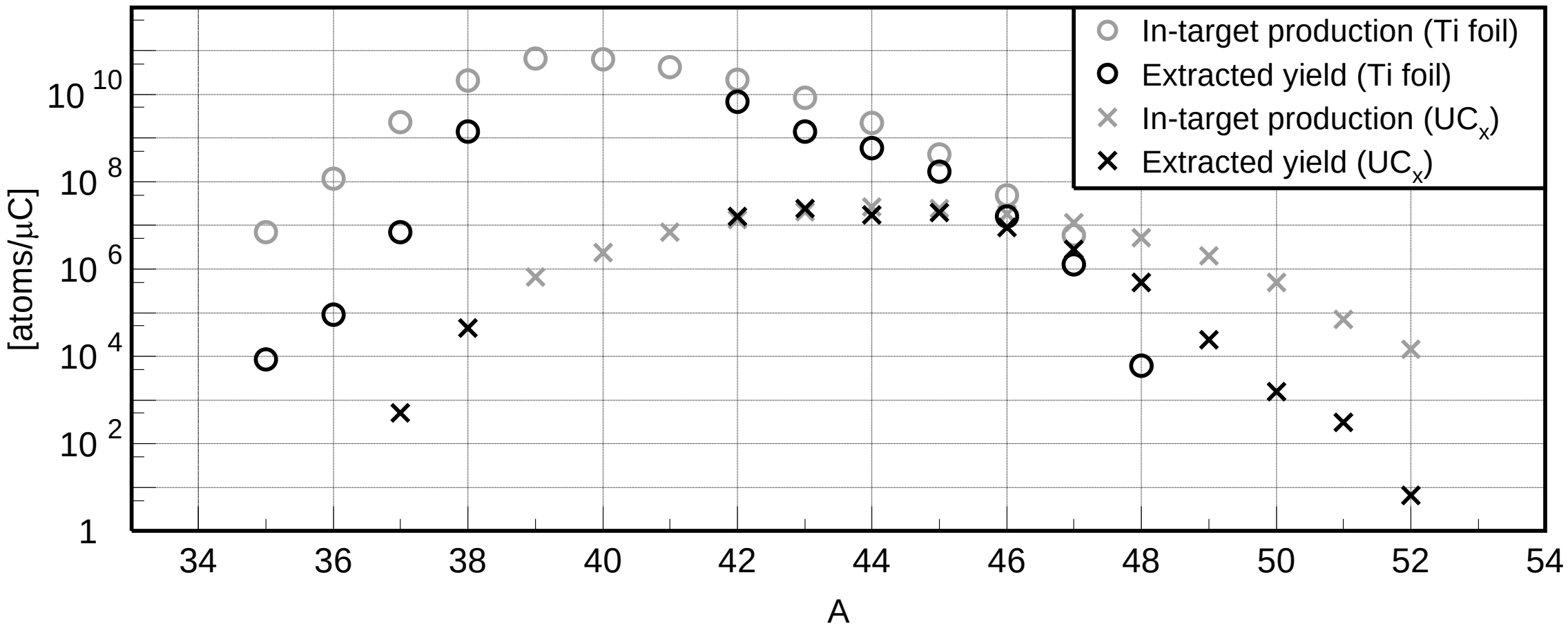

**Figure 10: Comparison of the K isotopic production rates in Ti and $UC_x$ targets calculated using ABRABLA with the ISOLDE yields of these nuclides extracted from the same targets using a W-surface ion source.**

The comparison of potassium isotopic production rates in titanium and uranium-carbide targets calculated using ABRABLA with the ISOLDE yields of these nuclides extracted from the same targets using a tungsten surface ion source is shown in figure 10. The production rates are higher in the titanium target up to $A$=46. However, the distribution of the production rates in uranium carbide is broader and shifted towards the neutron-rich side, because fission contributes significantly to the production of potassium in uranium. Uranium carbide also seems to provide faster extraction so that its yields are closer to the in-target production rates than those from the titanium target.

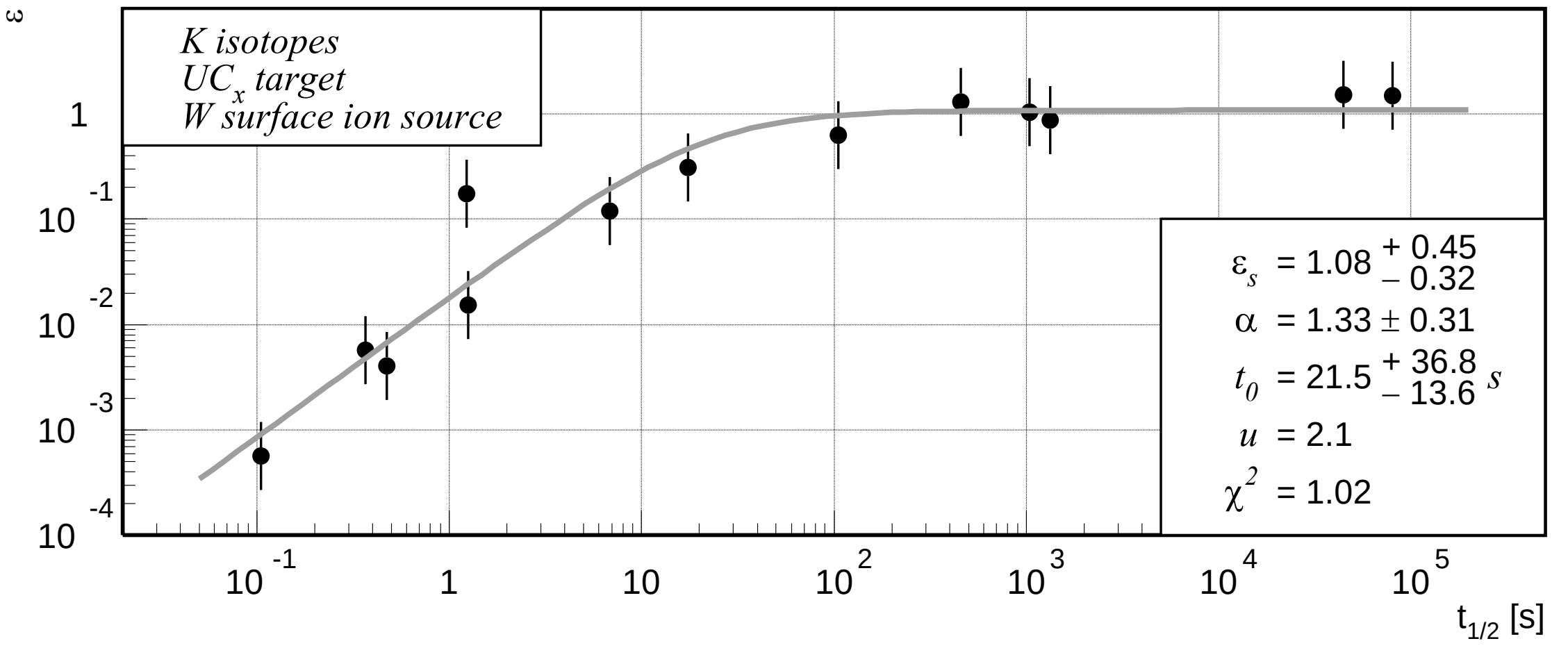

**Figure 11: The fit to the data for the K isotopes in a $UC_x$ target with a W surface ion source.**

Figure 11 represents the fit to the data for the K isotopes produced by the SC beam in a 13 g/cm$^2$ uranium-carbide target and extracted using a tungsten surface ion source. The extraction efficiency in the long half-life limit appears as 108%. Strictly speaking, a value higher than 100% is unphysical, but this estimate should be regarded taking into account its uncertainty range. The exponent of the power-function behavior for the short half-lives is within its uncertainty from the theoretical value, and the transition between the two regions occurs at half-lives of the order of tens of seconds.

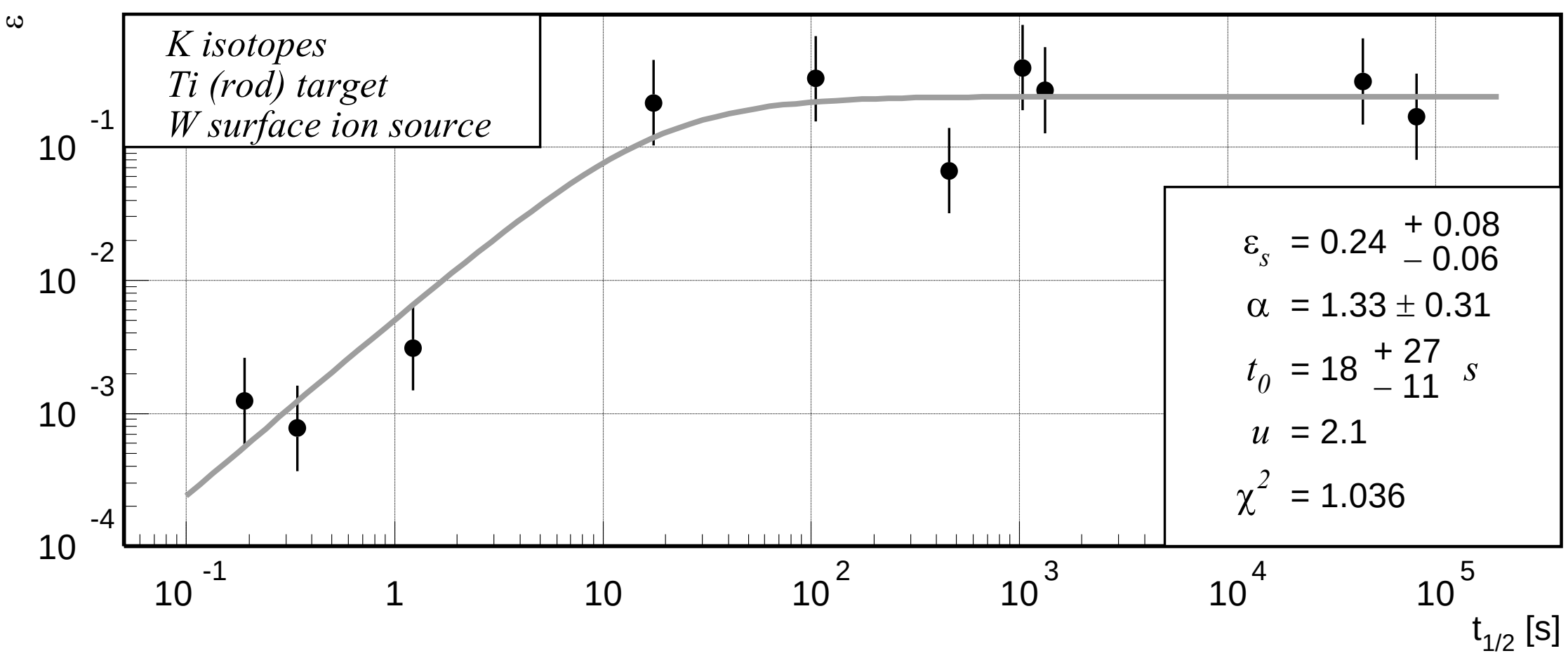

**Figure 12: The fit to the data for the K isotopes in a Ti target with a W surface ion source.**

Figure 12 represents the fit to the data for the K isotopes produced by the SC beam in a titanium rod target and extracted using a tungsten surface ion source. The extraction-efficiency value in the long half-life limit is consistent with the value obtained for sodium in the same target. The exponent of the power-function behavior for the short half-lives is within its uncertainty from the theoretical value, and the transition between the two regions occurs at half-lives of the order of tens of seconds.

### *6.1.3 Rubidium and cesium production in uranium carbide*

In the interaction of protons with uranium, up to energies of at least 1 GeV, the dominant mechanism for the production of rubidium and cesium is spallation followed by fission [61]. In the fission of uranium, rubidium and cesium are conjugate fragments and their elemental production cross sections are of the same size. When fission is preceded by spallation, the actual fissionning nucleus can have smaller atomic number than uranium, and the production is slightly shifted towards lighter elements. Thus, in the interaction of high-energy protons with uranium, the production of rubidium is increased with respect to that of cesium. Figure 13 represents a comparison of rubidium and cesium isotopic production cross sections in the reaction of 1 GeV protons with uranium measured at GSI-FRS, and at the LNPI synchrocyclotron of the Khlopin Radium Institute by Belyaev et al. [61], as well as those calculated using ABRABLA for the same reaction at 1 GeV and at 600 MeV. In all the datasets the integral of the rubidium distribution is more than 30% higher than that of cesium.

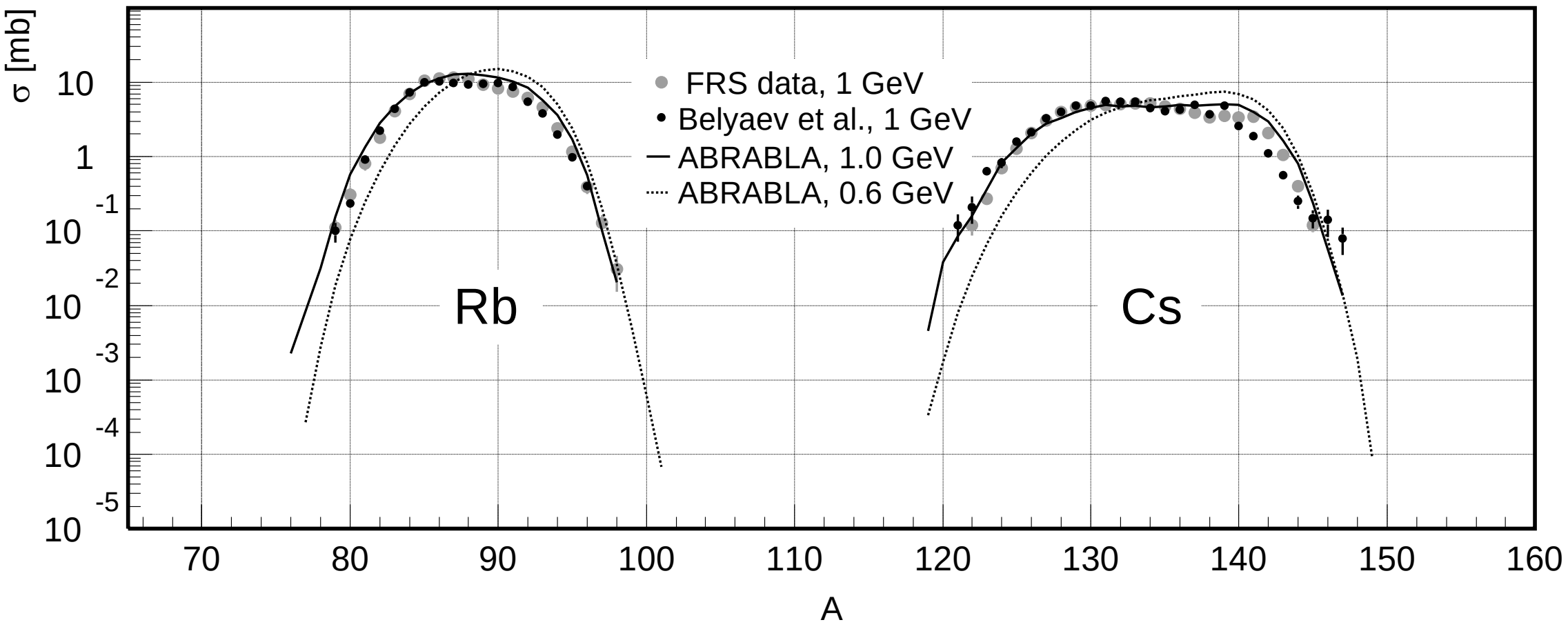

**Figure 13: Comparison of Rb and Cs isotopic production cross sections in the reaction of protons with uranium at 1 GeV measured in inverse kinematics at GSI-FRS (gray points), measured by Belyaev** et al**. at the Khlopin Radium Institute and calculated using ABRABLA (solid line). ABRABLA results at 600 MeV are also shown (dotted line)**

However, the yields cited for cesium in the ISOLDE database are by one order of magnitude higher than those of rubidium [1]. Rubidium being the lighter of the two elements, one would not expect its extraction efficiencies to be lower than those for cesium. One possible explanation could be that the rubidium and cesium data have been obtained in two different sets of measurements, under different experimental conditions. In the figure 14 we compare the database yields for rubidium and cesium with another set of measurements by Bjørnstad et al. at a later date [32]. The ratio of the rubidium to the cesium yields in the data from Bjørnstad et al. reflects more closely the ratio of the measured cross sections, but the integral cesium yields still seem to be about two times higher than those of rubidium.

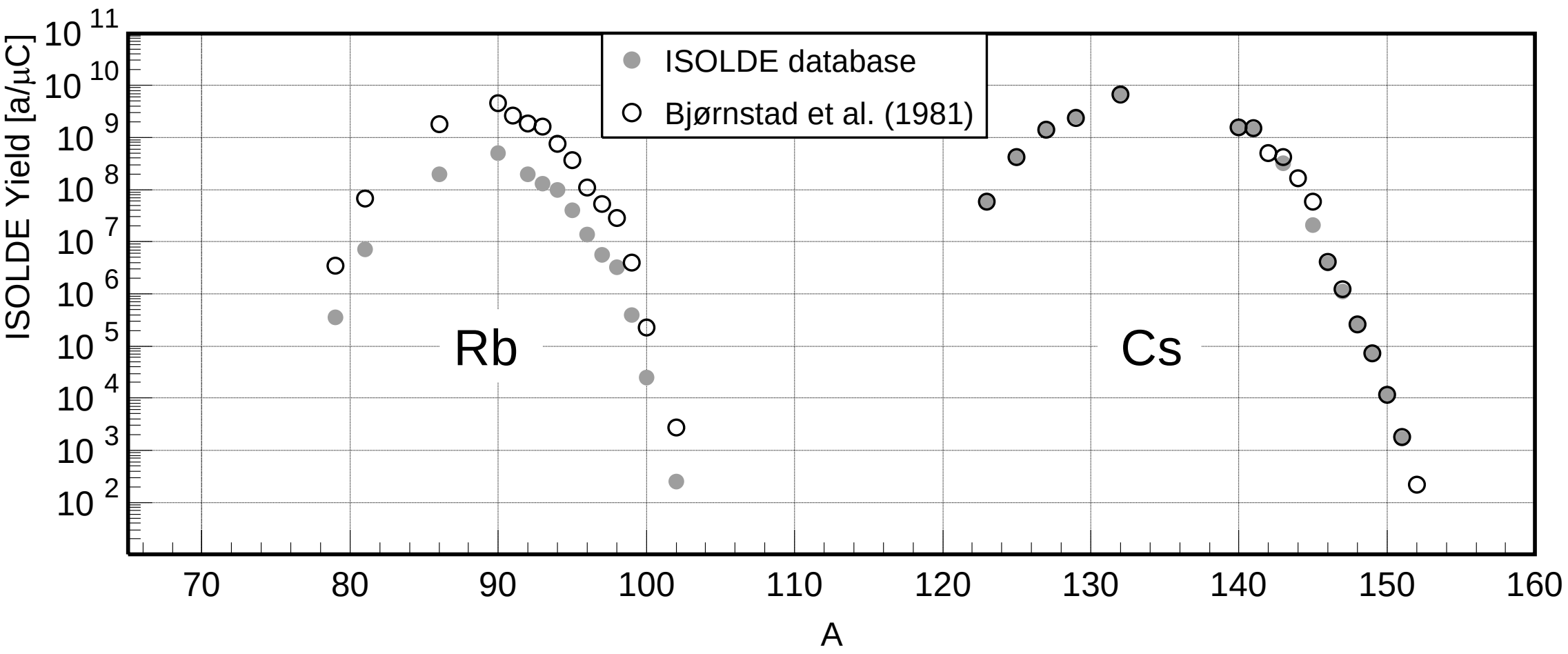

**Figure 14: Comparison of ISOLDE yields for Rb and Cs from $UC_x$ as found in [1] (grey points) and in [32] (empty points)**

As rubidium and cesium yield data by Bjørnstad et al. have been obtained in a consistent set of measurements, we have decided to use them instead of the data from the database. The target used by Bjørnstad et al. contained 16.4 g of uranium per $cm^2$.

Figure 15 represents the fit of the extraction-efficiency dependence on the half-life to the data for the rubidium isotopes in a uranium-carbide target with a W surface ion source. The value of the exponent of the power-function behavior is lower than expected, but it is established on rather few data points. Besides, the three most neutron-rich, and shortest-lived, isotopes ($A$ = 98-100) that are seen in the figure 14 are not shown in the figure 15 because their apparent efficiencies are way too high for that region of the half-lives, probably because of the systematic uncertainties of the underlying data and the corresponding calculations that can both be large so far from stability.

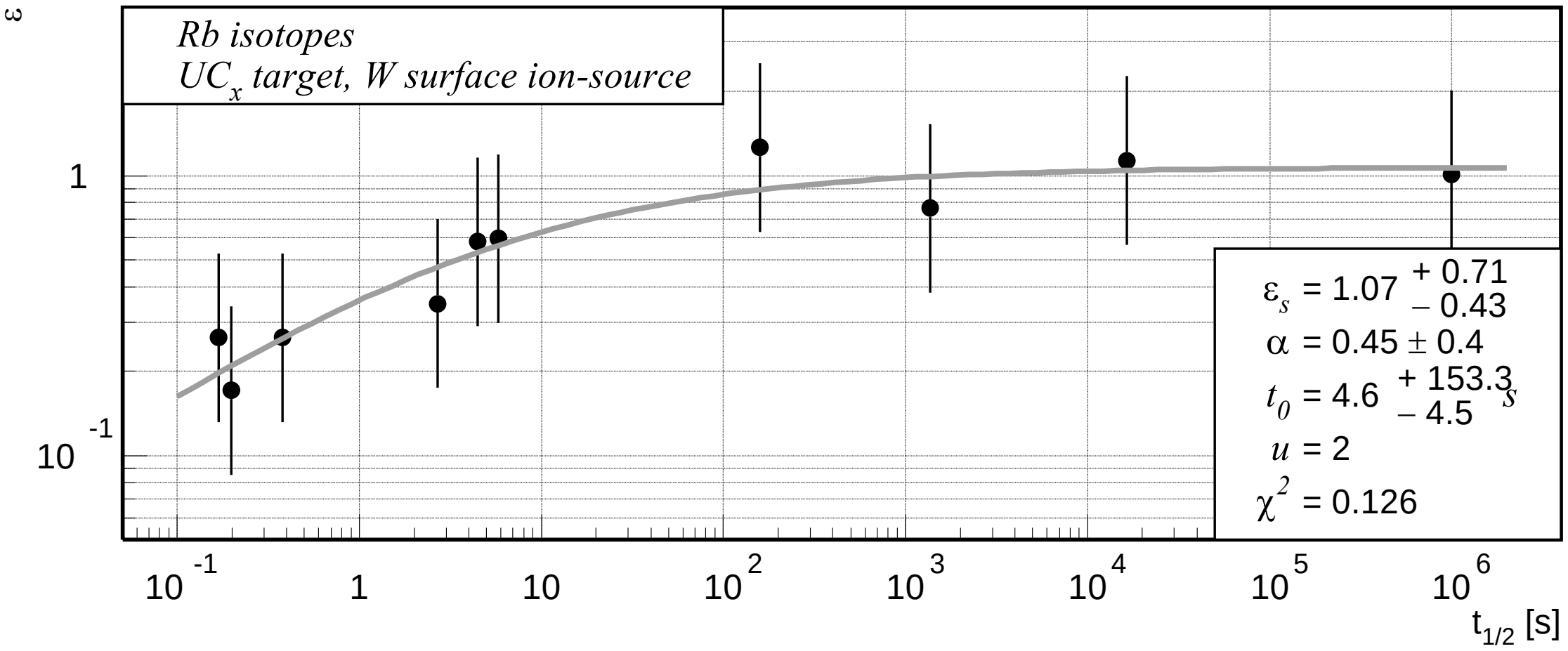

**Figure 15: The fit to the data for the Rb isotopes in a UC$_x$ target with a W surface ion source. The ISOLDE yields were taken from the work of Bjørnstad** et al. **[32] Some of the most neutron-rich isotopes (98-100) that are seen in the figure 14 are not shown because their apparent efficiencies are way too high for that region of the half-lives, probably because of the uncertainties of the underlying data and the corresponding calculations that can both be large so far from stability.**

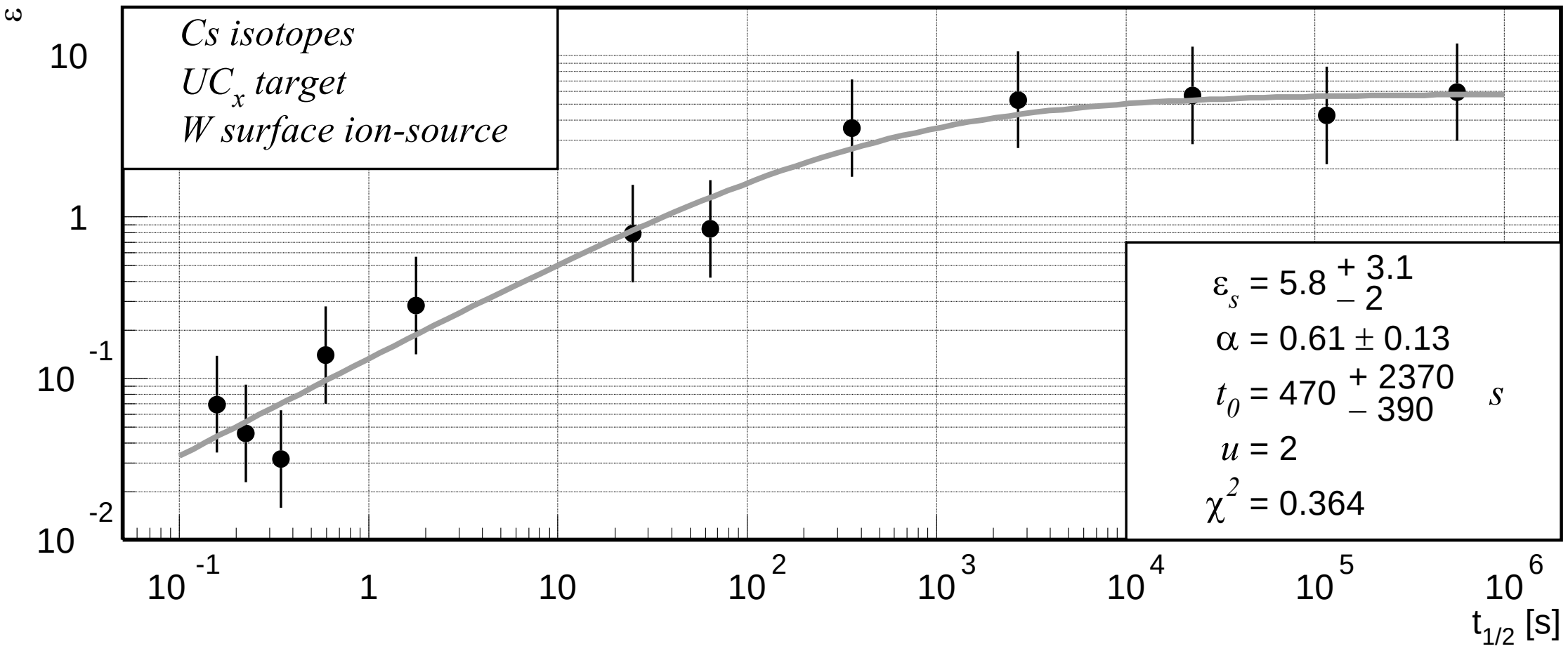

**Figure 16: The fit to the data for Cs isotopes in a UC$_x$ target with a W surface ion source. Some of the most neutron-rich isotopes (149-151) that are seen in the figure 14 are not shown because their half-lives are unknown.**

Figure 16 represents the fit of the extraction-efficiency dependence on the half-life to the data for the cesium isotopes in the same target – ion-source system. Yields of the longest-lived cesium isotopes are several times higher than what one would expect from the cross sections, assuming 100% extraction efficiency, so that the fitted saturation-efficiency parameter has an unphysical value of 580%. Significant contribution from side feeding is excluded. The most long-lived isotope, $^{132}$Cs, for which the apparent efficiency is the highest (600%), is shielded. According to ABRABLA results, the cross sections for the production of precursors to other long-lived cesium isotopes are several times lower than those for the corresponding cesium isotopes.

The five most long-lived isotopes in the figure 16 are also the most neutron-deficient ones present in the database. Since the overproduction is the most pronounced in that region, a possible explanation could be that its origin is in the spallation-evaporation reactions induced by the proton beam in some of the materials directly surrounding the target. The target container is made of tantalum, but the production cross section for $^{132}Cs$ in tantalum is three orders of magnitude smaller than in uranium. Significant contribution could be made only in a material with an atomic number much closer to that of cesium. Uranium-carbide targets may contain lanthanide impurities because of the chemical similarity between lanthanides and actinides. However, these impurities usually do not exceed the permil range [60]. Yields of the cesium isotopes up to $^{132}Cs$ might be overestimated by wrongly assigning indium activity to cesium [60]. It is difficult to explain so high yields of neutron-deficient cesium isotopes, and we conclude that, for this reason, these results for cesium in uranium carbide should be regarded with caution.

### *6.1.4 Rubidium from thorium carbide*

Figure 17 represents the efficiencies for rubidium isotopes in a thorium-carbide target with a tungsten surface ion source. Yield data exist only for isotopes with half-lives of several seconds and longer. No efficiency decrease is observed for the shortest-lived among these, which indicates fast release. A constant function was fitted, and the obtained efficiency value is close to 100%.

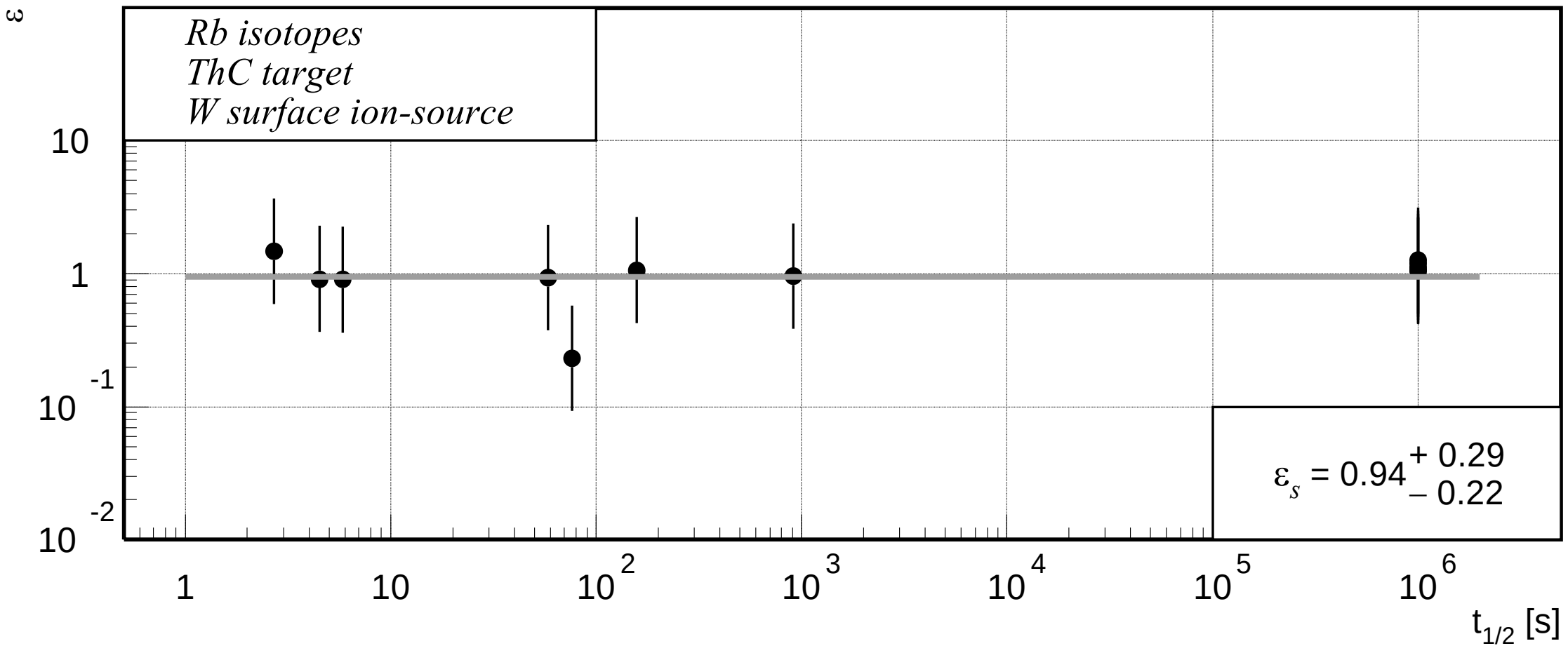

**Figure 17: Rubidium efficiencies in a ThC target with a W surface ion source. Yield data exist only for isotopes with half-lives of several seconds and longer. No efficiency decrease is observed for the shortest lived among these. A constant function was fitted.**

### *6.1.5 Neutron-deficient rubidium and cesium from spallation-evaporation targets*

In spallation-evaporation reactions, the cross sections for the production of nuclides close to the target, by removal of a relatively small number of nucleons, are very high. Suitable spallation-evaporation targets for the production of neutron-deficient rubidium and cesium are, respectively, niobium and lanthanum. Peaks of the rubidium and cesium isotopic yield distributions from these targets are about one order of magnitude higher than the peaks of the respective distributions from the uranium carbide, and shifted to the neutron-deficient side.

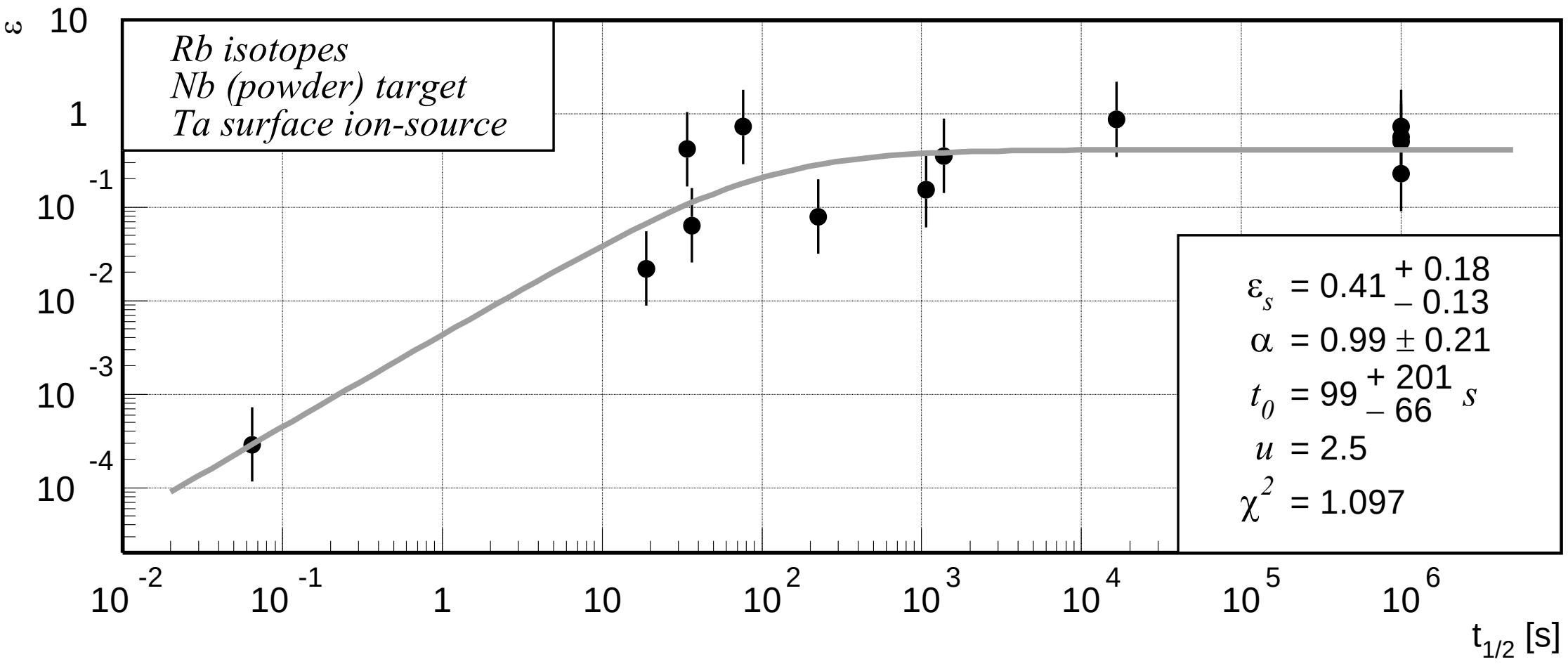

**Figure 18: The fit to the data for the Rb isotopes in a Nb powder target with a Ta surface ion source.**

Figure 18 represents the fit to the data for the rubidium isotopes in a niobium metal powder target with a tantalum surface ion source. The target thickness was 50 g/cm$^2$. The efficiency in the limit of long half-lives is around 40%.

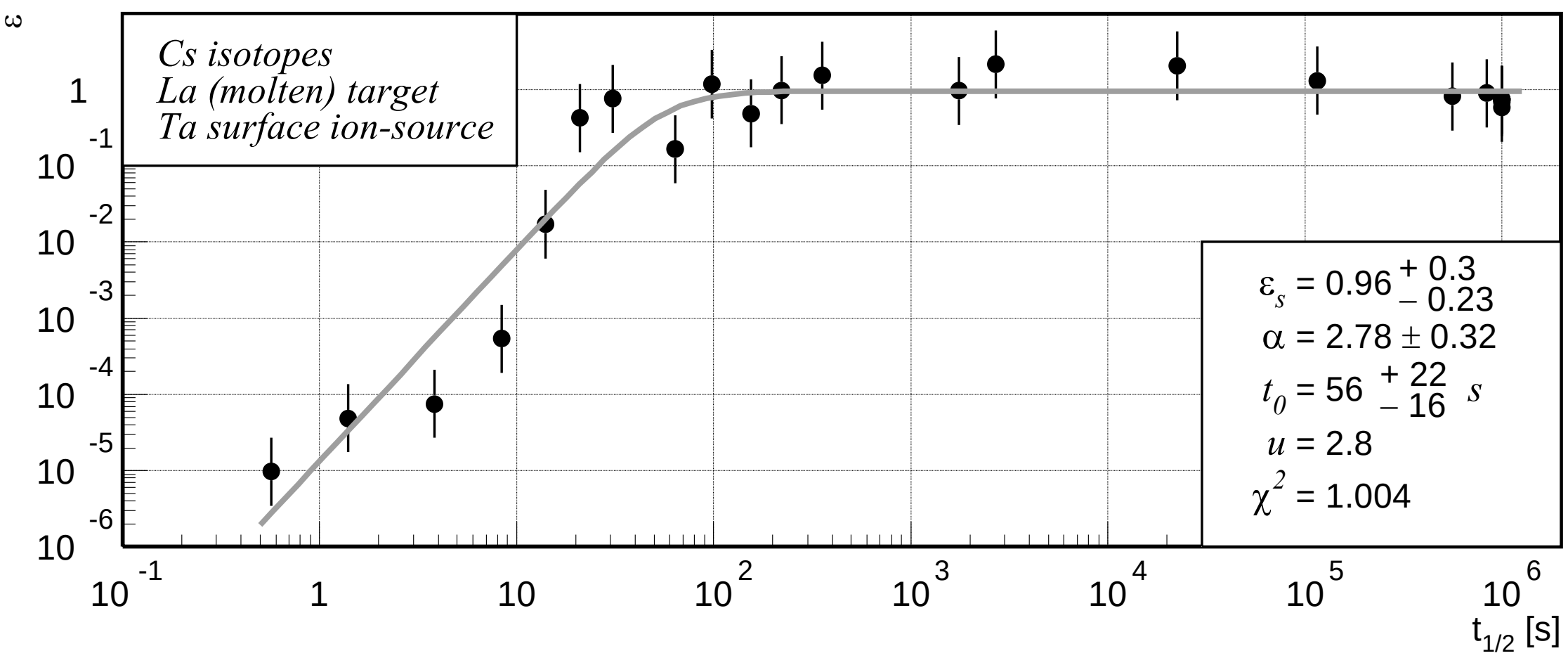

**Figure 19: The fit to the data for Cs isotopes in a La molten metal target with Ta surface ion source.**

Figure 19 represents the fit to the data for cesium isotopes in a lanthanum molten metal target with a tantalum surface ion source. The target thickness was 140 g/cm$^2$. One notices a very steep efficiency drop with short half-lives – the value of the parameter $\alpha$ is 2.8.

### *6.1.6 Francium*

Figure 20 represents the fit to the francium data in uranium carbide. The target contained 10 g of uranium per cm$^2$. The saturation efficiency is 56%, the exponent of the power-function behavior with short half-lives is within the uncertainty range from the theoretical value of 1.5, and the transition from the power-function to constant behavior takes place around $t_0$ = 17 s.

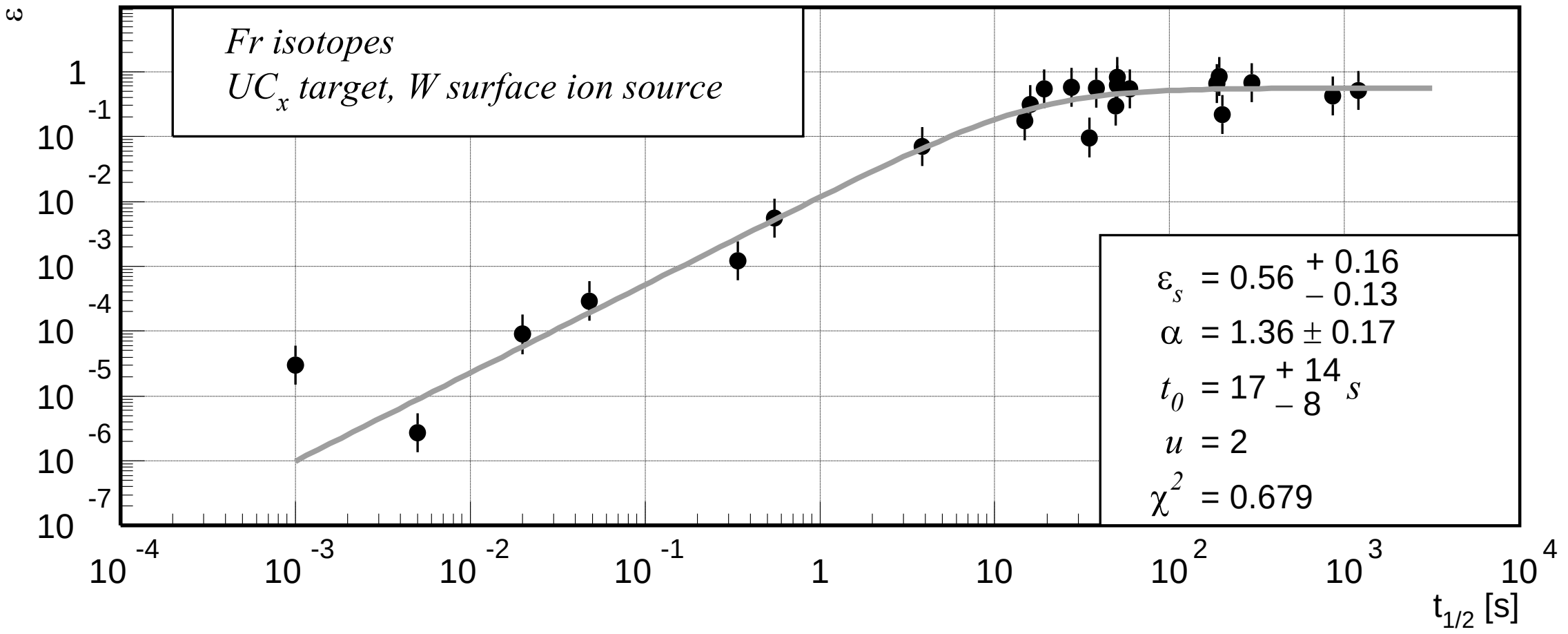

**Figure 20: The fit to the data for Fr isotopes in a UC$_x$ target with a W surface ion source. The point at 1 ms ($^{218}$Fr) was excluded from the fit (see explanation in the text).**

The efficiencies of the shorter-lived isotopes show a smooth power-function behavior for half-lives down to several milliseconds. Only $^{218}$Fr with 1 ms ground-state half-life seems to lie above that trend by two orders of magnitude. One of the possible physical reasons for that is that a significant part of the $^{218}$Fr nuclei is produced in the 22 ms metastable state at 86 keV. This state decays by emission of α particles with energies very close to those emitted from the ground state. Indeed, the extraction efficiency for $^{218}$Fr is of the same magnitude as that for the isotopes with half-lives of the order of tens of milliseconds. Therefore, the point corresponding to $^{218}$Fr has been excluded from the fitting procedure.

### *6.1.7 Side feeding of francium*

Important side-feeding of francium can potentially come from the α decay of actinium. Actinium does not diffuse from the target because it is chemically similar to uranium. It decays to francium in the target. The α-decay precursors of all francium isotopes with $A \leq 216$ are produced with much lower cross sections than the corresponding francium isotope. $^{216,217}$Fr are not present in the ISOLDE yield database because they are too short-lived. The production cross sections for $^{222,223}$Ac are, respectively, two and three times higher than that for the primary production of their α-decay daughters, $^{218,219}$Fr. The α-decay precursors of francium isotopes with $220 \leq A \leq 224$ are very long-lived. Their influence could be observed only in a very long run with one target. Actinium isotopes heavier than 229 mass units are not α emitters.

## *6.2 Alkaline earths*

The ionization potential of the alkaline earth elements is low enough for them to be ionized by means of surface ionization. However, in this way, additional chemical separation is necessary in order to eliminate the accompanying alkalis, especially on the neutron-deficient side, where the production cross sections of the latter are higher than that for the corresponding alkaline earths. This is often achieved by controlled addition of a fluoride gas, because of the tendency of the alkaline earths to form molecular ions of the type $MeF^+$. Then the mass separation is adjusted to select the mass of the molecular ion formed by the desired alkaline-earth isotope [29]. The three heaviest elements of this group are the most volatile and the most efficiently produced [33].

### 6.2.1 *Magnesium*

Figure 21 represents the fit to the data for magnesium isotopes in a tantalum foil target using a plasma ion source. The thickness of the target was 122 g/cm$^2$. For some of the isotopes, two different, but close, yield values are quoted in the yield database. In such cases, both values were used. The efficiency reaches 4% for long half-lives

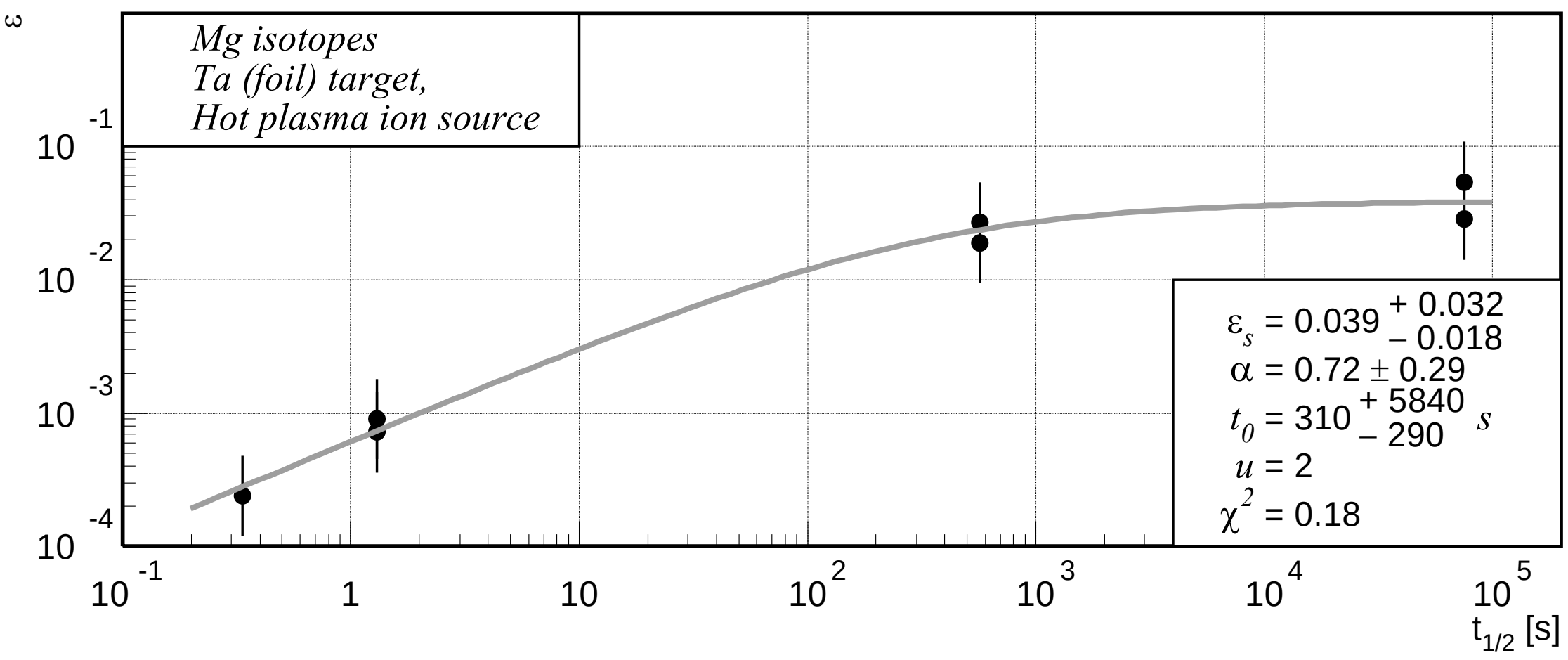

**Figure 21: The fit to the data for Mg isotopes in a Ta target using a plasma ion source**

### 6.2.2 *Calcium*

Figure 22 represents the fit to the calcium data in a titanium target using a tungsten surface ion source with a $CF_4$ leak. The use of the fluoride gas leak made it possible to eliminate the potassium contamination of $^{37,38,39}$Ca beams. However, as there are only four yield data points, the uncertainties of the fitted parameters are rather large. The efficiency for $^{47}$Ca, with a half-life of 4.5 days is 1.5%. The thickness of the target was 34 g/cm$^2$.

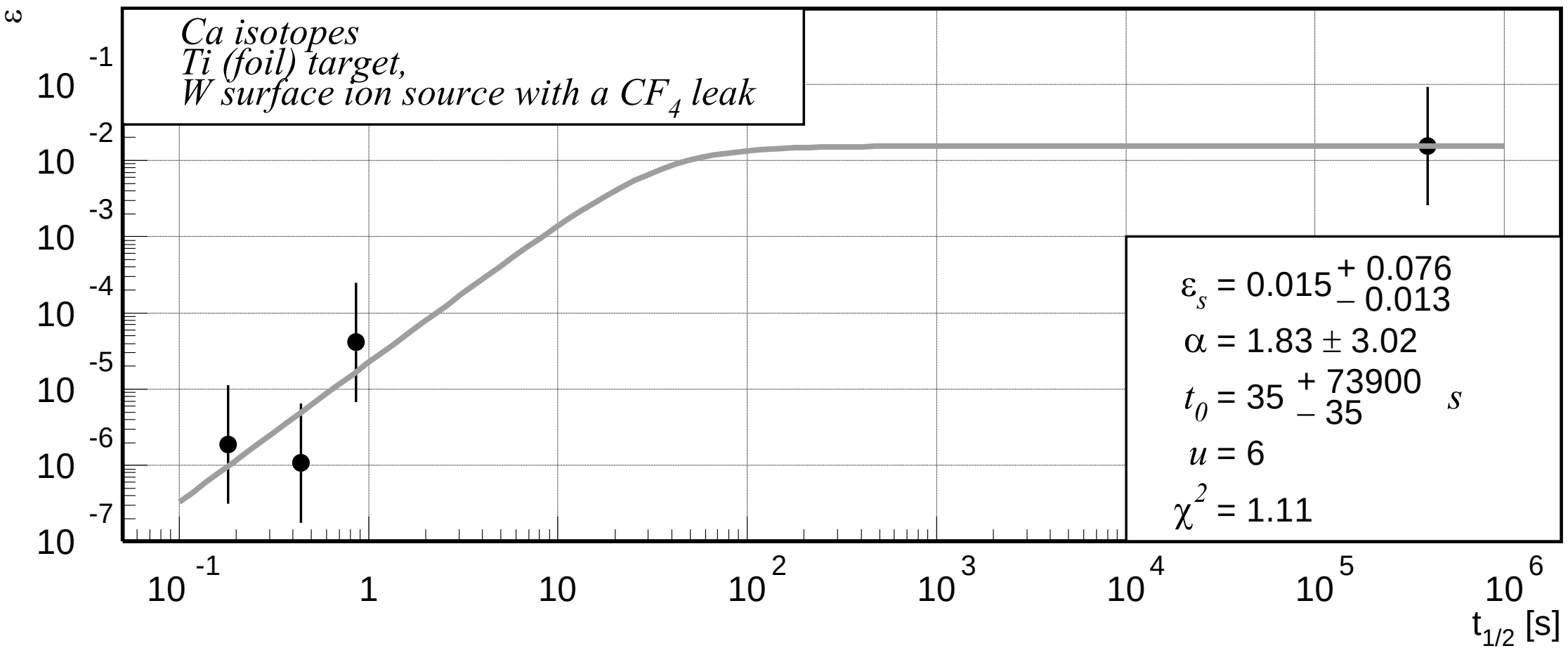

**Figure 22: The fit to the Ca data in a Ti target using a W surface ion source with a $CF_4$ leak**

### 6.2.3 Strontium

Figure 23 represents the fit to the efficiencies for strontium in a niobium foil target of 40 g/cm$^2$. The use of the $CF_4$ gas made it possible to eliminate the contamination from rubidium on the neutron-deficient side. The points are rather scattered, and the uncertainties of the fitted parameters are large. The efficiency reaches 10% for long half-lives.

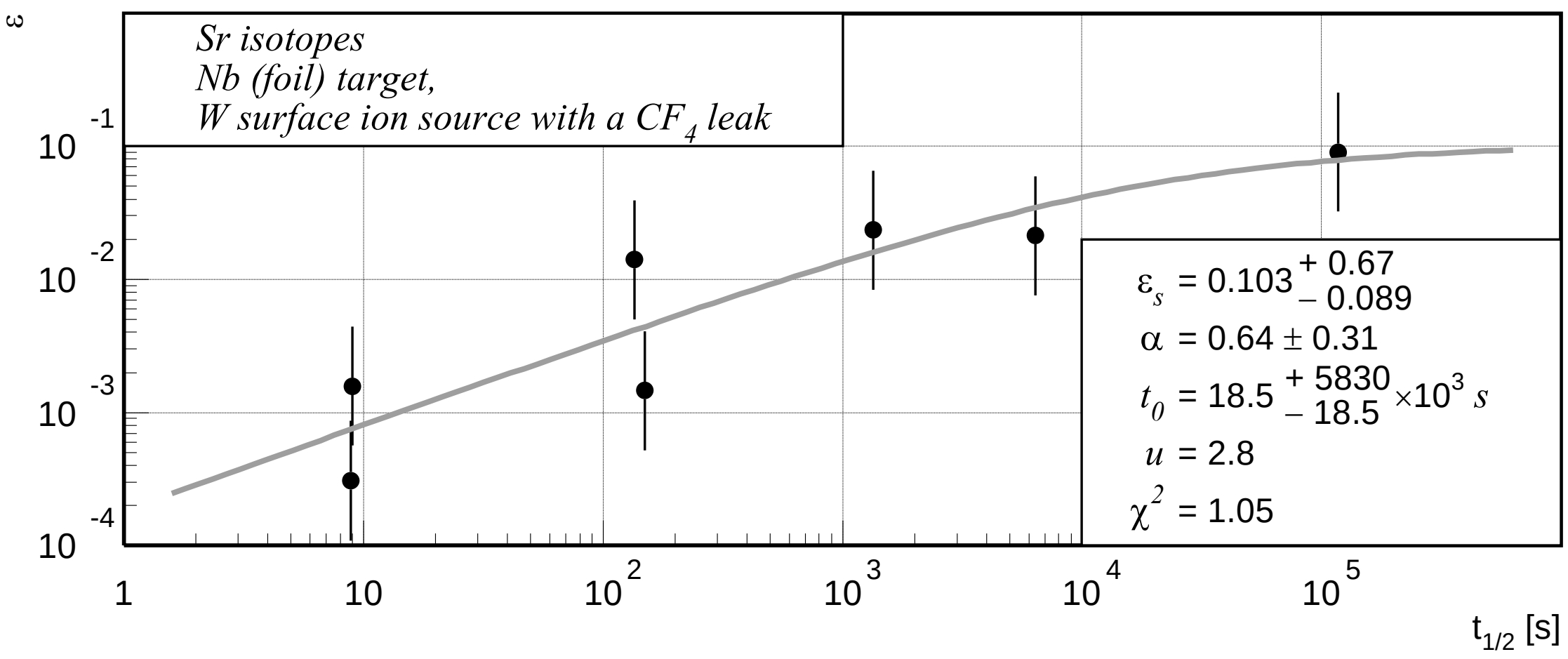

**Figure 23: The fit to the Sr data in a Nb target with a W surface ion source and a $CF_4$ leak**

Figure 24 represents the fit to the efficiencies for strontium in a uranium-carbide target with 85 g/cm$^2$ of uranium. Only stable and neutron-rich isotopes are present. The efficiency reaches 13% for long half-lives.

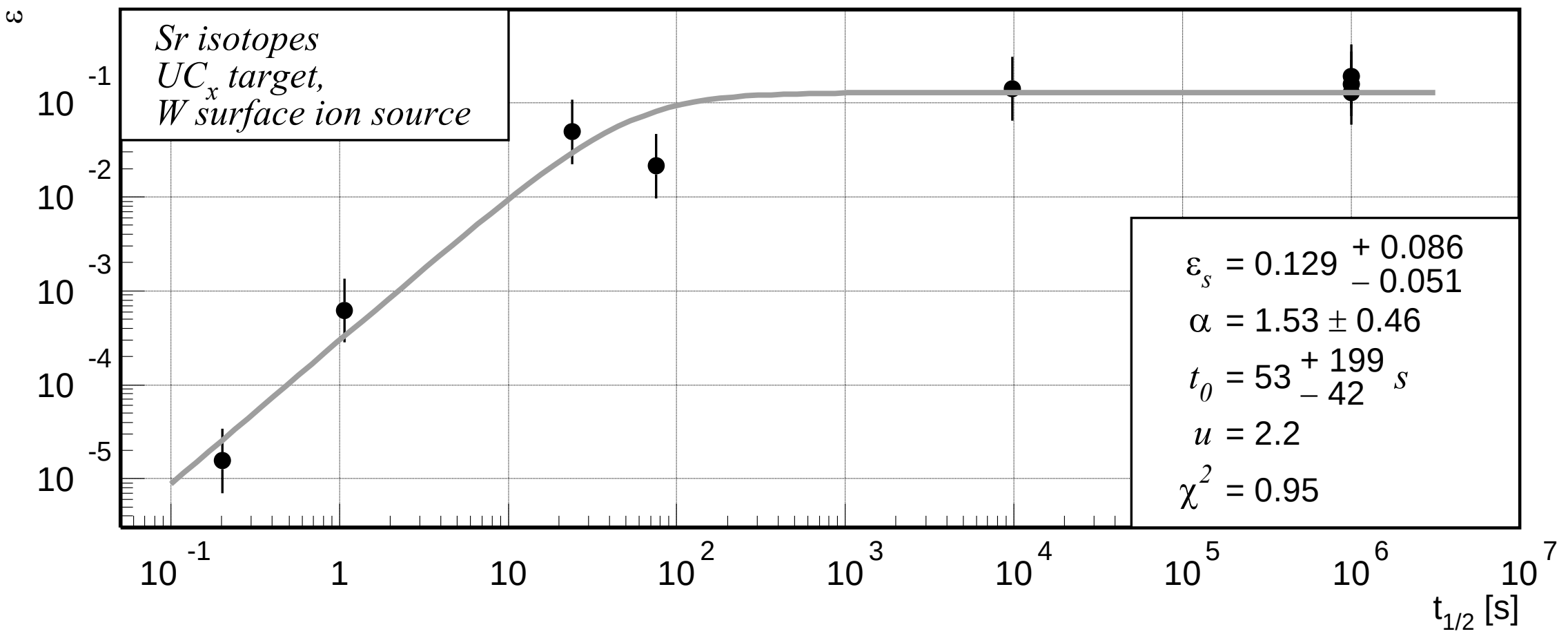

**Figure 24: The fit to the Sr data in a $UC_x$ target with a W surface ion source**

### 6.2.4 Barium

Figure 25 represents the fit to the barium efficiencies in a molten lanthanum target of 122 g/cm$^2$, with a tungsten surface ion source. There are many points corresponding to relatively long-lived isotopes. The efficiency reaches about 40% in the long half-life limit. Unfortunately, there are no yield data for isotopes with half-lives shorter than 100 s, which makes the estimation of the $\alpha$ and $t_0$ parameters rather uncertain.

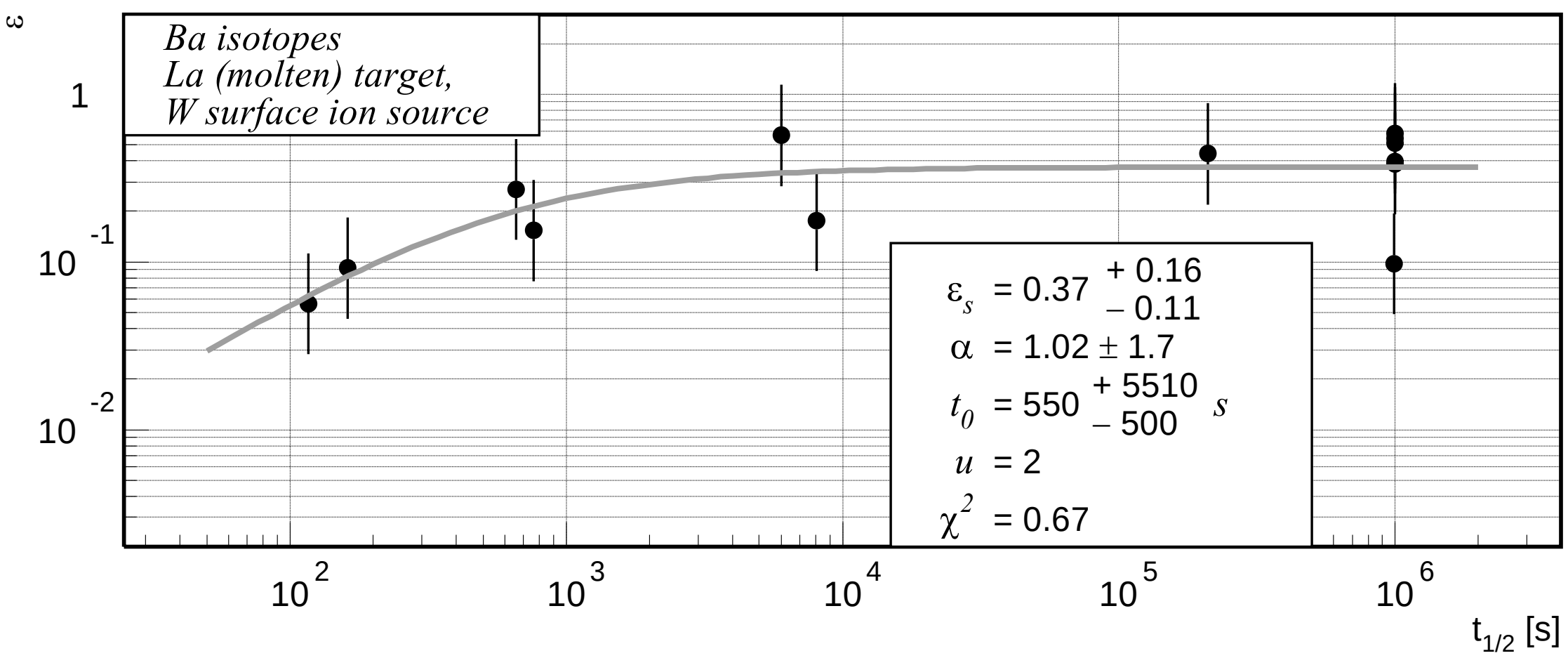

**Figure 25: The fit to the Ba efficiencies in a molten La target with a W surface ion source**

Figure 26 represents the fit to the barium data in a 15 g/cm$^2$ uranium-carbide target with a tungsten surface ion source.

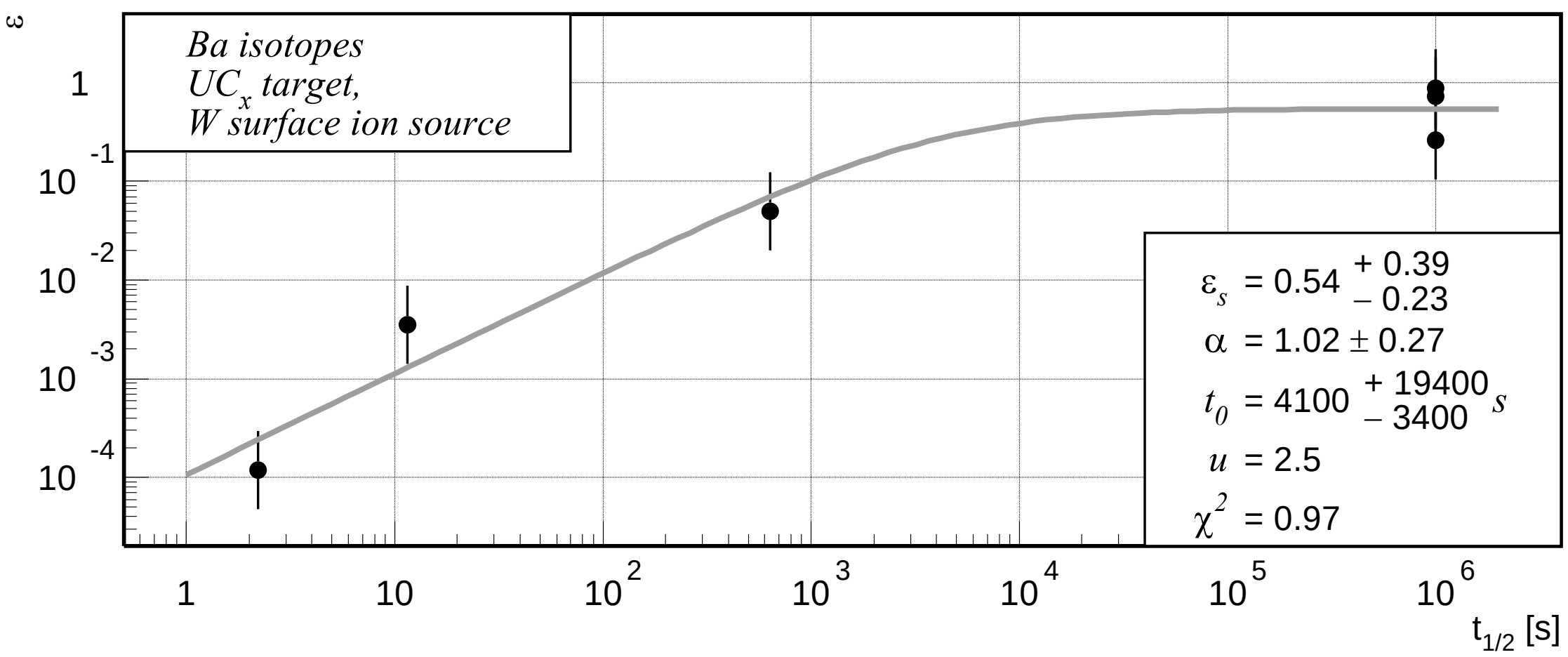

**Figure 26: The fit to the Ba data in a $UC_x$ target with a W surface ion source**

In uranium carbide, the efficiency for barium seems to be significantly higher than that for strontium. This resembles the case of cesium and rubidium in the same type of target, but there are too few data points for a more thorough discussion. In the case of barium, data are missing on the neutron-deficient side, which might be a consequence of the difficulties with the isobaric contamination from cesium.

### 6.2.5 *Radium*

Figure 27 represents the fit to the radium data in a 55 g/cm$^2$ thorium carbide target with a tungsten surface ion source. There are few data points that refer to long-lived isotopes. The fitted efficiency value for long half-lives is about 7%.

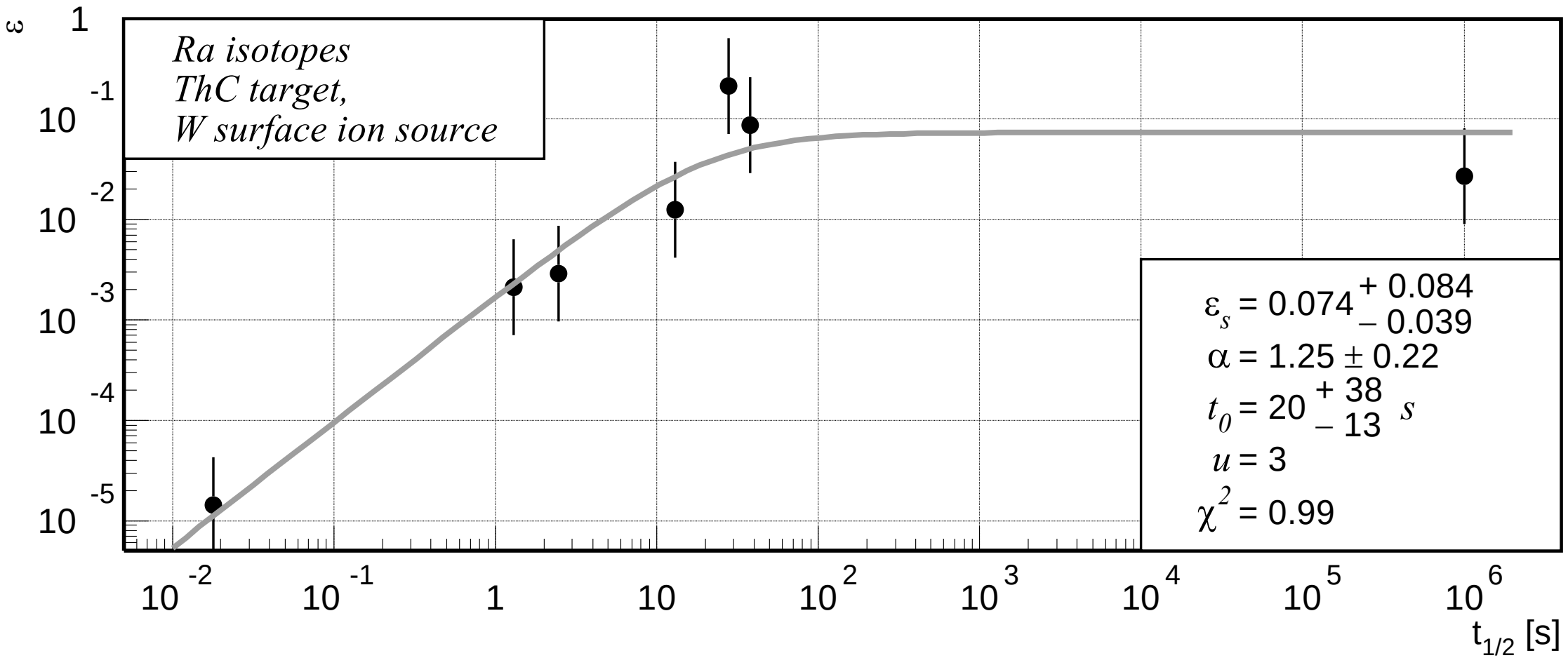

**Figure 27: The fit to the Ra data in a ThC target with a W surface ion source**

Figure 28 represents the fit to the radium data in a 13 g/cm$^2$ uranium-carbide target with a tungsten surface ion source. The fitted efficiency value in the limit of long half-lives appears to be about 240%. The reason for such an unphysical value is unknown. Radium is rather close in mass to uranium and cross-section calculations are reliable.

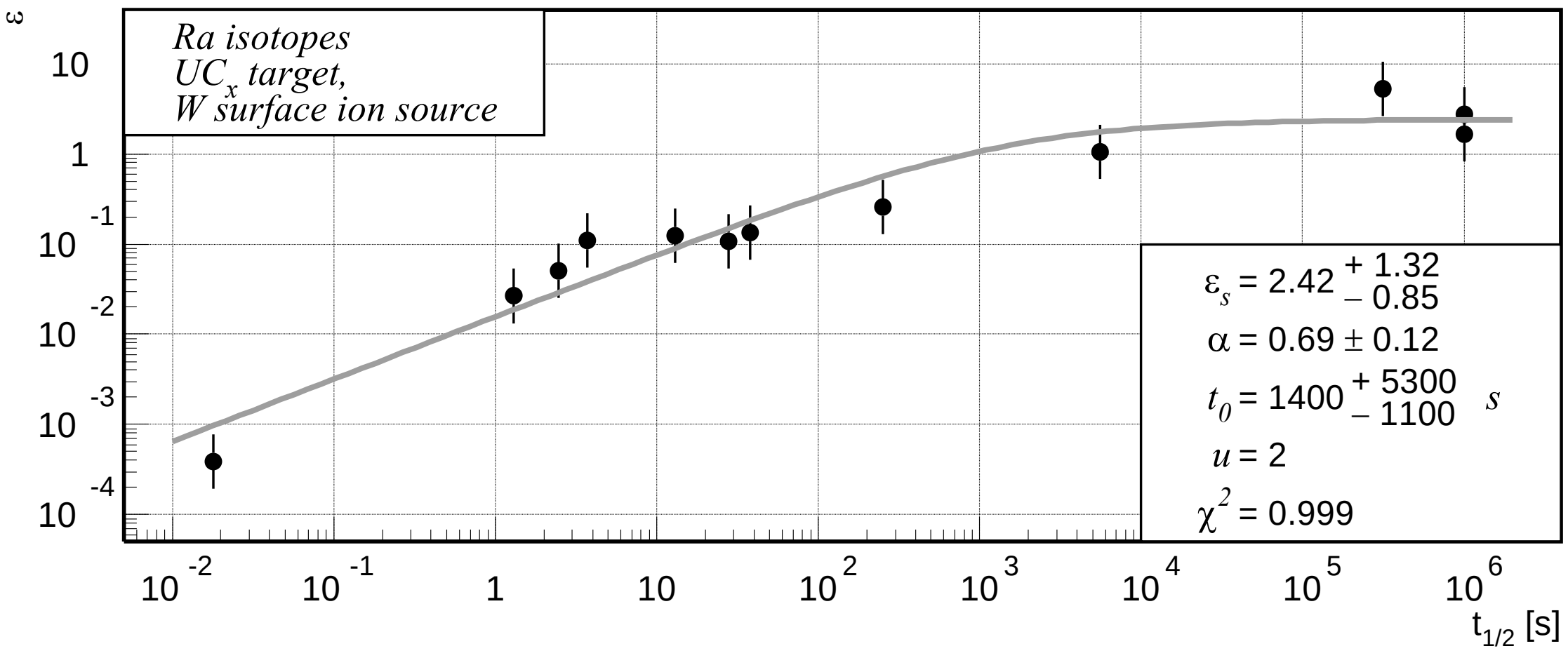

**Figure 28: The fit to the Ra data in a $UC_x$ target with a W surface ion source**

### 6.2.6 *Alkaline earths – summary*

The yield data on the different alkaline earths often include only few isotopes. In such cases, although crude estimates of the parameters are made, their uncertainties are large and it is difficult to give a consistent overview of their values. The efficiency values for long half lives in all types of targets generally tend to be higher for medium-mass elements (strontium and barium), than for lighter ones (magnesium and calcium).

## 6.3 *Group 2B*

The group 2B elements are zinc, cadmium and mercury. The volatility of these elements allowed making their beams rather early [33]. The efficiency values for zinc isotopes in all kinds of targets are too scattered for the function 14 to be fitted (see sect. 7) but the results for cadmium in uranium carbide and mercury in molten lead seem to be quite reliable.

### *6.3.1 Cadmium*

Figure 29 represents the fit to the cadmium data in uranium carbide. The thickness of the target was 13.6 g/cm$^2$. The efficiency reaches 1% for half-lives of the order of seconds. For shorter half-lives, the efficiency follows the power law dependence on the half-life with the exponent typical for solid matrices.

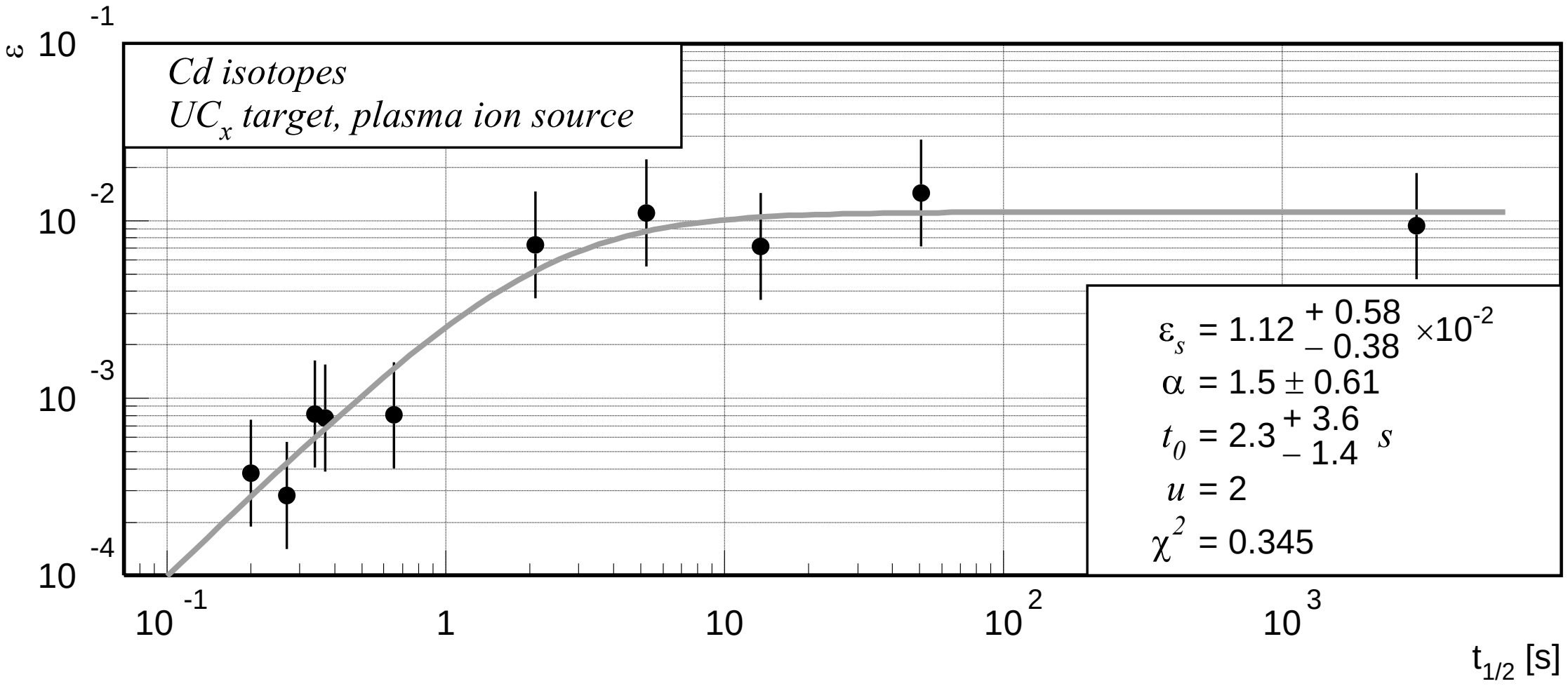

**Figure 29: The fit to the data for Cd isotopes in a UC$_x$ target with a plasma ion source.**

### *6.3.2 Mercury*

Figure 30 represents the fit to the mercury data in molten lead. The target thickness was 170 g/cm$^2$. The yield data [1,29,32] cover a wide range of masses and half-lives, and the efficiency curve is very smooth. For long half-lives, the efficiency reaches 8%.

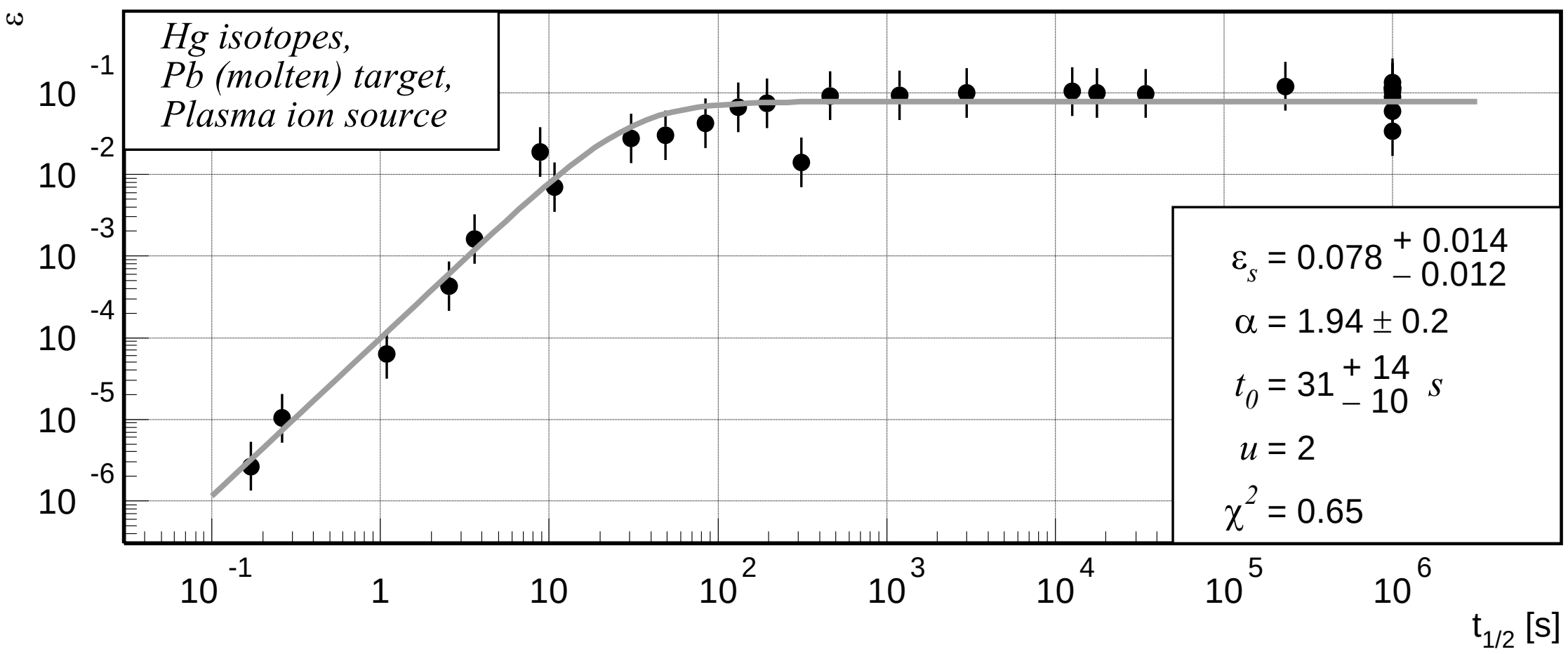

**Figure 30: The fit to the data for Hg isotopes in a molten Pb target with a plasma ion source.**

## *6.4 Halogens*

Due to their high electron affinity, these elements can be extracted with chemical selectivity using negative surface ionization. They are volatile at relatively low temperatures. [32,62,63]

### *6.4.1 Chlorine*

The figure 31 represents the fit to the data for chlorine isotopes in a 10 g/cm$^2$ uranium-carbide target with a negative surface ion source. The efficiency value in the limit of long half-lives is 5%.

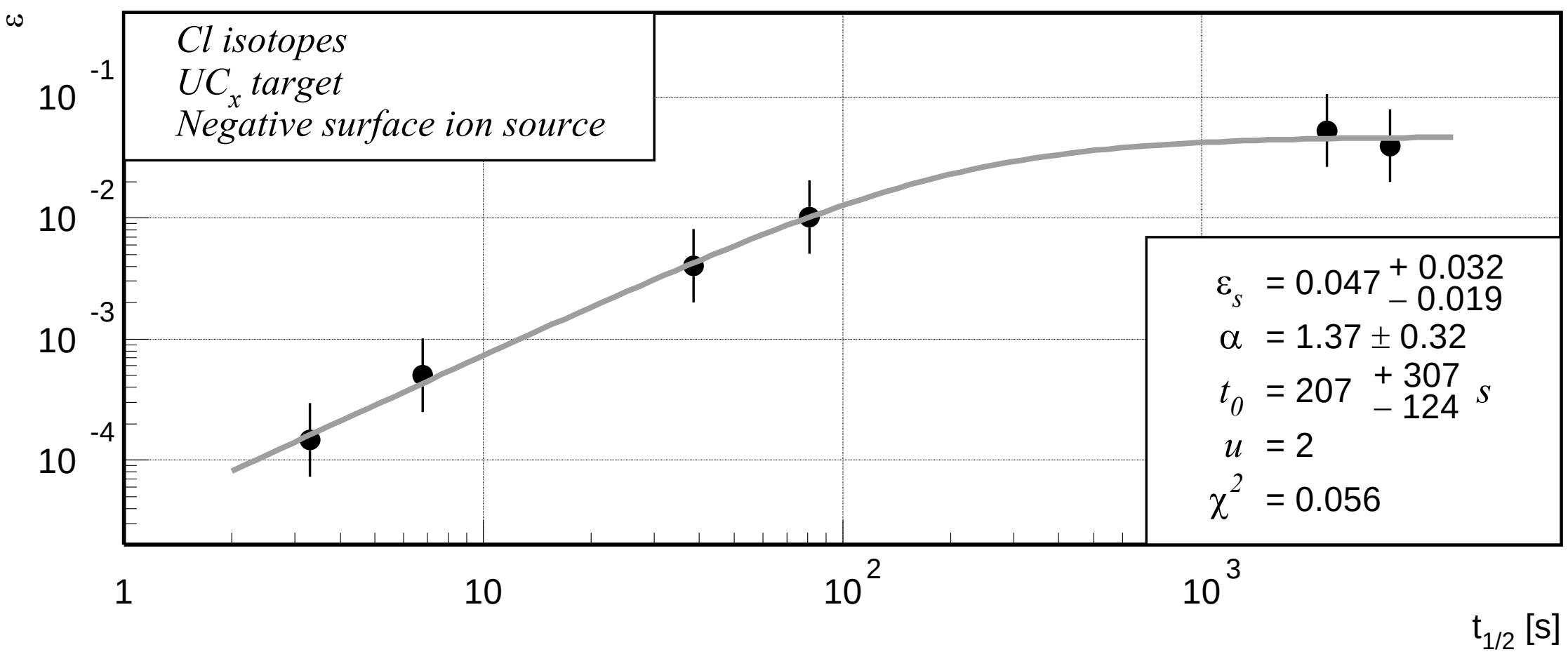

**Figure 31: The fit to the data for Cl isotopes in a $UC_x$ target with a negative surface ion source.**

The yields of chlorine from a thorium-oxide target with 46 g/cm$^2$ of thorium are proportionally higher than those from a 10 g/cm$^2$ uranium-carbide target, except for the most neutron-rich isotopes, because of their short half-life. The diffusion from the uranium-carbide matrix is faster, which allows for better efficiency with short-lived isotopes. Indeed, the yield of $^{42}$Cl (half-life 6.8 s) is 10 times higher from uranium carbide. Figure 32 represents the fit to the data for chlorine isotopes from thorium oxide. The slower diffusion from thorium oxide with respect to uranium carbide is reflected in the higher value of the exponent $\alpha$ of the power-function behavior for the short half-lives.

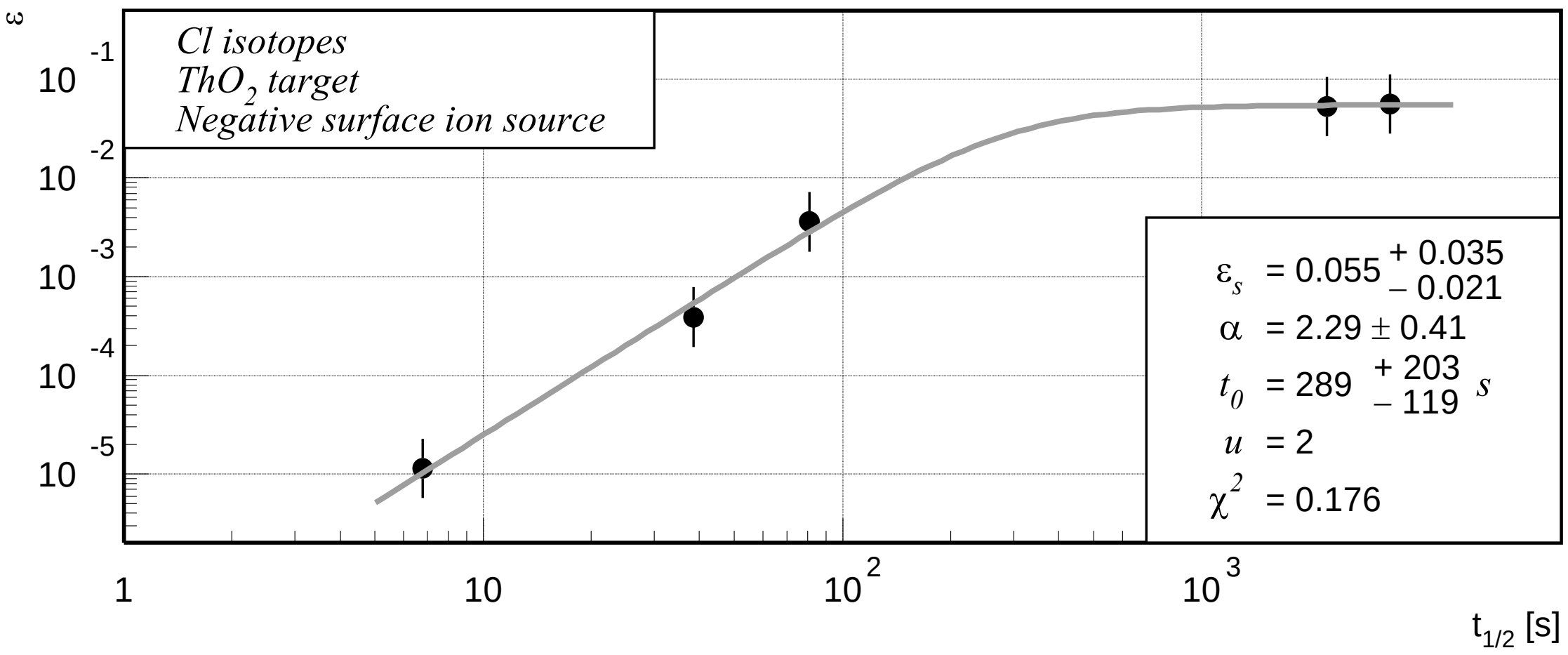

**Figure 32: The fit to the data for Cl isotopes in a $ThO_2$ target with a negative surface ion source.**

Figure 33 represents the fit to the data for chlorine isotopes in a tantalum/niobium mixed-powder target with a negative surface ion source. The total target thickness was 88 g/cm$^2$. The saturation efficiency is 11%. $^{34}$Cl with 1.5 s ground-state half-life is also produced in a metastable state with 32 minutes half-life. However, the yields in the metastable state are more than one order of magnitude lower than those of the isotopes with comparable half-lives, e.g. $^{38,39}$Cl [1]. Therefore, we conclude that only a small fraction (< 10%) of $^{34}$Cl is produced in the metastable state, and the ground state yield is not significantly affected by it.

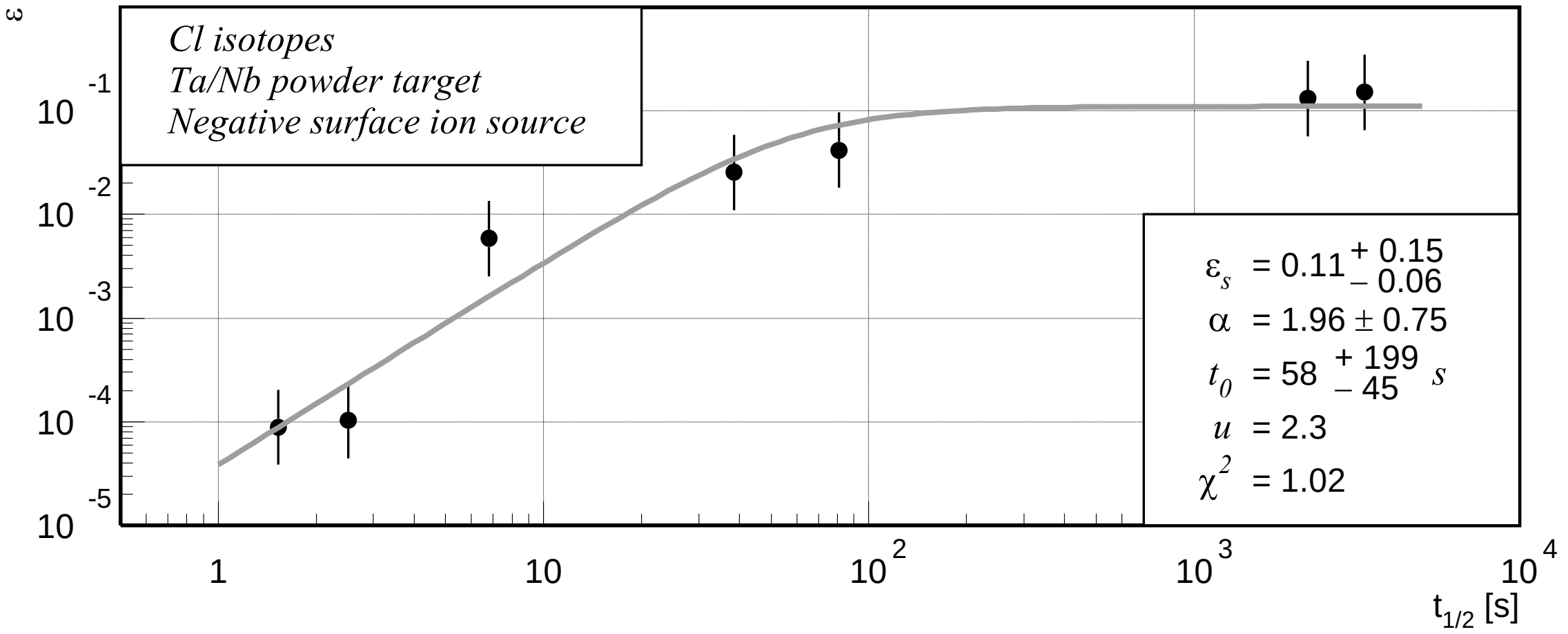

**Figure 33: The fit to the data for Cl isotopes in a Ta/Nb powder target with a negative surface ion source**

### *6.4.2 Bromine*

Figure 34 represents the fit to the data for bromine isotopes in a 13 g/cm$^2$ uranium-carbide target with a negative surface ion source. The efficiency in the limit of long half-lives is 4%.

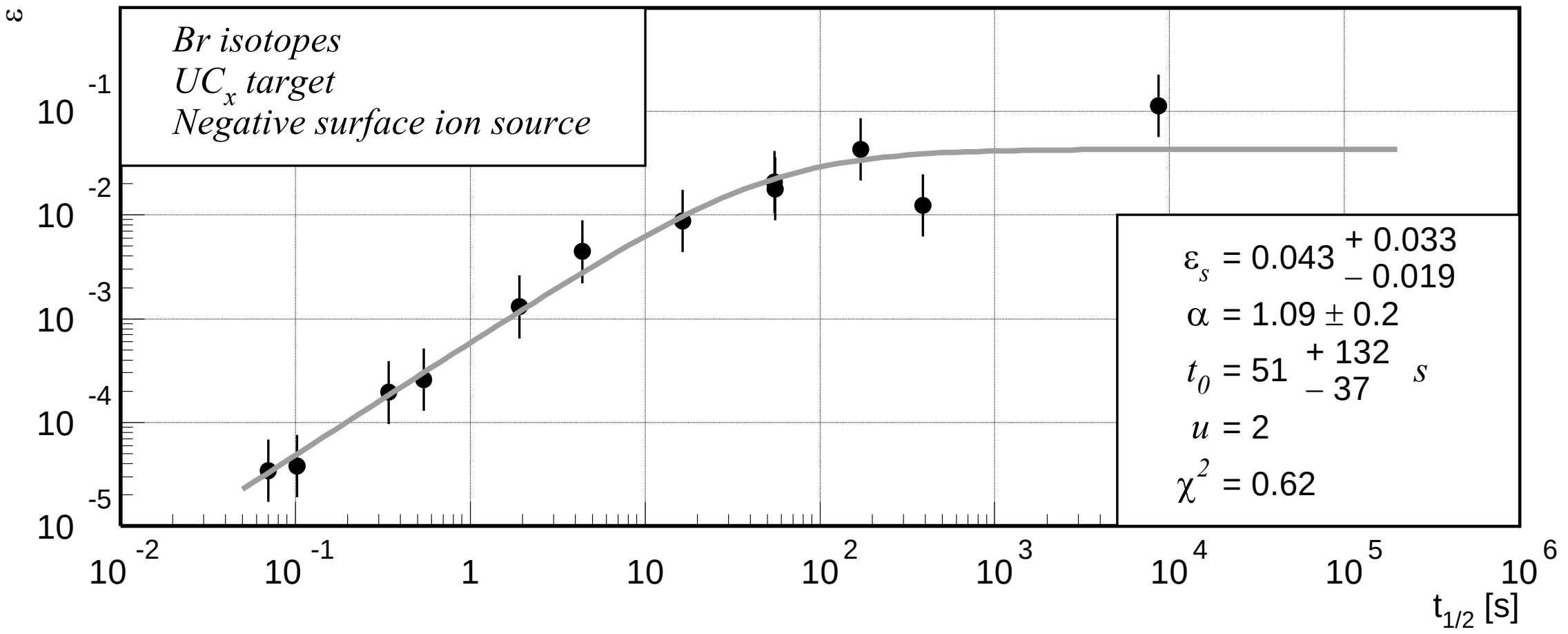

**Figure 34: The fit to the data for Br isotopes in a $UC_x$ target with a negative surface ion source.**

Figure 35 represents the fit to the data for bromine isotopes in a 46 g/cm$^2$ thorium-oxide target with a negative surface ion source. The saturation efficiency is around 3%.

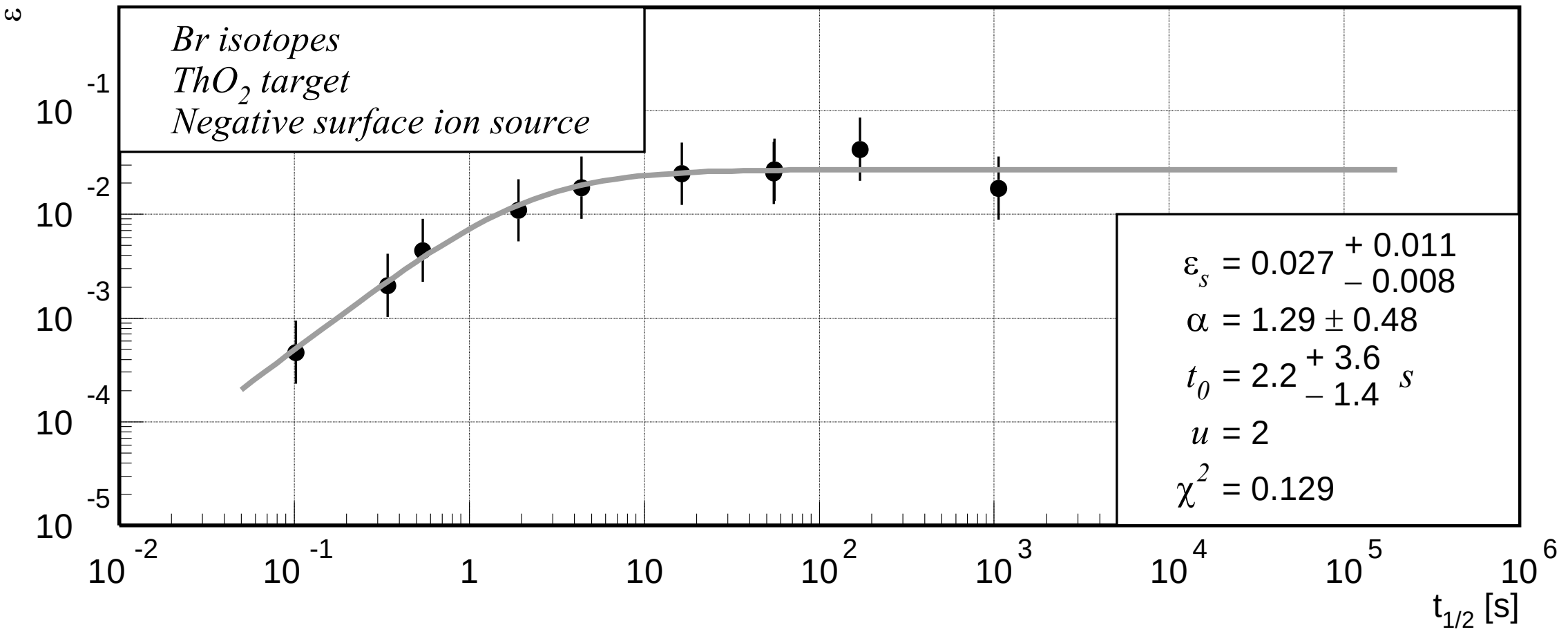

**Figure 35: The fit to the data for Br isotopes in a $ThO_2$ target with a negative surface ion source.**

Figure 36 represents the fit to the bromine data in a niobium metal-powder target with a negative ion source. The thickness of the niobium target was 85 g/cm$^2$, and the energy of the protons behind the target is around 450 MeV. At this energy bromine production cross sections shift very slightly toward the neutron-rich side, the cross sections for individual nuclides changing by up to a factor of two. Therefore, it is safe to integrate the production rates using their mean values.

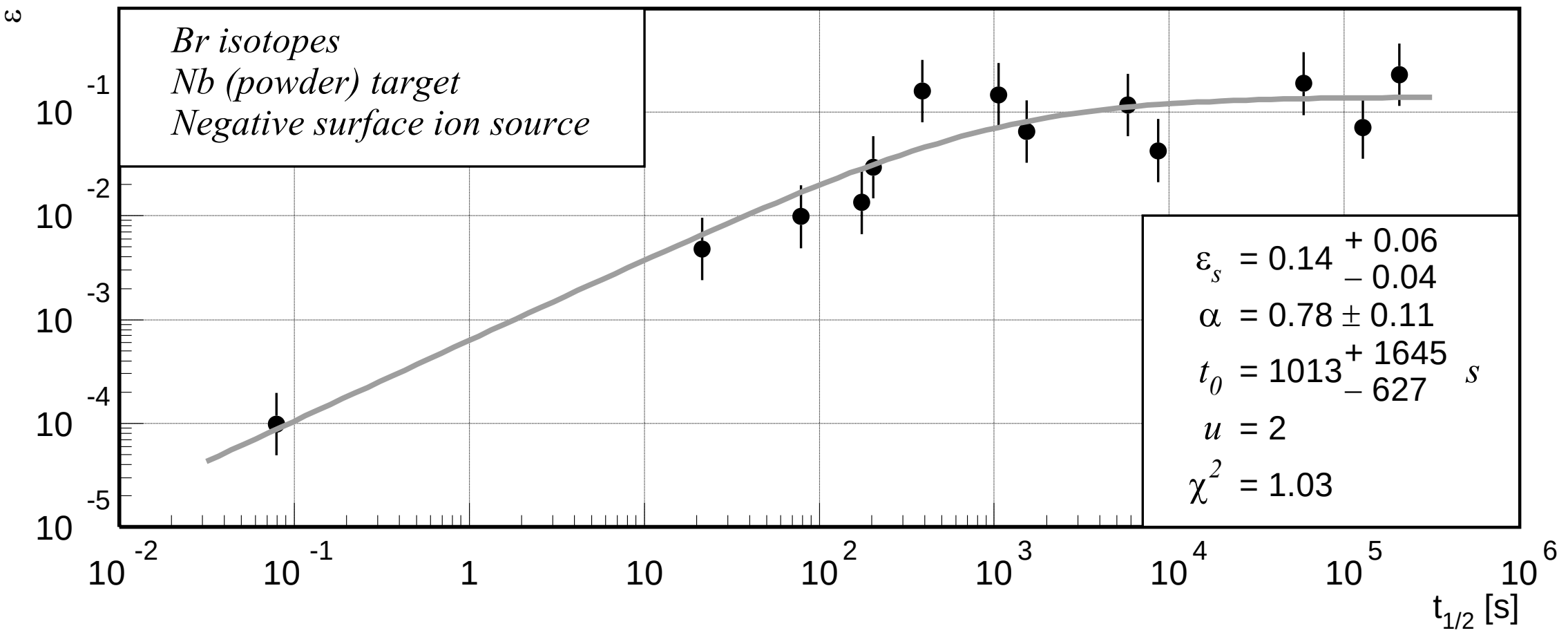

**Figure 36: The fit to the data for Br isotopes in a Nb target with a negative surface ion source.**

The saturation efficiency for bromine in niobium is higher than in uranium carbide or thorium oxide.

### *6.4.3 Iodine*

Figure 37 represents the fit to the data for iodine isotopes in a 10 g/cm$^2$ uranium-carbide target with a negative surface ion source. The saturation efficiency is surprisingly high in comparison to that for bromine and chlorine from the same target. A limit was imposed on the saturation-efficiency parameter while fitting so that it could not exceed unity. However, as yield data exist only for few relatively long-lived isotopes, it is not clear whether they reflect the overall tendency. The iodine isotopic yield distribution exhibits a rather unexpected shape and magnitude in comparison with ABRABLA calculations (see figure 38). Therefore, the results presented here for iodine should be regarded with caution.

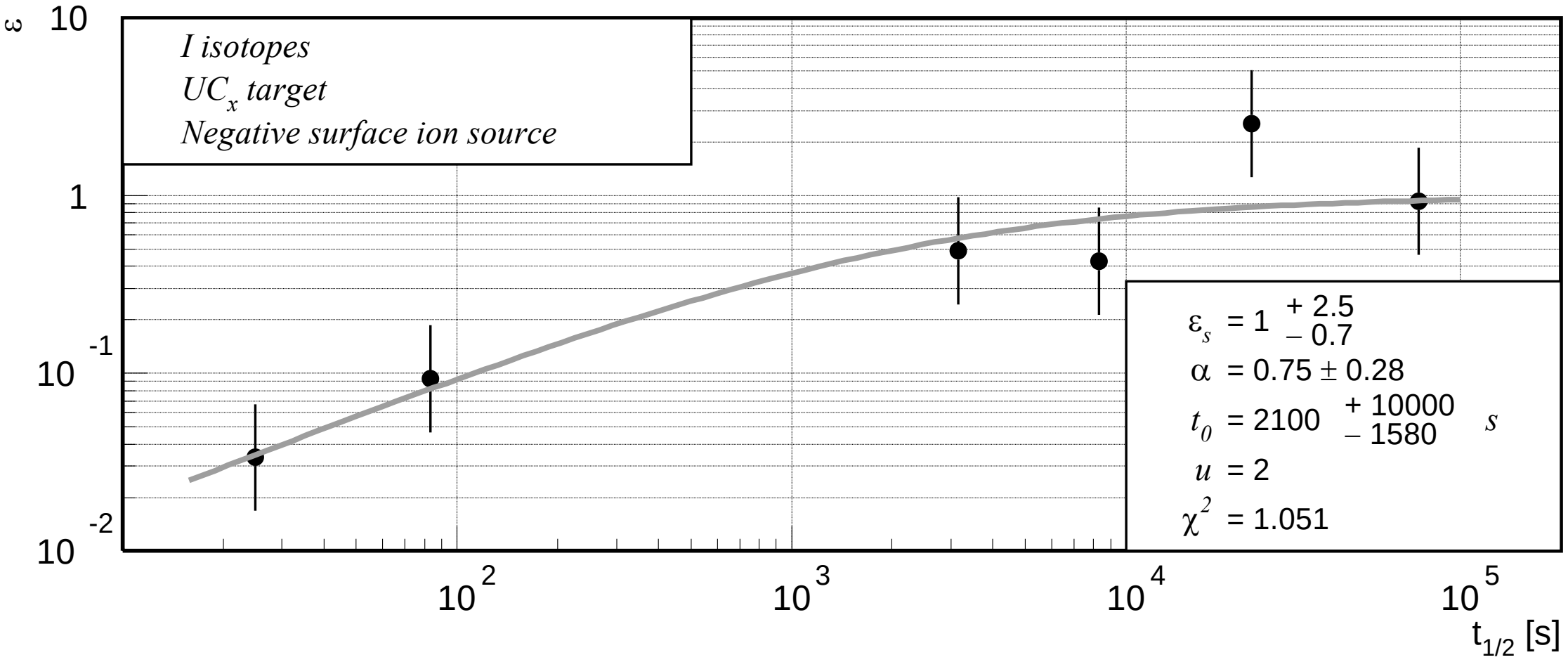

**Figure 37: The fit to the data for I isotopes in a $UC_x$ target with a negative surface ion source.**

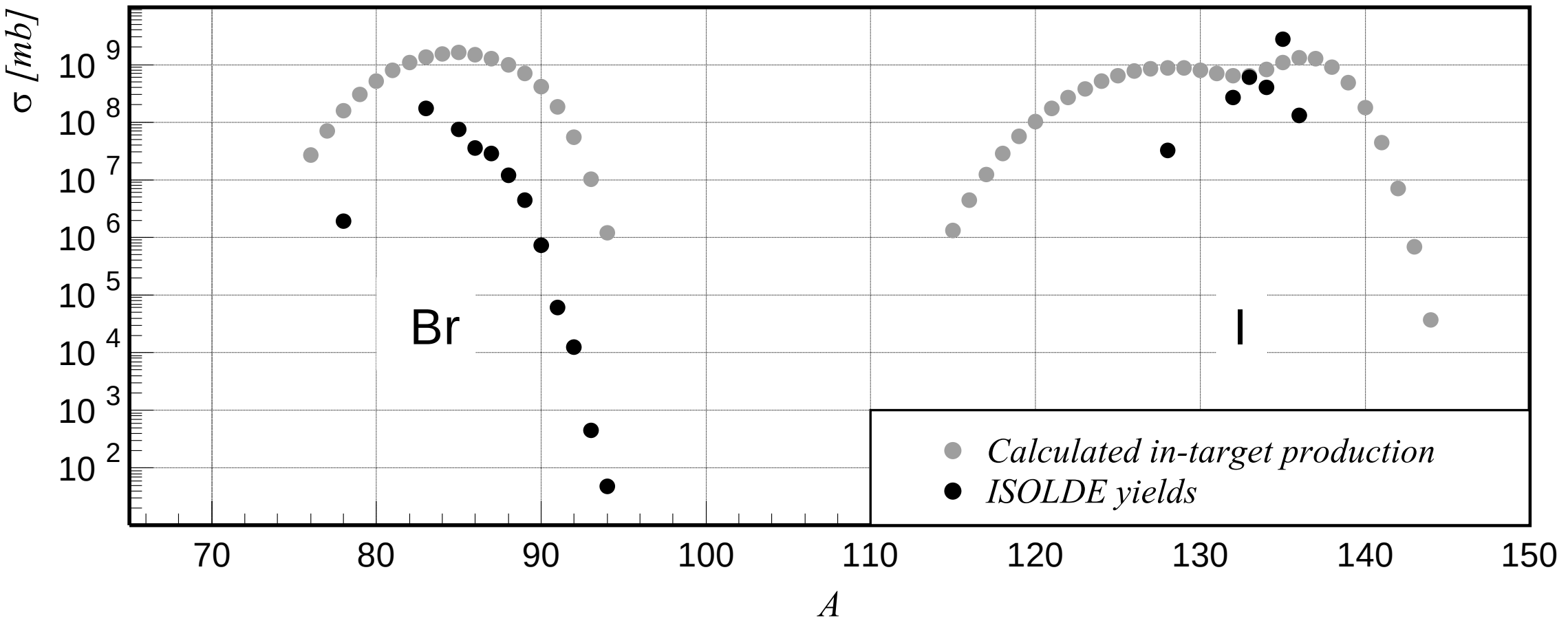

**Figure 38: Comparison of calculated in-target rates and ISOLDE yields for Br and I isotopes in the UCx target.**

Figure 39 represents the fit to the data for iodine isotopes in a 46 g/cm$^2$ thorium-oxide target with a negative surface ion source. The efficiencies for the neutron-deficient isotopes appear systematically lower than those for the neutron-rich isotopes of a comparable half-life. Particularly, the points corresponding to the two most neutron-deficient isotopes that are present in the ISOLDE database, $^{116,117}$I, are remarkably lower than the trend set by the isotopes of similar half-life. These are plotted as empty points in the graph. The reason for this might be the following: In the neutron-deficient region of this mass range, the production cross sections depend very strongly on the incident proton energy [61]. As $^{116,117}$I are situated at the outermost slope of the neutron-deficient part of the isotopic distribution, that makes it difficult to precisely calculate the cross sections.

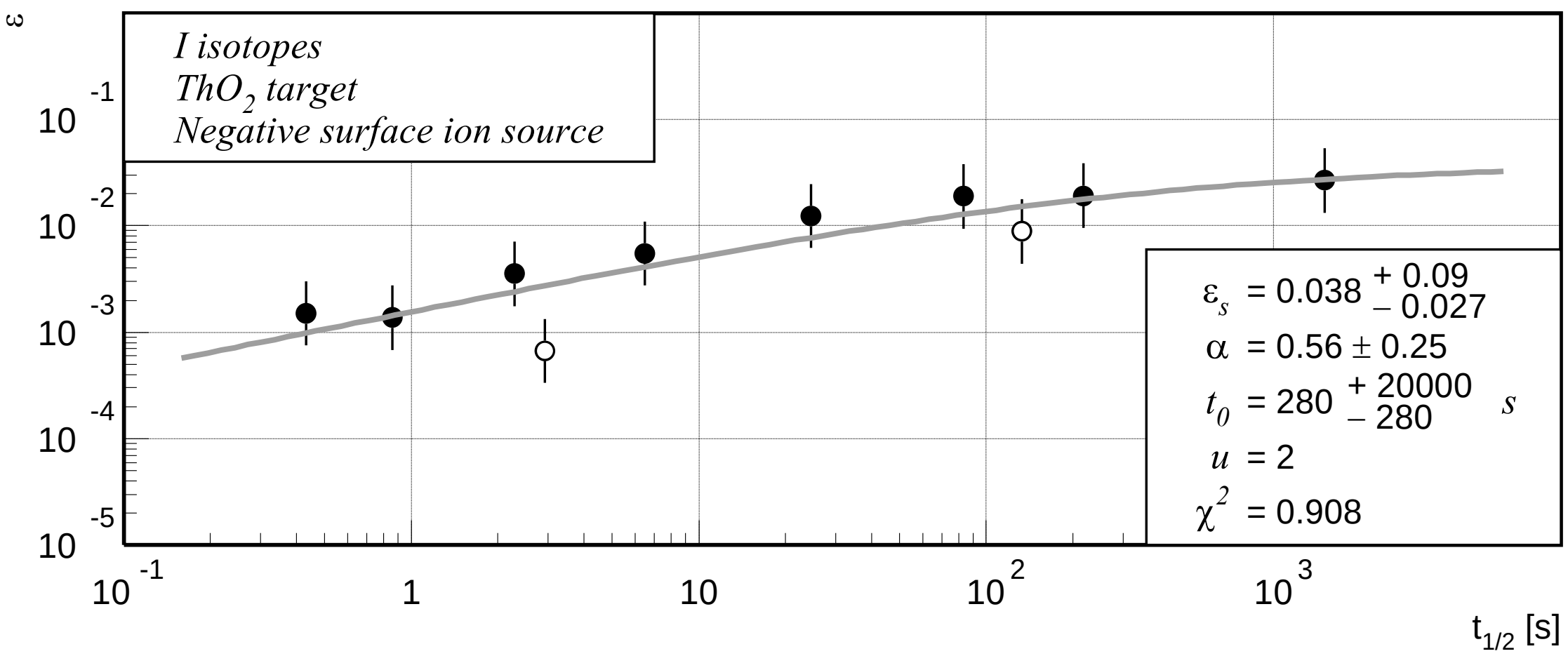

**Figure 39: The fit to the data for I isotopes in a $ThO_2$ target with a negative surface ion source. The two most neutron-deficient isotopes, $^{116,117}$I, are plotted as empty points.**

Figure 40 represents the fit to the data for iodine isotopes in a 19 g/cm$^2$ barium-zirconate target with a negative surface ion source. The saturation efficiency value is 0.4%.

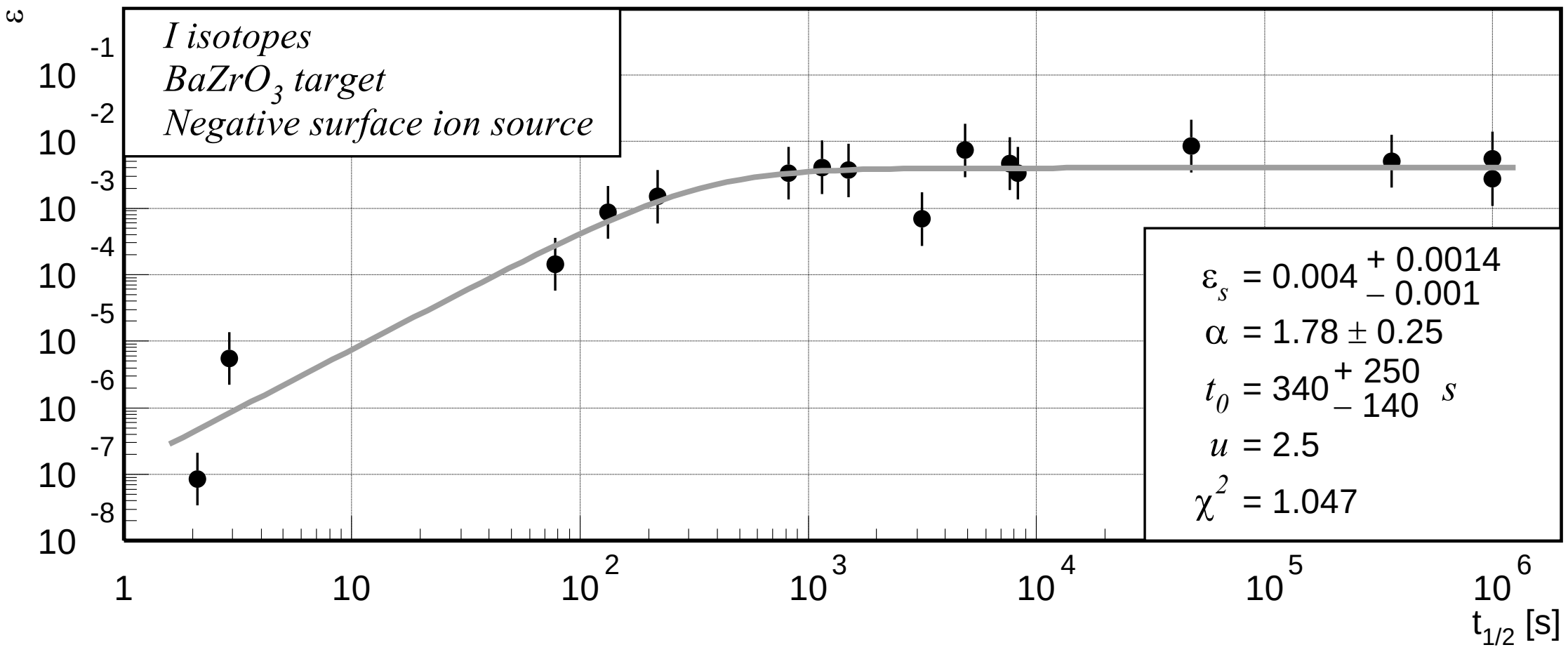

**Figure 40: The fit to the data for I isotopes in a $BaZrO_3$ target with a negative surface ion source.**

### *6.4.4 Astatine*

Figure 41 represents the fit to the data for astatine isotopes in a 47 g/cm$^2$ thorium-oxide target with a negative surface ion source. There are two astatine isotopes for which the side feeding by α decay from francium and actinium can be easily accounted for. The α-decay precursors of $^{210,211,212}$At are $^{214,215,216}$Fr that are produced with cross sections several times higher than those of the corresponding astatine isotopes. Besides, the half-lives of $^{215,216}$Fr are so short that we can safely assume that they decay to astatine before diffusing out of the target. The half-life of $^{214}$Fr is 5 ms. In uranium carbide, the fraction extracted before decay is of the order of $10^{-5}$ (see figure 20). The fraction extracted from thorium carbide is probably of the same order, or smaller. Therefore, we have assumed that it decays 100% to $^{210}$At before leaving the target. A smaller, but also important contribution comes from the cascade of α decays from $^{218,219,220}$Ac through $^{214,215,216}$Fr. $^{220}$Ac has a longer half-life than $^{214}$Fr but due to its chemical properties that are similar to thorium, we assume that it does not diffuse out of the target. We have included these contributions into our estimated in-target production rates. For comparison, the extraction efficiencies for $^{210,211,212}$At that would be obtained without this correction are shown by empty points in the figure 41. Of course, this is far from a complete treatment of the side feeding, but is a good example of the possible size of the effect.

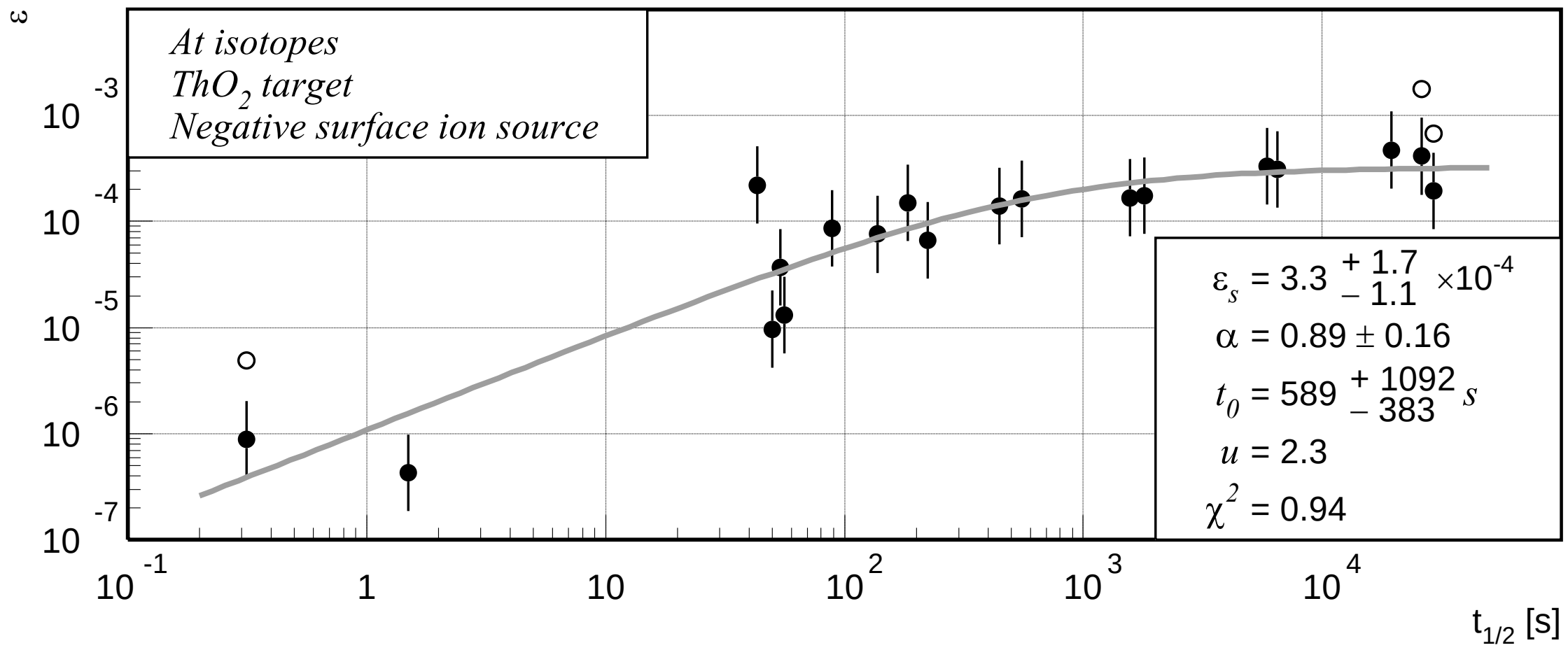

**Figure 41: The fit to the data for At isotopes in a $ThO_2$ target with a negative surface ion source. Empty points represent values obtained for the respective isotopes without correction for the side feeding from Fr and Ac.**

### *6.4.5 Halogens - summary*

The extraction efficiencies for halogens seem to be significantly lower than those for the alkalis in all types of targets. That probably indicates a lower efficiency of negative ion sources. The difference is the most pronounced with the uranium-carbide target, which could, in addition, reflect unfavorable chemical properties of the uranium carbide for the diffusion of the halogens.

## ***6.5 Noble gases***

The noble gases are the easiest elements to produce by the ISOL method because they are the least reactive of all the elements and are volatile even at room temperature. They have been produced at ISOLDE using a wide variety of targets. Heavier materials based on fissile elements, like thorium carbide, were used for more neutron-rich isotopes and lighter materials for the neutron-deficient ones [31]. Extraction using plasma ion sources with cooled transfer lines allows for chemical selectivity through condensation of less volatile elements. Targets based on alkaline-earth oxides with such ion sources show particularly fast extraction properties [34].

### *6.5.1 Neon*

Yield data exist for neon in two alkaline-earth oxide targets, magnesium-oxide and calcium-oxide. Both targets were used with plasma ion sources with cooled transfer lines. As expected, the tendencies in neon extraction efficiencies from these two systems are similar. The fast extraction is reflected in the fact that the saturation value is reached for half-lives of the order of seconds.

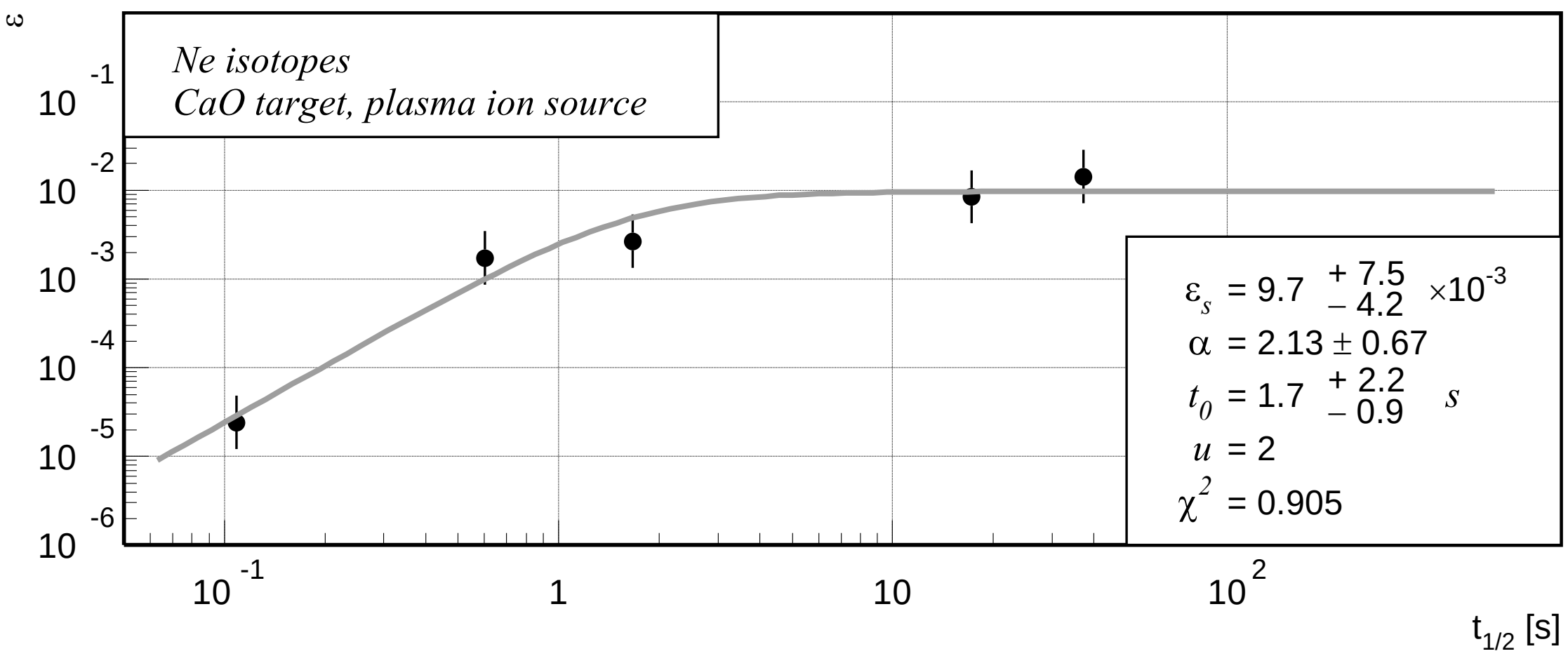

**Figure 42: The fit to the data for Ne isotopes in a CaO target with a plasma ion source and a cooled transfer line.**

Figure 42 represents the fit to the data for neon isotopes in a 5.4 g/cm$^2$ calcium-oxide target with a plasma ion source and a cooled transfer line. The efficiency value in the limit of long half-lives is around 1%.

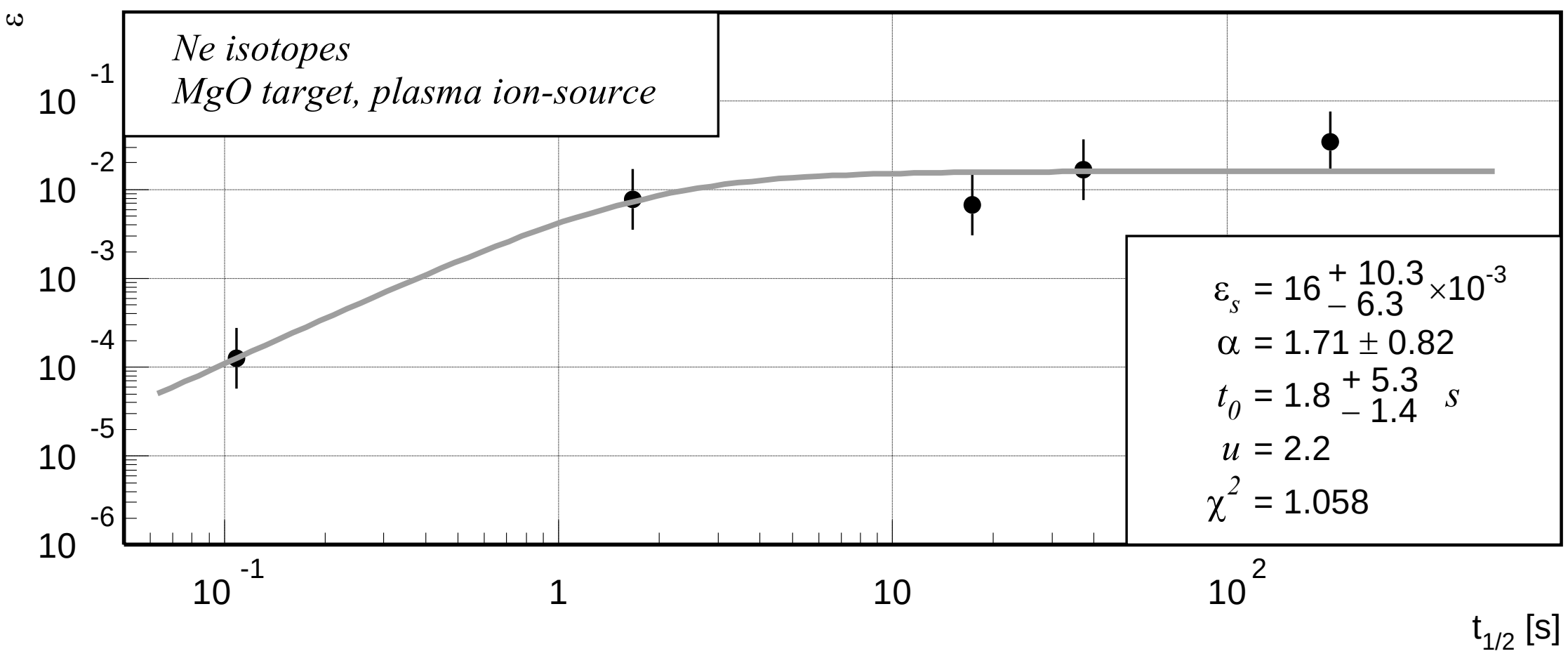

**Figure 43: The fit to the data for Ne isotopes in a MgO target with a plasma ion source.**

Figure 43 represents the fit to the data for neon isotopes in a 3 g/cm$^2$ magnesium-oxide target with a plasma ion source and a cooled transfer line.

### *6.5.2 Argon*

Figure 44 represents the fit to the data for argon isotopes in a 5 g/cm$^2$ calcium-oxide target with a plasma ion source and a cooled transfer line.

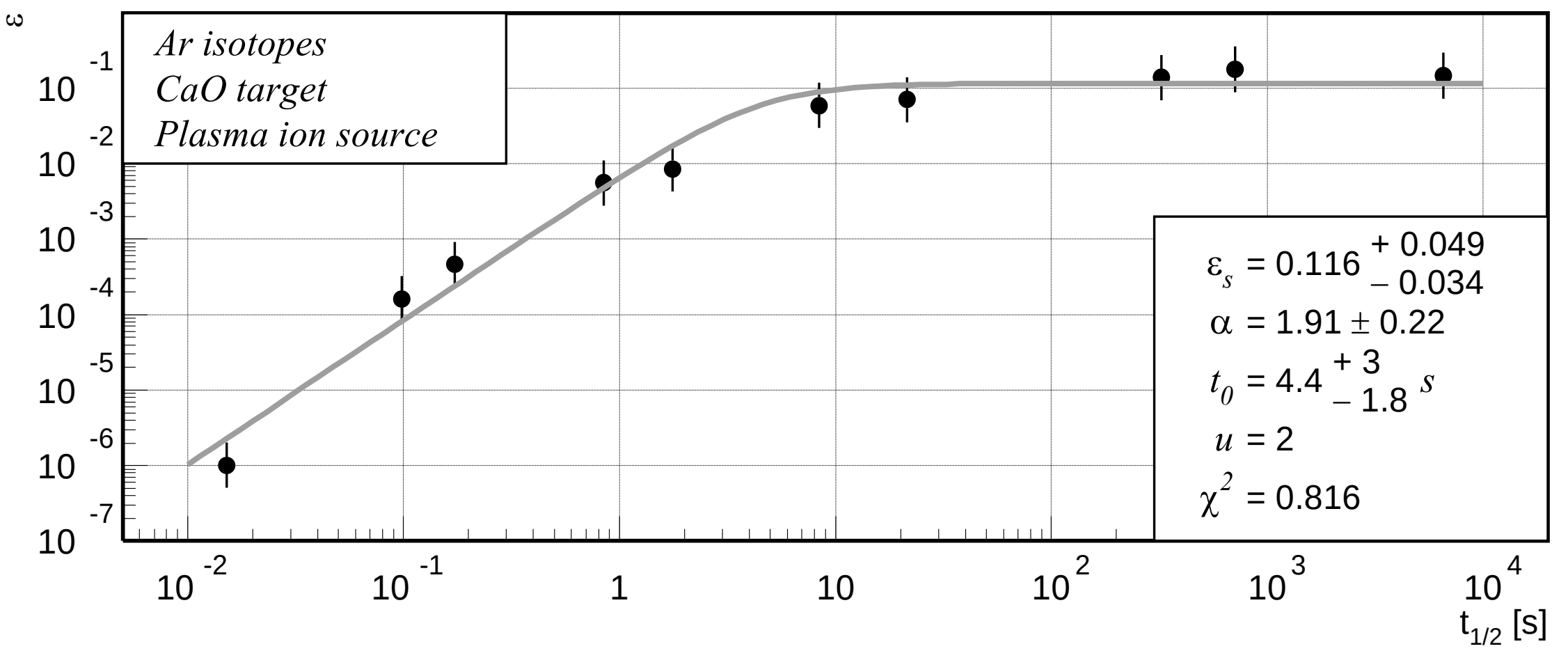

**Figure 44: The fit to the data for Ar isotopes in a CaO target with a plasma ion source and a cooled transfer line.**

### *6.5.3 Krypton*

Figure 45 represents the fit to the data for krypton isotopes in a 55 g/cm$^2$ thorium-carbide target with plasma ion source and a cooled transfer line.

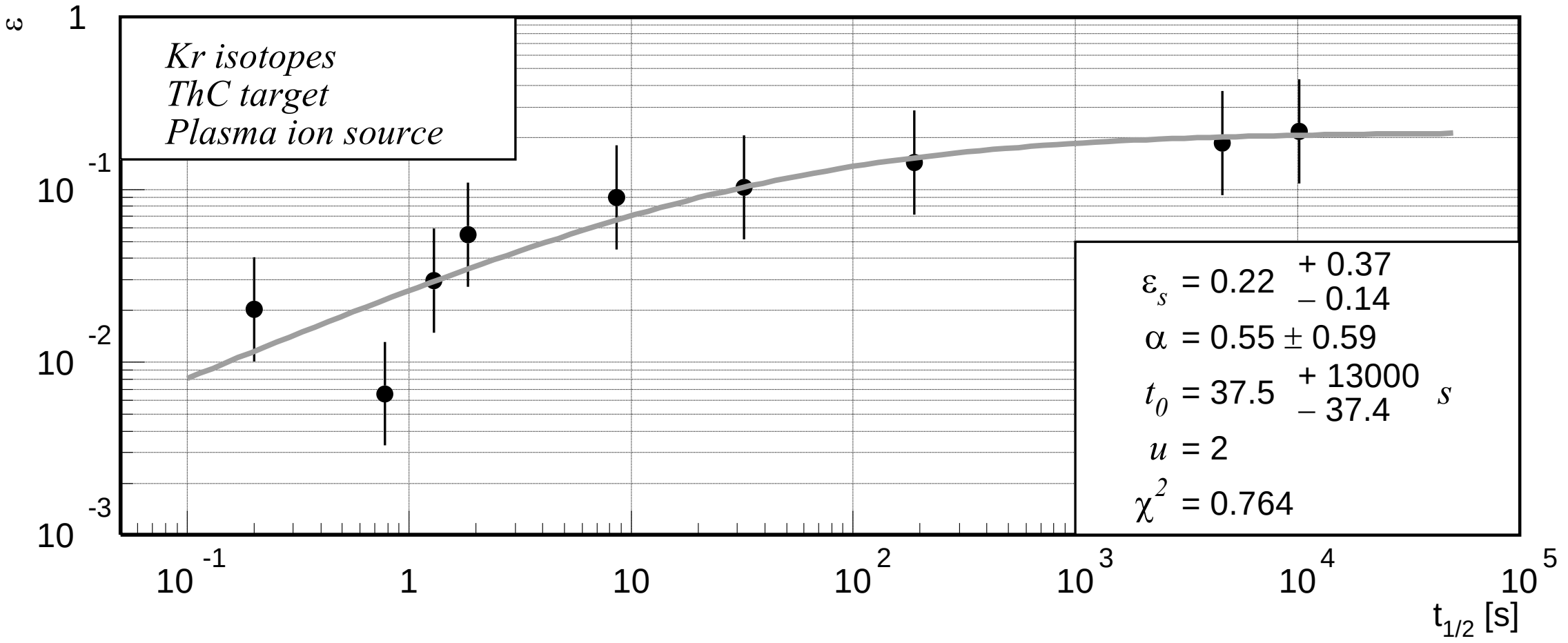

**Figure 45: The fit to the data for Kr isotopes in a ThC target with a plasma ion source and a cooled transfer line.**

### 6.5.4 Xenon

Figure 46 represents the fit to the data for xenon isotopes in a 55 g/cm$^2$ thorium-carbide target with a plasma ion source and a cooled transfer line. Xenon and krypton are conjugate fragments in the fission of thorium, in a similar way as cesium and rubidium in the fission of uranium. Therefore, their production rates can be expected to be comparable, which is indeed the case in our calculations. However, as in the case of cesium, the yields of the xenon neutron-deficient isotopes are surprisingly high in comparison with the expected production rates. The yield of $^{122}$Xe is even one order of magnitude higher than the calculated in-target production rate, and stands far out of the yields of other isotopes with comparable half-life or comparable mass number. The upper limit was set on the saturation efficiency parameter while fitting so that it could not exceed unity.

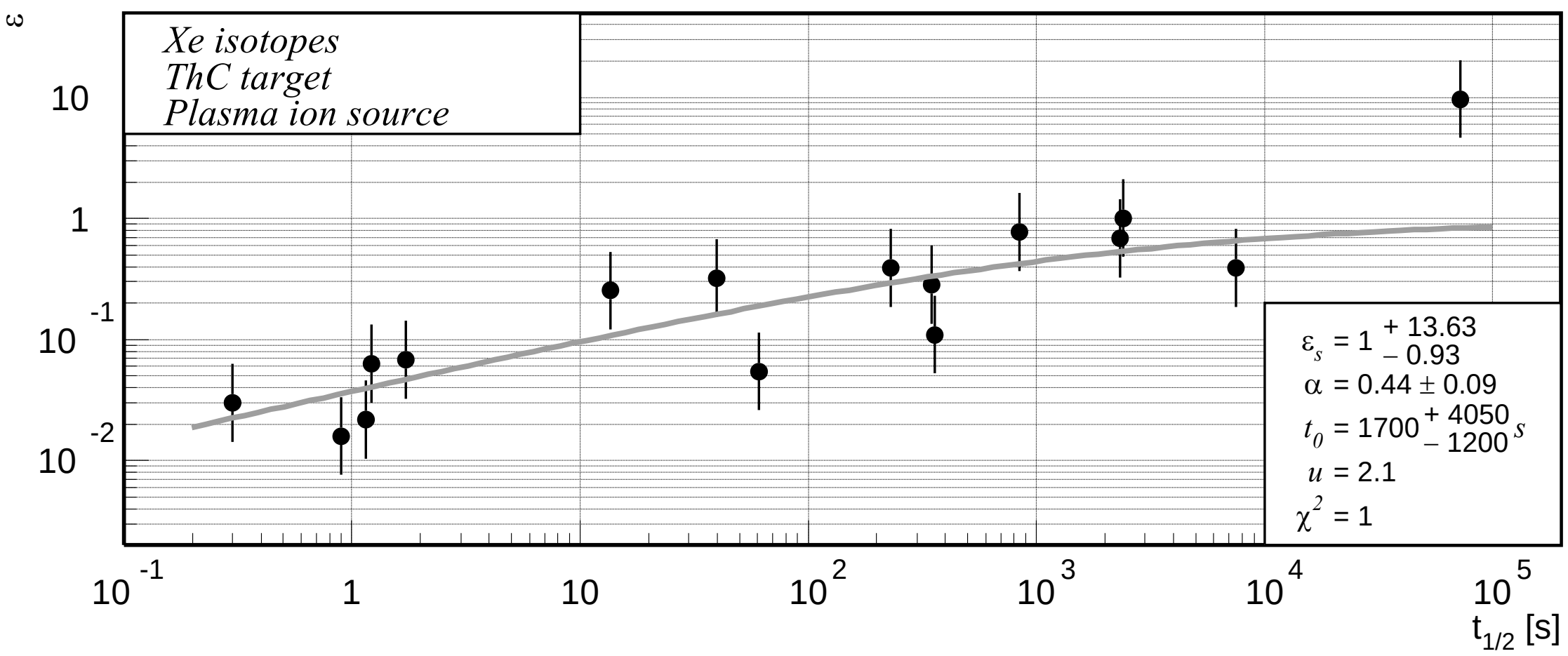

**Figure 46: The fit to the data for Xe isotopes in a ThC target with a plasma ion source and a cooled transfer line.**

## 7. Cases where the tendency is difficult or impossible to establish

When studying the ratio of the ISOLDE final yields to the in-target production rates, with particular attention to its dependence on the isotope half-life, one encounters difficulties arising from several sources of systematic uncertainties, the magnitude of which would be difficult to estimate:

- difficulties to precisely calculate cross sections at the steep outer slopes of isotopic distributions far off stability
- isobaric and molecular contaminations of ISOLDE beams
- side feeding
- unknown fraction of nuclides of a certain type that are produced in isomeric states

We have already seen how these uncertainties and their combinations can affect extraction efficiencies for individual isotopes in different isotopic chains. Often in these cases we can still obtain rough estimates of the parameters of the half-life dependence of the extraction efficiency, but sometimes the situation is such that we have reasons to doubt that the obtained values are correct within their uncertainty ranges.

There are also cases where it was impossible to describe the dependence of the extraction efficiency on the isotope half-lives by the function 14. For illustration, we include here several such cases.

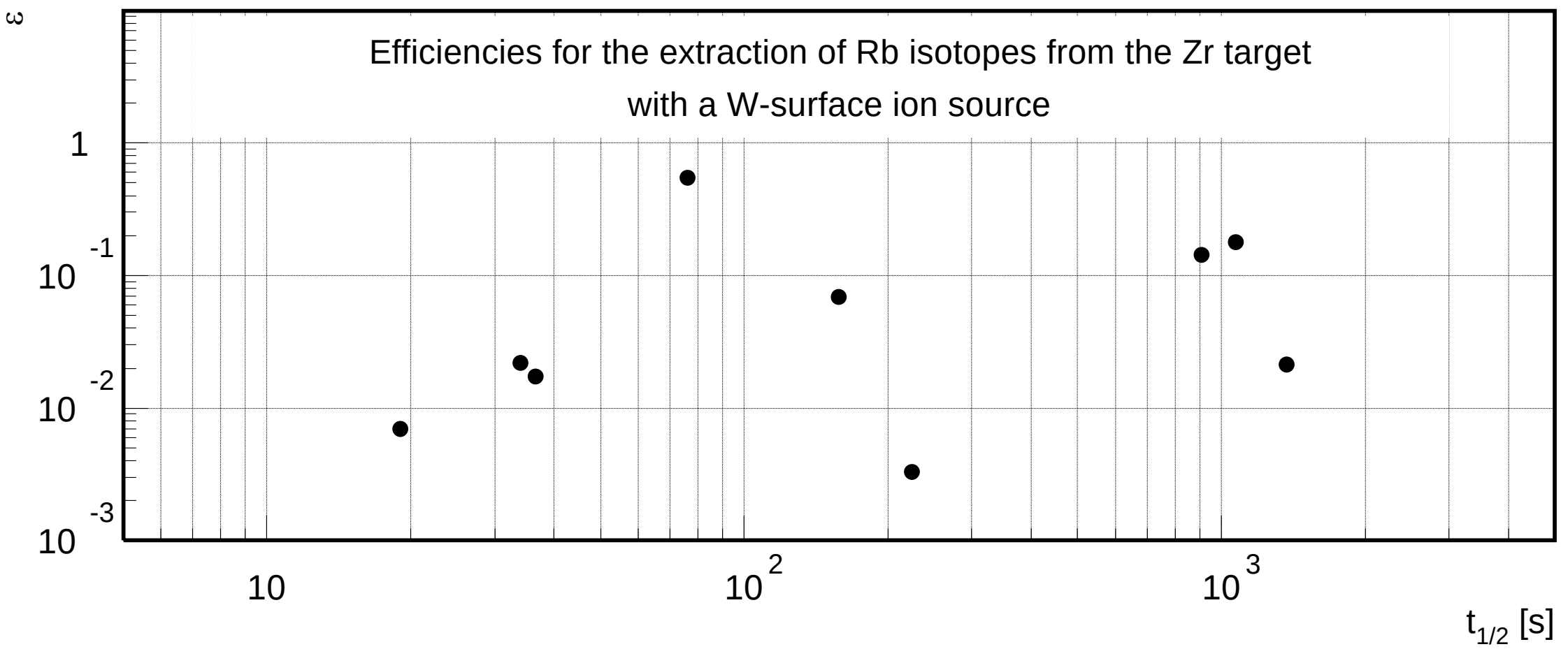

**Figure 47: Overall extraction efficiencies for Rb isotopes from the Zr target with a W-surface ion source.**

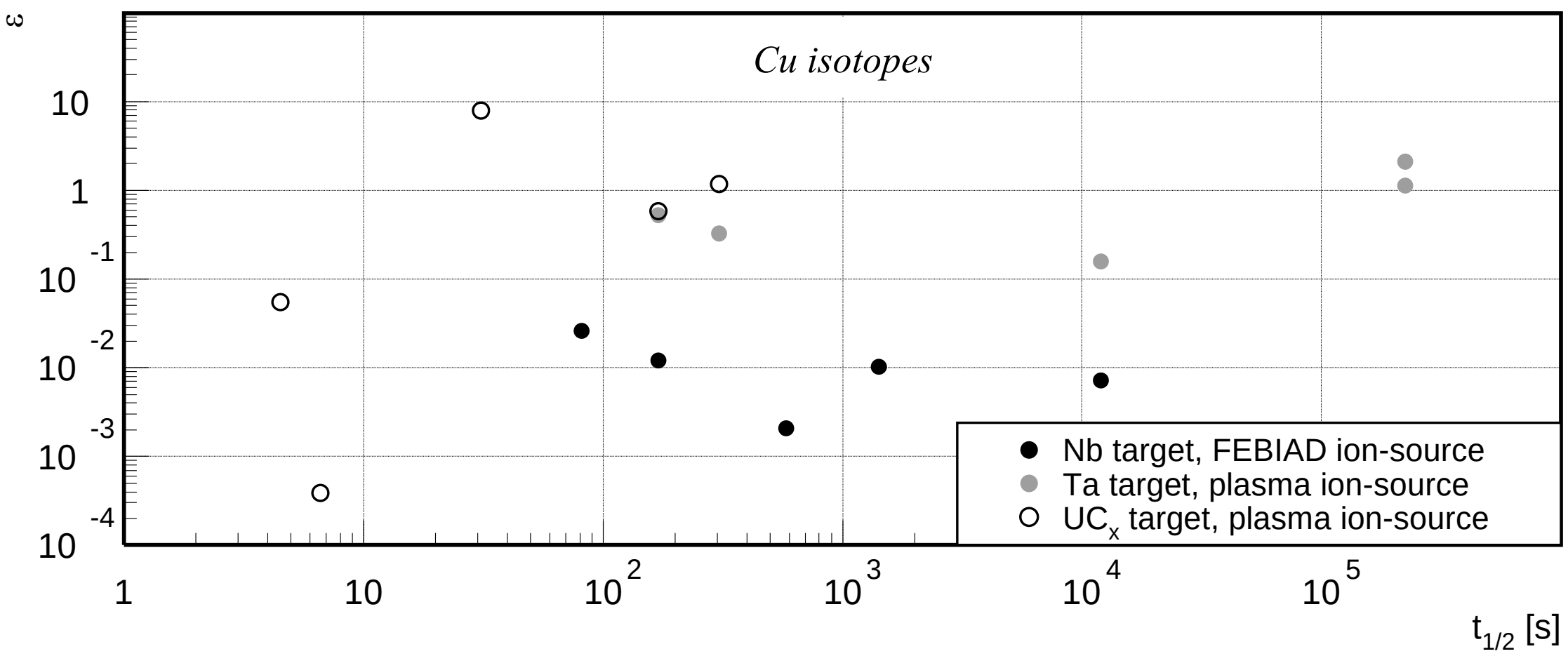

**Figure 48: Overall extraction efficiencies for Cu isotopes from several different targets. Note that the data for each of the targets cover only a limited range in isotopic half-lives.**

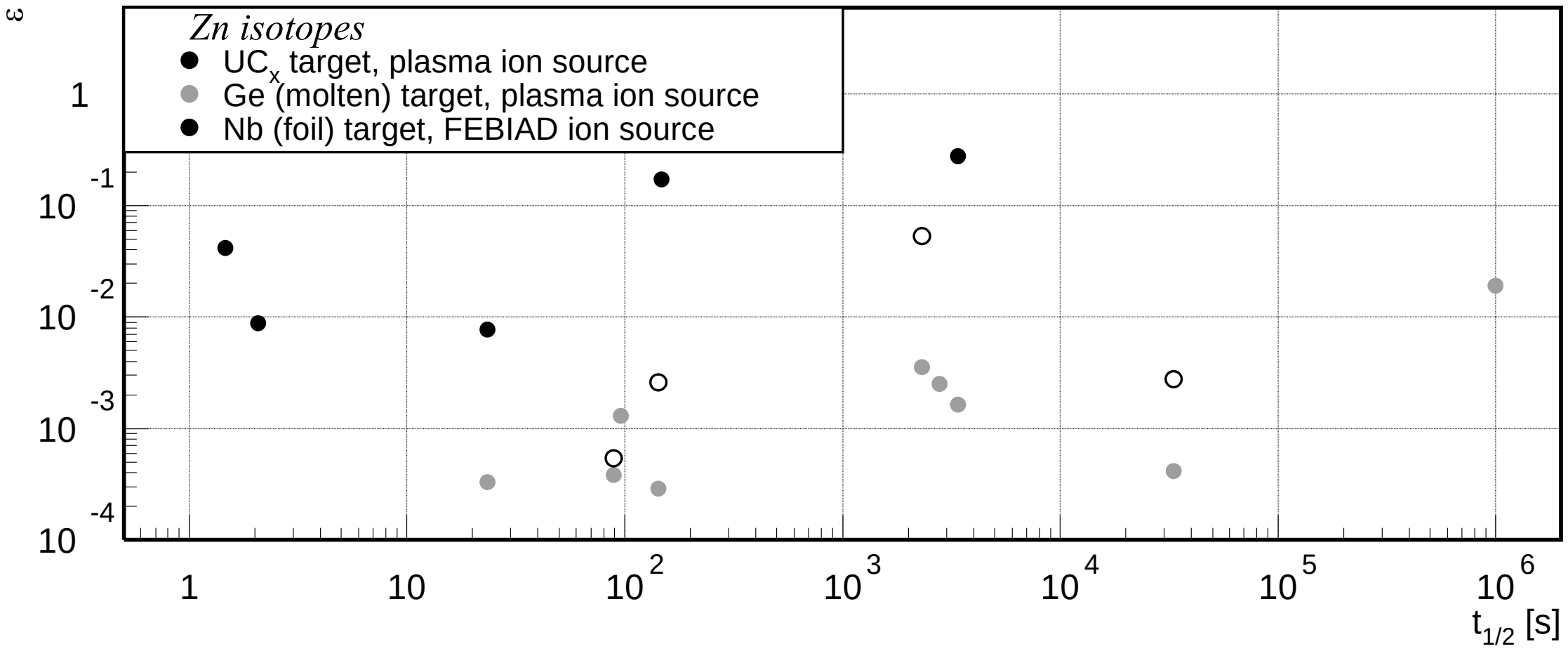

**Figure 49: Overall extraction efficiencies for Zn isotopes from several different targets.**

Other cases where the efficiency function could not be determined include:

- Manganese in niobium, tantalum and uranium carbide
- Zinc in niobium and uranium carbide
- Gallium in niobium, tantalum and uranium carbide
- Arsenic in zirconium oxide and uranium carbide
- Bromine in the thorium/niobium mixed powder target
- Rubidium in zirconium
- Strontium in zirconium
- Yttrium in niobium
- Silver in tantalum and uranium carbide
- Indium in tantalum and uranium carbide
- Tin in tantalum and uranium carbide
- Antimony in uranium carbide
- Xenon in lanthanum, lanthanum carbide and lead
- Cesium in thorium carbide
- Radon in thorium carbide and thorium oxide

For most of the elements listed here, the extraction efficiency seems uncorrelated with the isotopic half-life regardless of the target. This might indicate difficulties with chemical separation of these elements from isobaric contaminants, or difficulties with nuclide recognition in yield measurements.

## 8. Conclusions

In this work we have compared measured ISOLDE yields with new and high-quality information on the nuclide production cross sections in proton-induced reactions. This procedure reveals important properties of the nuclide extraction efficiencies, which is important for the predictions of the beam intensities at future ISOL facilities. It was found that the extraction efficiency and the isotope half-life are well correlated. This correlation follows a general pattern in cases of many different elements extracted form different targets. The dependence of the extraction efficiency on the isotope half-life can be parameterized using an analytical function with three parameters that depend on the element being extracted, as well as on the target-ion source system. These parameters summarize most important properties of the extraction efficiency for the isotopic chain and the target – ion source system in question, namely, the efficiency value in the limit of long half-lives, the exponent of the power-function behavior for the short half-lives and the characteristic value of the half-life around which the transition from the power-function to the constant behavior takes place.

The value of the extraction efficiency in the limit of long half-lives obtained using the method employed in this work directly indicates overall losses that occur apart from the decay losses. These can include losses due to chemical reactions, leaks or atoms escaping without being ionized. This parameter can be reliably estimated in most cases, especially when there is information on the yields of a large number of long-lived isotopes. For alkalis it is usually very high, often reaching 100%. For lighter alkaline earths it is in the percent range, and several tens of percent for heavier ones. For halogens it is usually in the percent range, sometimes going significantly below it, which reflects difficulties with negative ionization. For noble gases, it starts from 1% for neon, grows for heavier elements up to 100% for xenon.

The parameters $t_0$ and $\alpha$ describe the effect of the decay losses with short half-lives. These losses can reduce the yields of very short lived isotopes by several orders of magnitude. The values of $t_0$ and $\alpha$ seem to be very sensitive to systematic uncertainties stemming from the sources outlined in the previous section. The parameter $t_0$ ranges from the order of one second to several hundred seconds. It is the shortest for alkalis and noble gases, indicating fast extraction of these two groups. The exponent $\alpha$ is usually between 1 and 2. As expected, low-density targets, like carbides, exhibit faster extraction, and high-density ones, like liquid targets, are slower.

We have seen that, in some cases, it was impossible to parameterize the dependence of the yield-to-in-target-production ratios on the isotope half-lives by the function 14. However, the large number of cases where the proposed parameterization works well, as well as the basic understanding of the extraction process and of the decay losses, led us to conclude that this parameterization indeed reflects the normal behavior of the yield-to-in-target-production ratios with varying isotopic half-lives in the absence of large systematic errors.

This study provides quantitative estimates of the essential characteristics of the overall extraction efficiencies for a wide variety of isotopic chains from different target and ion-source systems. This information is helpful for completing our understanding of the efficiencies of the ISOL method, particularly from the practical point of view. For example, the magnitude of the overall losses of nuclides in processes like chemical reactions, sticking to the walls or condensation is difficult to measure or to estimate independently while, on the other hand, the most important information for practical applications is their overall effect that is revealed by this kind of study. Besides, the information on the behavior of the overall efficiencies with short half-lives can help identifying the issues that need most attention in the process of target and ion-source development. This kind of study is, in principle, applicable across the entire table of elements and for all target and ion-source systems.

## 9. Acknowledgments

We acknowledge the financial support of the European Community under the FP6 "Research Infrastructure Action – Structuring the European Research Area" EURISOL DS Project Contract no. 515768 RIDS. The EC is not liable for any use that may be made of the information contained herein.

We are grateful to Ulli Köster for very helpful remarks and discussion about specific ISOLDE data points.